\documentclass[11pt]{article}
\usepackage[dvips]{graphicx}
\usepackage{latexsym}
\def\Comment#1{}
\newcommand{\bean}{\begin{eqnarray*}}
\newcommand{\eean}{\end{eqnarray*}}
\newcommand{\gapproxeq}{\lower
.7ex\hbox{$\;\stackrel{\textstyle >}{\sim}\;$}}
\newcommand{\lapproxeq}{\lower
.7ex\hbox{$\;\stackrel{\textstyle <}{\sim}\;$}}

\newcommand\lsim{\mathrel{\rlap{\lower4pt\hbox{\hskip1pt$\sim$}}
    \raise1pt\hbox{$<$}}}
\newcommand\gsim{\mathrel{\rlap{\lower4pt\hbox{\hskip1pt$\sim$}}
    \raise1pt\hbox{$>$}}}
\newcommand{\ba}{\begin{array}}
\newcommand{\ea}{\end{array}}
\newcommand{\nn}{\nonumber}

\newcommand{\be}{\begin{equation}}
\newcommand{\ee}{\end{equation}}
\newcommand{\bear}{\begin{eqnarray}}
\newcommand{\eear}{\end{eqnarray}}
\newcommand{\tab}{\hspace*{0.5cm}}

\newcommand{\ket}{\,\rangle}
\newcommand{\bra}{\langle \,}
\newcommand{\eqn}[1]{(\ref{#1})}
\newcommand{\cO}{{\cal O}}
\newcommand{\bel}[1]{\be\label{#1}}

\newcommand{\mA}{\mathcal{A}}
\newcommand{\mB}{\mathcal{B}}

\newcommand{\mN}{\mathcal{N}}

\newcommand{\Frac}[2]{\frac{\displaystyle #1}{\displaystyle #2}}
\newcommand{\Int}{\displaystyle{\int}}

\newcommand{\Oint}{\displaystyle{\oint}}
\begin{document}
\thispagestyle{empty}
\begin{titlepage}
\begin{center}
\hspace*{10cm} {\bf  } \\
\hspace*{10cm} {\bf   }\\
\vspace*{2.75cm} 
\begin{Large}
{\bf 
Spin--1 Correlators  at Large $N_C$: 
\\
Matching OPE and Resonance Theory up to $\cO(\alpha_s)$}
 \\[2.4cm]
\end{Large}
{ \sc J.J. Sanz-Cillero } \\[0.8cm]

{\it 
%Institut de Physique Nucl\'eaire
Groupe Physique Th\'eorique, IPN Orsay
   \\[0.5cm]
Universit\'e Paris-Sud XI,  91406 Orsay, France
%
%
%IN2P3, IPN--Orsay
%   \\[0.5cm]
%     IPN - 15, rue Georges Clemenceau
%    \\[0.5cm]
%     91406 ORSAY CEDEX (FRANCE)
   \\[0.5cm]
e-mail: cillero@ipno.in2p3.fr
\\[0.5cm]}

\today  
\vspace*{2cm}
\begin{abstract}
\noindent
The relation between the quark--gluon description of QCD 
and the hadronic picture is studied up to order $\alpha_s$.  
The analysis of 
the spin--1 correlators is developed  
within the large $N_C$ framework. 
Both   representations are shown to be 
equivalent in the euclidean domain, where the Operator
Product Expansion is valid.   
By considering different models for the hadronic 
spectrum at high energies, 
one is able to recover the  $\alpha_s$ running in the correlators, to
fix the $\rho(770)$ and $a_1(1260)$ couplings, and to produce a prediction for
the values of the condensates.  
The Operator Product Expansion is improved by the large $N_C$ resonance theory,
extending its range of validity.  
Dispersion relations are employed in order to study the minkowskian region  and 
some convenient sum rules,    
specially sensitive to the resonance structure of
QCD, are worked out.   
A first experimental estimate of these sum  rules     
allows a cross--check of 
former  determinations of the QCD parameters and helps to discern and
to discard some of the considered hadronical models.  
Finally, the truncated resonance theory and 
the  Minimal Hadronical Approximation arise as a natural 
approach to the full resonance theory, not as a model.

\end{abstract}
\end{center}
\vfill

\eject
\end{titlepage}

\pagenumbering{arabic}

\parskip12pt plus 1pt minus 1pt
\topsep0pt plus 1pt
\setcounter{totalnumber}{12}

\section{Introduction}
\tab From many evidences, 
Quantum Chromodynamics (QCD) has been shown to be the proper theory 
for the strong interactions~\cite{QCDorigin,QCDPich}. 
The Operator Product Expansion (OPE)  has 
resulted a very powerful and successful 
instrument to describe  the amplitudes in the  
domain of deep euclidean 
momenta~\cite{OPEWilson,OPE,Shifman,OPEVincenzo,piaqq,narisonpid,SRFriot,
SRBijnens,YndurainAGG}. 
However, in the low energy region, 
the theory in terms of quarks and gluons becomes highly non perturbative  and 
these degrees of freedom  get confined within complex hadronic states. 
Likewise, the extrapolation of the euclidean OPE information to the range of
minkowskian momenta is highly non-trivial.

In the large $N_C$ limit 
--being $N_C$ the number of colours--, QCD suffers 
large simplifications ~\cite{NC1,NC2,NC3}.  
This limit of QCD   will be denoted as $QCD_\infty$ and  
it  turns out to be a very useful tool to understand many features in real QCD,  
providing an alternative power counting to describe the hadronic interactions.  Taking $N_C\to\infty$,
keeping $\alpha_s N_C$ fixed, there exists a systematic expansion of the
$SU(N_C)$  gauge theory in powers of $1/N_C$, which for  $N_C=3$ provides a
good quantitative approximation scheme to the hadronic world. 
Assuming  confinement at $N_C\to\infty$,    
$QCD_\infty$ is equivalent to a theory  with an infinite
number of hadronic states where  
the processes are then given by the tree-level 
exchange of an infinite number of resonances.

In this paper, we study  the spin--1   correlators,
\be
(q_\mu q_\nu\, - \, q^2 g_{\mu\nu}) \, \Pi_{_{XY}}(q^2)\, = \, 
i\, 
\Int dx^4 \, e^{iqx} \, \bra T\{J_{_{X}}(x)_\mu \, 
J_{_{Y}}(0)_\nu^\dagger \}\ket 
\, ,
\ee  
with $X/Y=V,A$ and the currents $J_{_V}^\mu=\Frac{1}{2}\bar{u}\gamma^\mu
u-\Frac{1}{2}\bar{d}\gamma^\mu d$ and  
$J_{_A}^\mu=\Frac{1}{2}\bar{u}\gamma^\mu \gamma_5 u
-\Frac{1}{2}\bar{d}\gamma^\mu \gamma_5 d$. 
Only the  sector of light quarks $u/d/s$ will be considered and  
we will work under the chiral and large $N_C$ limits. 
We will analyse the $V+A$ and $V-A$ combinations, 
$\Pi_{_{LL}}=\Pi_{_{VV}}+\Pi_{_{AA}}$ and  
$\Pi_{_{LR}}=\Pi_{_{VV}}-\Pi_{_{AA}}$ respectively.

Dispersion relations are nowadays a widely employed method 
to relate the theoretical OPE results in the euclidean domain 
with the available  experimental data $\frac{1}{\pi}$Im$\Pi(t)^{^{exp}}$ 
in the positive energy
region~\cite{SRorigin,
Weinbergsumrules,Adler,DRsumrules,DRgaussian}. 
In {\bf Section~2}, a 
pair of alternative sum-rules are presented, 
providing a comparison and cross-check of former dispersive 
determinations like  Laplace or pinched sum-rules.   
\\
\tab  
We  introduce first the usual moment integrals
$\mA^{^{(n)}}(Q^2)$, which give a largest weight to the low
energy region ($t\ll Q^2$), suppressing the high energy range.
However, through the  introduction of the Legendre
polynomials into our sum rules, we may build some particular combinations of the
moment integrals, namely $\mB^{^{(k)}}(Q^2)$,  which 
enhance both low and high energies ($t\ll Q^2$ and $t\gg Q^2$)  
and produce 
a stronger suppression on the intermediate region. Hence, we are able to 
use at the same time information from the experimental 
data (low momenta) and perturbative QCD  
(expected to work at $t\to \infty$). 
\\
\tab 
On the other hand, we will also consider the average  
$\frac{1}{\pi}$Im$\overline{\Pi}(z)$
of the spectral function $\frac{1}{\pi}$Im$\Pi(t)$ through some  rational 
distributions $\xi_a$, peaked around $t\sim z$ with a given dispersion 
$(\Delta t)_{_{\xi_a}}$ which suppresses the outer regions. 
This is an analogous procedure to the Gaussian sum-rules~\cite{DRgaussian}   
and, in the limit $(\Delta t)_{_{\xi_a}}\to 0$, the 
average would recover the value of the amplitude at $t=z$. 
The advantage of our distributions $\xi_a$  
is that they only depend on the first moment integrals, 
still under theoretical control; the influence of higher moments is killed.  
Unfortunately, although one may prove  that 
$\frac{1}{\pi}$Im$\overline{\Pi}(z)$ follows an OPE-like power behaviour, 
narrower and narrower distributions require a 
more  precise knowledge of the higher dimension condensates and their anomalous
dimensions.  
In addition, the appearance of duality violating 
terms that cannot be analytically expanded around 
${Q^2\equiv-q^2\to +\infty}$ 
may yield observable contributions that are 
dropped off by the OPE~\cite{dualviolation,Cataduality}.

In {\bf Section~3}, former OPE calculations are revisited under the perspective
of these sum rules.  
An analysis of the amplitudes in purely perturbative QCD (pQCD)   
is also performed, with all the condensates and duality violating terms  
set to zero.   
The range of validity of the OPE and pQCD is reduced to the first moments,
diverging once we go to higher orders. Matching  the averaged 
correlator $\frac{1}{\pi}$Im$\overline{\Pi}_{_{LR}}(z)^{^{OPE}}$   
to the experimental one requires accurate information about the condensates of
high dimension. Nonetheless, 
in the $V+A$  case, one finds that pQCD seems to work fine for energies up to 
$z\sim 1$~GeV$^2$, pointing out the more reduced 
impact from the OPE condensates in this channel.  
The  phenomenological analysis of the experimental 
data~\cite{Aleph,CLEO,OPAL,Amendolia,Novo2000,PIVV}   
in order to determine 
the OPE parameters ($\alpha_s$ and the condensates) is relegated to a next
work. An alternative derivation seems relevant since there is  
still some controversy on the values of the higher dimension  
condensates~\cite{OPEVincenzo,piaqq,narisonpid,SRFriot,SRBijnens}.

In {\bf Section~4}, we study large $N_C$ QCD and its manifestation 
into a meson theory with an infinite number of narrow-width resonances  
($R\chi T^{^{(\infty)}}$).  
The $\chi$ denotes that our hadron theory must be built up  chiral 
invariant in order to ensure the right low energy 
dynamics~\cite{WE:79,chpt,therole,quantumloops}, 
although  this detail is not relevant for
the present work.   
First of all, a resonance theory dual to QCD must recover the free-quark
logarithm in $\Pi_{_{LL}}(q^2)$ and the $1/Q^{2m}$ OPE structure in
$\Pi_{_{LR}}(q^2)$~\cite{NC2,Cataduality,EspriuRegge,
toymodel,LVS,Beanecutoff,Periscutoff,PerisRegge,matchingDR,711}. 
The novelty of this work is to
introduce the conditions required  
to recover the $\cO(\alpha_s)$ running in $\Pi_{_{LL}}(q^2)$.  
Reproducing  the $\alpha_s^2$ 
logarithmic dependence of the condensate anomalous dimensions is,
however, 
a rather complicate  problem that goes beyond this work.     
We will make the identification $QCD_\infty=R\chi T^{^{(\infty)}}$ 
since the resonance theory recovers the OPE in the euclidean domain 
providing, in addition, further information of QCD. 
For instance, Duality  violating terms $\exp{\left[-\rho\,\, Q\right]}$ 
lacking in the OPE ($Q\equiv \sqrt{-q^2}$) 
can be handled and the positive~$q^2$ range  becomes accessible.
\\
\tab 
Once the analysis is taken up to $\cO(\alpha_s)$, 
one is aware that two different  
energy regimes must be considered; the spectral function
$\frac{1}{\pi}$Im$\Pi(t)$ will be split into a perturbative part with $t$
greater than some separation scale $t_p\sim 2$~GeV$^2$, responsible of
the pQCD behaviour, and a non-perturbative part with $t<t_p$, essential
to recover the right OPE $1/Q^{2m}$ structure.  
In the resonance picture, a similar splitting is required. 
The infinite resonance summation  in the spectral function is also  
separated into a perturbative and a non-perturbative  sub-series.  
The perturbative sub-series is fixed by pQCD, once a model for the asymptotic 
spectrum of meson masses $M_n^2$ is assumed. Due to $\cO(\alpha_s)$ corrections,
the  parameters of the light resonances  in the non-perturbative range 
may suffer important variations with
respect to the asymptotic behaviour of the spectrum. 
They will be fixed through a
short-distance matching to the OPE.
\\
\tab 
In {\bf Section~4.3}, some available model for the resonance
mass spectrum are 
studied~\cite{Cataduality,EspriuRegge,Son,
Com,toymodel,LVS,Beanecutoff,Periscutoff,PerisRegge},  
getting a set of predictions for the
$\rho(770)$ and $a_1(1260)$ parameters, together with the OPE condensates of
dimensions four and six in $\Pi_{_{LL}}(q^2)$ and $\Pi_{_{LR}}(q^2)$ 
respectively.  
Through a five-dimensional model~\cite{Com},  
we exemplify how the resonance models  
implicitly include the OPE information together with the duality violating
terms. A more exhaustive 
analysis have been recently done for other models in Ref.~\cite{Cataduality}.
\\
\tab 
In {\bf Section~4.4}, the Minimal Hadronical Approximation 
(MHA) at large 
$N_C$~\cite{KPdR,therole} arises
naturally as a low energy theory of $R\chi T^{^{(\infty)}}$ where 
the infinite series of mesons is truncated.  The 
lightest resonance parameters encode the $1/Q^{2m}$ 
information coming from the
larger mass states.  A very successful phenomenology already exists at large
$N_C$~\cite{therole,spin1fields,PI:02,3puntos,3puntosandchiral,
extrachiral1,extrachiral2,multipuntosextra}. 
This framework has allowed the developing of  
robust calculations at next-to-leading order in
$1/N_C$~\cite{quantumloops,CP:01,Natxoprepara,BGT:98,
ChR:98,Meissnerloops,HLS}, achieving a good control of the 
final state interactions~\cite{PI:02,
anchura,GP:97,PP:01,JOP,Palomar,ND,GPP:04,PaP:01}.

In {\bf Section~5}, the dispersion relations developed before are applied to
$R\chi T^{^{(\infty)}}$.  Through the usual moment integrals
$\mA^{^{(n)}}(Q^2)$   we show the
equivalence with pQCD
and the OPE. The real improvement of $R\chi T^{^{(\infty)}}$
with respect to to the  OPE  appears manifestly through the 
$\mB^{^{(k)}}(Q^2)$ sum rules.  The physical components 
$\mB^{^{(k)}}(Q^2)^{^{exp}}$ 
oscillate as $k$ grows, damping off beyond some $k$, whereas the OPE yields a
divergent non-oscillating behaviour. The large $N_C$ resonance theory 
naturally reproduces the oscillation although it never vanish since the states
own zero-widths. However, one finds a pretty good agreement with the
phenomenology for the first components, where the damping is still not present. 
This allows considering in {\bf Section~6} 
the averaged amplitudes
$\frac{1}{\pi}$Im$\overline{\Pi}(z)$ and the exploration of the different models
for the spectrum. We are  actually sensitive to the asymptotic
behaviour of  the  mass spectrum $M_n^2$, being some hadronical models more
favoured by the phenomenology.

The paper is, therefore, separated in three differentiated parts:
In Section~2, we introduce the theoretical tools. In Section~3, we revisit 
the general features of QCD within the OPE and pQCD frameworks. 
Finally, Sections~4,~5~and~6 are devoted to the study of the large $N_C$
resonance description and its connection with the experimental data and the OPE.   
In  Section~7, the results are summarised and some final
conclusions are extracted.

\section{Dispersion relations in QCD correlators}
\tab
The perturbative calculation of the vector correlator at lowest order 
in  the $\alpha_s$ expansion, $\cO(\alpha_s^0)$, is provided by the free--quark
loop.  In dimensional regularization one has:  
\bel{eq.PIasin1}
\Pi_{_{VV}}(q^2)\, = \, - \,  \Frac{N_C}{24 \pi^2}
\, \left[\lambda_\infty \,  + \, \ln{\Frac{-q^2}{\mu^2}} \right] 
\, + \, \, \cO(\alpha_s)\, 
\ee
with the logarithmic ultraviolet divergence 
$\lambda_\infty(\mu)=\Frac{2\mu^{d-4}}{d-4}+\gamma_E-\ln{4\pi}
%= \frac{2}{d-4}+\gamma_E-\ln{4\pi}+\ln{\mu^2}+\cO(d-4)
$, being
$\gamma_E\simeq 0.5772$ the Euler constant and   $\mu$ the energy scale
introduced in the renormalization procedure.

One useful way to get rid of the renormalization ambiguity 
and the ultraviolet divergences is through the Adler-function~\cite{Adler}. 
\be
\mA_{_{VV}}(Q^2) \, = \, - \, \Frac{d}{d\ln{q^2}} \Pi_{_{VV}}(q^2)\, , 
\ee
with $Q^2\equiv -q^2$. 
This function carries the whole  information of the vector correlator, 
which can
be reconstructed through
\be
\Pi_{_{VV}}(q^2)\, = \, \Pi_{_{VV}}(q_0^2) \, 
- \, \displaystyle{\Int_{z=-q_0^2}^{z=-q^2}} \Frac{dz}{z} \, 
\mA_{_{VV}}(z) \, \, ,
\ee
once a given renormalization prescription $\Pi_{_{VV}}(q_0^2)$ is provided.

\subsection{Moments  $\mA^{^{(n)}}(Q^2)$ of a correlator}
\tab
In general, for any given correlator $\Pi(q^2)$, 
it is possible to consider a set of more general  moment integrals with a 
larger number of derivatives~\cite{DRsumrules}:
\be
\mA^{^{(n)}}(Q^2)\,\,\,\equiv \,\,\, 
\Frac{1}{n!}\, \left(-q^2\right)^n \,\, \left[ \Frac{d}{dq^2}\right]^n 
\Pi(q^2)\, ,
\ee
where by construction one includes the correlator 
$\mA^{^{(0)}}(Q^2)=\Pi(q^2)$ and 
$\mA^{^{(1)}}(Q^2)=\mA(Q^2)$ the  usual Adler function.

These functions are related with the imaginary part of the correlator through
the dispersion relations
\bel{eq.dispersiveA}
\mA^{^{(n)}}(Q^2) \,\,\,=\,\,\, 
\Int_0^\infty \, \Frac{dt \,\, Q^{2n}}{(t+Q^2)^{n+1}}\,\, 
\Frac{1}{\pi} \mbox{Im}\Pi(t)
\,\, , 
\ee
with a large enough  number of subtractions so the integral
is convergent ($n\geq 1$ for the vector and axial correlators).

Eq.~\eqn{eq.dispersiveA} can be written in a slightly different way through 
the change of variable $x=\Frac{t-Q^2}{t+Q^2}$: 
\bel{eq.dispersiveB}
\mA^{^{(n)}}(Q^2) \,\,\,
=
\,\, \Int_{-1}^{+1}\, dx \,\, \Frac{(1-x)^{n-1}}{2^n} \,\, \, 
\cdot \,\,\, \, 
\Frac{1}{\pi}\mbox{Im}\Pi\left[Q^2\left(\Frac{1+x}{1-x}\right)\right] 
\,\, ,
\ee
with $n=1,2,...$

The moment integrals of the spin--1 correlators with $n\geq 1$    
are therefore  physical quantities, free of ultraviolet divergences;  
on the contrary to the  correlator, the $\mA^{^{(n)}}(Q^2)$ are finite and 
renormalization scale independent.

Eq.~\eqn{eq.dispersiveB} shows that the moment integrals 
are simply the projections of the function 
\be
\sigma_{z}(x)\,\,\, \equiv \,\,\,   
\Frac{1}{\pi}\mbox{Im}\Pi\left[z\left(\frac{1+x}{1-x}\right)\right]\, ,
\ee 
in the different directions of the non-orthogonal basis  
of polynomials 
$\left\{p_n(x)\equiv\frac{1}{2^{n}}(1-x)^{n-1}\right\}_{n=1}^\infty$  
of the Hilbert space of real functions $L^2(-1,+1)$,  
with the scalar product 
$\bra f\, |\, g\ket=\Int_{-1}^1dx \,\, f(x) \,\,  g(x)$: 
\be
\mA^{^{(n)}}(z)\,\, \,\,=\,\,  \,\,
\bra \,\, p_n \,\, | \,\, \sigma_{z} \,\, \ket   
\,\, .
\ee

\subsection{Orthonormal decomposition $\mB^{^{(k)}}(Q^2)$ of the correlator}
\tab
The non-orthogonal basis $\{p_l(x)\}$ is not very convenient in order
to recover the absorptive part of the correlator. 
We can rewrite the observable $\sigma_z(x)$ in terms of the orthonormal basis 
provided by the Legendre polynomial $P_k(x)$  ($P_1(x)=1, \, P_2(x)=x ...)$: 
%$\left\{g_k(u)\equiv \frac{\sqrt{2k-1}}{(1+u)}
%P_{k-1}\left(\frac{1-u}{1+u}\right)\right\}_{k=1}^\infty$. 
\be
g_k(x) \,\,\,\, \equiv \,\, (-1)^{k-1}\,\, 
\sqrt{\Frac{2k-1}{2}}\,\,  P_{k-1}(x) 
\,\, \,\,=\,\, \,\, 
\sum_{l=1}^k  \, M^{^{\mB\mA}}_{kl}\, 
%\Frac{(1+x)^{l-1}}{2^l} 
\,\,p_l(x)\, ,
\ee
with $k=1,2,...$ and 
related to the former basis $\left\{ p_l(x)\right\}_{l=1}^{\infty}$ 
through some given constants $M^{^{\mB\mA}}_{kl}$. 
The vectors of the new basis obey 
$\Int_{-1}^{+1} dx \,\,  g_{m}(x)\,\, g_n(x)\, =\,
\delta_{m,n}$. 
This provides for the imaginary part of the correlator the spectral
decomposition
\bel{eq.impispectral}
\sigma_z(x)\, \, = 
% \, \Frac{1}{\pi}\mbox{Im}\Pi\left[z\left(\frac{1-x}{1+x}\right)\right]\, \, = 
%\,\,
%\left(\Frac{2}{1+x}\right)^{n_c-1}
%\,\,\, \left\{
 \sum_{k=1}^\infty 
\,\mB^{^{(k)}}(z) \,\, 
\cdot \,\,
g_k(x)
% \left[\sqrt{2k-1}/2 \, P_{k-1}\left(\frac{z-t}{z+t}\right)\right]
%\, \right\} 
\, ,
\ee
%with $x=(z-t)/(z+t)$ already replaced, and 
with the different components  given by the projections 
\bel{eq.cambio}
\mB^{^{(k)}}(z)\,\,\,\, 
=\,\,\,\, \bra \,\,g_k\,\,|\,\,\sigma_{z}\,\,\ket 
\,\,\,\,=\,\, \,\,\sum_{l=1}^k \,M^{^{\mB\mA}}_{kl} \, \,\,
\mA^{^{(l)}}(z) \,\, ,
\ee
where the $\mB^{^{(k)}}(z)$ depend just on the lowest 
moments. They can be calculated as well through the dispersion relation
\be
\mB^{^{(k)}}(z) \,\,\,\, =\,\,\,\, 
\Int_{-1}^1 \, dx \, \, g_k(x) \,\, \cdot \, \, \sigma_z(x) \,\,\,\, 
= \,\,\,\, (-1)^{k-1} \, \sqrt{\Frac{2k-1}{2}} \, 
\Int_0^\infty \, \Frac{2\, z \,\, dt}{(z+t)^2} \,\, 
\,\,P_{k-1}\left[ \frac{t-z}{t+z}\right] \,\, \cdot  \,\, 
\Frac{1}{\pi} \mbox{Im}\Pi(t) \,\, . 
\ee  

The components in the Legendre basis are bounded if the spectral function is
finite:
\be
\left| \mB^{^{(n)}}(z)\right| \quad \leq \quad 
\sqrt{\bra \,\, \sigma_z \,\,| \,\, \sigma_z\,\, \ket}
\,\, 
= \,\,\sqrt{\sum_{k=1}^\infty\left| \mB^{^{(k)}}(z)\right|^2 }
 \quad \leq \quad 
\sqrt{2} \,\,\, \mbox{max}\left\{|\sigma_z(x)|\right\} \, .
\ee

The difference with other dispersion relations  is  
the use of the Legendre polynomials to
pinch the dispersive integral. 
For the moments  $\mA^{^{(n)}}(Q^2)$  
one employs a weighted distribution which 
enhances the low energy region, decreases at the range  of 
intermediate momenta, and  vanishes at $t\to +\infty$.  
The  distribution for the components $\mB^{^{(k)}}(Q^2)$ is completely
different: The dispersion integral is enhanced 
both  around $t=0$ and $t\to +\infty$, introducing  
a strong suppression of the integrand at  intermediate energies around 
$t=Q^2$.  Since the experimental data only reaches up to some finite energy and
local duality is expected to work at very high energies,  this procedure allows
minimising the uncertainties due to the  absence of data at intermediate 
energies.  
The Legendre polynomials allows replacing the lacking data by
the pQCD minkowskian amplitude, 
reducing the impact of duality violations in the
transition  from the experimental data to  perturbative QCD.

\subsection{Spectral function reconstruction}
\tab
One extracts several conclusions from Eq.~\eqn{eq.impispectral}. 
First to notice is that the  expression is an identity for any $x$ and $z$,  
although   a partial knowledge on the moments  introduces wrong 
dependences.  The errors in  Eq.~\eqn{eq.impispectral}  
due to uncertainties  on the $\mB^{^{(k)}}(z)$ are smaller 
around $x=0$ ($t=z$) whereas   
large  fluctuations occur at the extremes of the
interval, $x=- 1$ and $x=1$  ($t=0$ and $t=+\infty$ respectively), 
where the Legendre polynomial reach their absolute maxima and minima,  
$P_k(\pm 1)=(\pm 1)^k$. 
Thus, the optimal point for  Eq.~\eqn{eq.impispectral} 
corresponds to $x=0$: 
\bel{eq.impispectral2}
\sigma_z(0)\,\,\,\, =\,\,\, \, \Frac{1}{\pi}\mbox{Im}\Pi(z)\, \,\,\, = 
\,\,\,\, \sum_{k=1}^\infty \,
\,\,\,\mB^{^{(k)}}(z) \,\, \, 
\cdot \,\,\, 
g_k(0)
\, ,
\ee
where one has $g_k(0)=0$ for $k$ even,   and  
$|g_k(0)|=\sqrt{\frac{2\,  k-1}{2}}\,  |P_{k-1}(0)|= 
\sqrt{\frac{2\,k-1}{2^{2k-1}}}
\frac{\Gamma(k)}{ \Gamma\left(\frac{k}{2}+\frac{1}{2}\right)^2} 
\in \left[\frac{1}{\sqrt{2}},\sqrt{\frac{2}{\pi}}
\right)$ 
for $k$ odd, 
with the signs provided by $(-1)^{\frac{k-1}{2}}$. $\Gamma(k)$ is the Euler
Gamma function.

The components $\mB^{^{(n)}}(z)$ related to a physical spectral function 
(which remains finite) become  smaller and
smaller at a certain $k$ since the norm $\bra \sigma_z\,|\, \sigma_z\ket$ is
bounded, and the series in Eq.~\eqn{eq.impispectral2} converges. 
For instance, 
the spectral function corresponding to the vector correlator 
in the free-quark limit 
($\Pi_{_{VV}}(q^2)=-\frac{N_C}{24\pi^2}\ln{\frac{-q^2}{\mu^2}}$) 
is easily reconstructed from its components 
$\mB_{_{VV}}^{^{(k)}}(z)=\frac{N_C\,
\sqrt{2}}{24\pi^2}\, \delta_{k,1}$.

The eventual knowledge of the components
$\mB^{^{(k)}}(z)$ at all orders in $k$ allows the exact recovering of the
spectral function at any energy, in particular at 
$q^2=z$. Actually, if one knows a large enough amount of components 
$\mB^{^{(k)}}(z)$, such that the remaining terms in the series of
Eq.~\eqn{eq.impispectral2} already converge, then it is possible to give an
estimate of $\frac{1}{\pi}$Im$\Pi(z)$. The truncation error would be 
provided by the size of the last components $\mB^{^{(k)}}(z)$.

The relation in Eq.~\eqn{eq.impispectral2} 
is stable, i.e., small variations on the 
$\mB^{^{(k)}}(z)$ produce tiny fluctuations on 
$\frac{1}{\pi}$Im$\Pi(z)$. This must not be
confused with the fact that tiny modifications on the value of the correlator at
$q^2=-z<0$ may produce (and produces) large instabilities on the tower of
components and, hence, on the time-like correlator.

\subsection{Averaged amplitudes}
\tab
In many situations the description of the amplitude in some energy range    
may be complicated from the theoretical point of
view. In these
cases, it is sometimes  
more convenient to consider the amplitude averaged through
some distribution  peaked around a given  energy, in the fashion of the
Gaussian sum rules~\cite{DRgaussian}.

We  will  perform  the   average   of   the  spectral  functions  
$\frac{1}{\pi}$Im$\Pi(t)$    in the $x$--space given by the change of variable 
${t=z\left(\frac{1+x}{1-x}\right)}$,  being  
  $z$ the energy of interest.  
The spectral function is then provided by   
${\sigma_{z}(x)=\frac{1}{\pi}}$Im${\Pi
\left[z\left(\frac{1+x}{1-x}\right)\right]}$. 
      The central point   $x=0$  of the averaging distribution $\xi_a(x)$   
corresponds to $t=z$. 
This mapping of $t$ allows a simpler analysis in terms of the moments. 
We consider the family of distributions
\be
\xi_a(x)\,\, \equiv \,\, 
\mN_a
\,\, (1 \, -\,x^2 )^\frac{a}{2} \, ,
\ee 
being $a>0$ an  even number 
and $\mN_a=\left[\frac{\Gamma\left(\frac{a}{2}+\frac{3}{2}\right)}{
\sqrt{\pi}\,\, \Gamma\left(\frac{a}{2}+1\right)}
\right]$ a constant that normalises the distribution to~1. 
This functions are centered at zero  
($\bra x\ket_{\xi_a}=0$) and have dispersion $(\Delta
x)^2_{\xi_a}=\frac{1}{a+3}$. Hence this distribution  covers the spectral
function  around
$t=z$ within an interval $\Delta t\simeq 2 \, z \, \Delta x$.

\begin{figure}[t!]
\begin{center}
\includegraphics[width=7.5cm,clip]{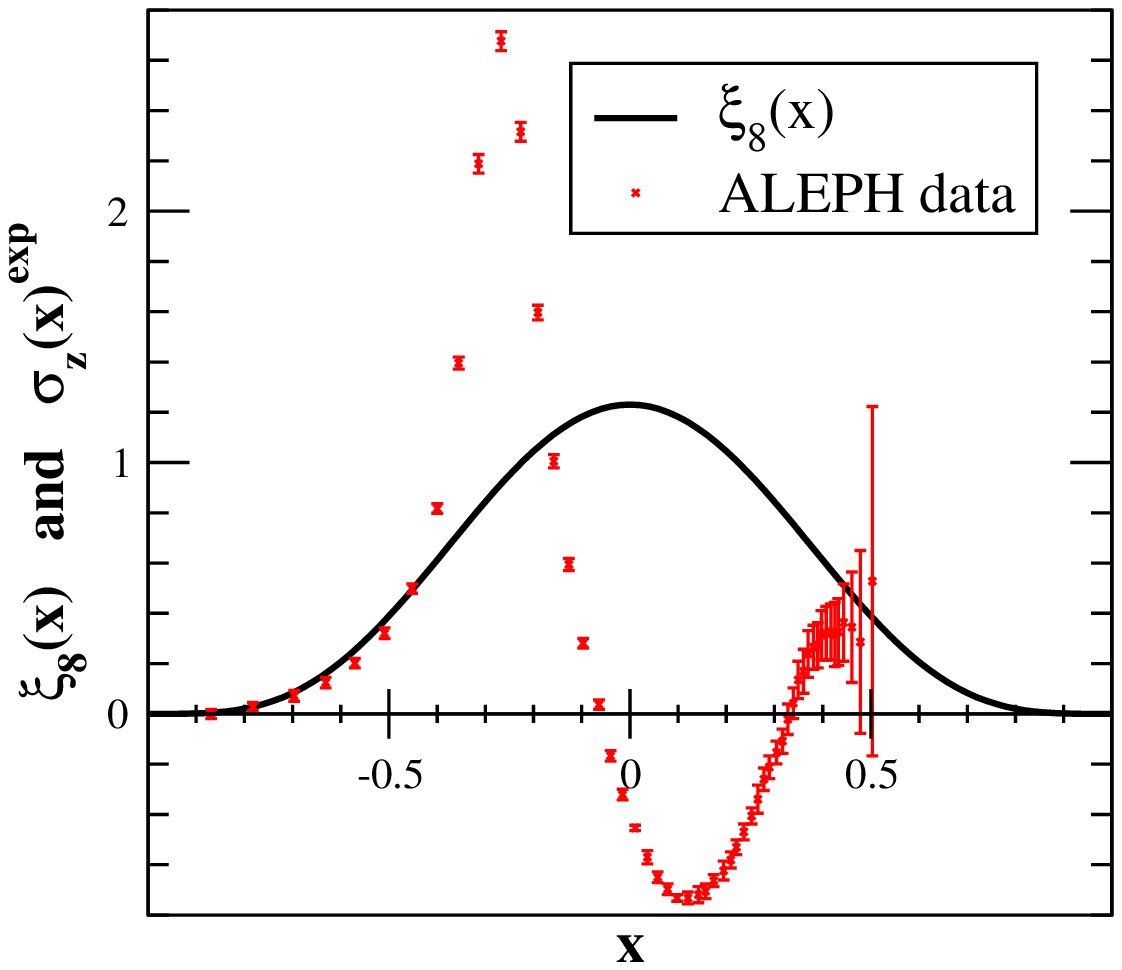}
\includegraphics[width=5.9cm,clip]{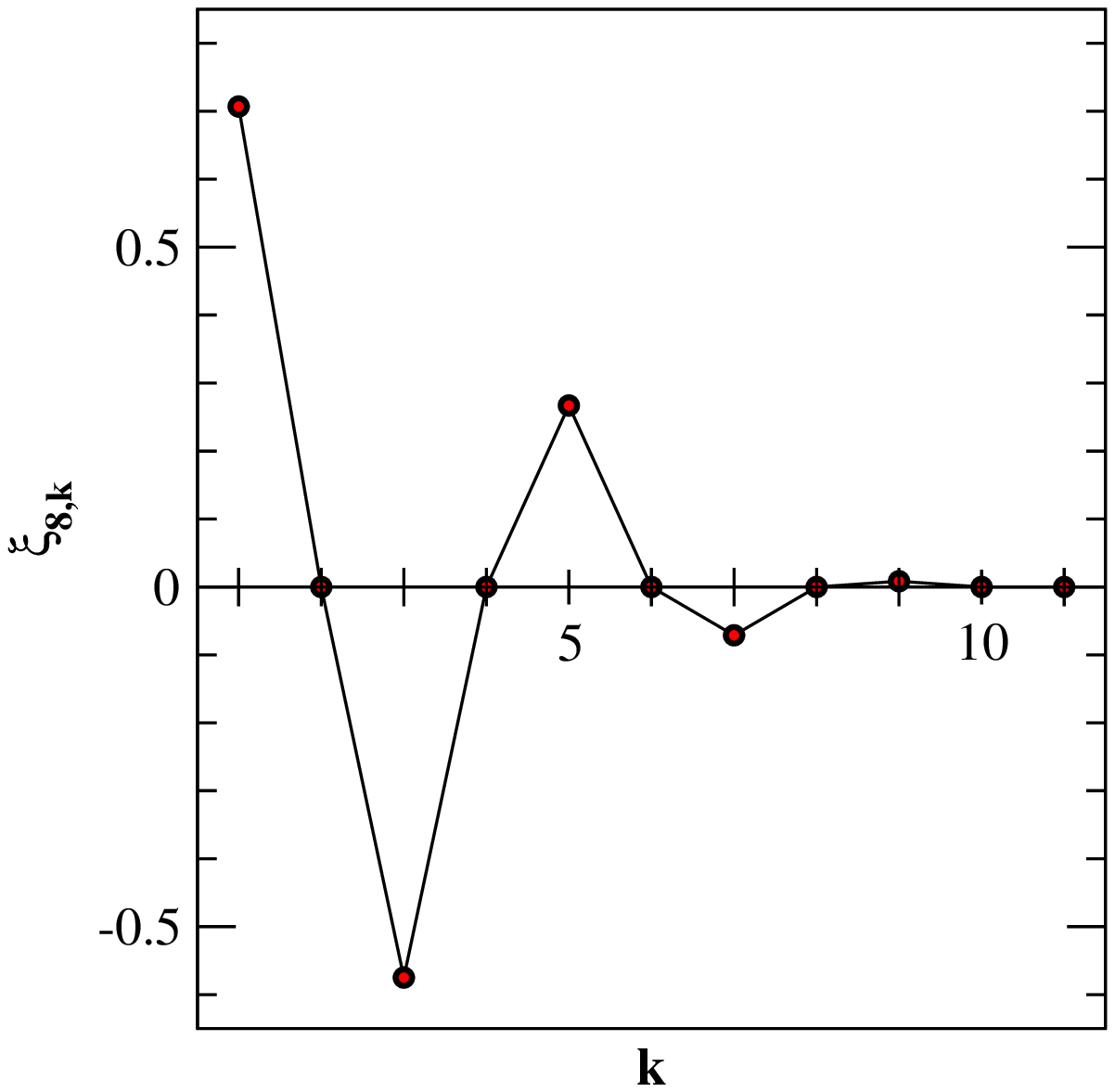}
\caption{Distribution  $\xi_a(x)$ for $a=8$. Its corresponding
components $\xi_{a,k}$ are shown  on the right-hand-side. 
The distribution is shown together with
the reference data $\sigma_z^{^{LR}}(x)^{exp}$~\cite{Aleph} for 
$z=1$~GeV$^2$.}
\label{fig.xi8}
\end{center}
\end{figure}

Since $\xi_a(x)$ is a polynomial of degree $a$, 
it accepts a decomposition in terms of the orthonormal basis of Legendre
polynomial $\left\{g_k(x)\right\}_{k=1}^\infty$:
\be
\xi_a(x)\,\,= \,\, \sum_{k=1}^{a+1} \, \xi_{a,k}\, \,\, g_k(x)\, ,
\ee
where $\xi_{a,1}=1/\sqrt{2}$ due to the normalization  
$\Int_{-1}^1  dx \, \,\, \xi_a(x)=1$, and 
the constant 
terms $\xi_{a,k}$ with even $k$ are zero due to the parity of $\xi_a(x)$. 
This distribution only  depends on the first $a+1$ Legendre polynomials.
In Fig.~\eqn{fig.xi8} we show the distribution and components of $\xi_8(x)$,
compared to the experimental data 
$\sigma_z^{^{LR}}(x)^{exp}$ for $z=1$~GeV$^2$~\cite{Aleph}. 
For the case with $a=8$ we have 
$\xi_{8,3}=-\frac{2\sqrt{10}}{11}, \, \xi_{8,5}=\frac{27\sqrt{2}}{143}, \, 
\xi_{8,7}=- \frac{2\sqrt{26}}{143}$ and~${\xi_{8,9}=\frac{7\sqrt{34}}{4862}}$.

The mean value of the spectral function  is defined through the average
\be
\Frac{1}{\pi}\mbox{Im}\overline{\Pi}(z)^{^{\xi_a}}\,\, \equiv \,\, 
\bra \sigma_z\ket_{\xi_a}\,\, = \,\, 
\Int_{-1}^{1} \, dx \, \, \sigma_z(x) \,\, \xi_a(x) \,  \, , 
\ee
where the orthonormal decomposition of the correlator in
Eq.~\eqn{eq.impispectral} yields 
\be
\Frac{1}{\pi}\mbox{Im}\overline{\Pi}(z)^{^{\xi_a}}\,\,= \,\, 
\sum_{k=1}^{a+1} \, \xi_{a,k}\, \,\, \mB^{^{(k)}}(z)\, \,  \, . 
\ee
Only the first $(a+1)$ moments are relevant for the 
amplitude averaged through $\xi_a$. 
This will be useful when our control on the high order components
is reduced.   

This procedure is analogous to the Gauss-Weirstrass transform of the
correlator,   
where the amplitude is averaged through a Gaussian
distribution~\cite{DRgaussian}. One would recover     
Im$\overline{\Pi}(z)^{^{\xi_a}}\to$Im$\Pi(z)$ in the limit when $a\to\infty$.    
Nonetheless, our theoretical control on the QCD components $\mB^{^{(k)}}(z)$ 
gets worse  as $k$ increases  and one
needs to go to high enough energies in order to make them reliable.

\section{Perturbative QCD and the operator product expansion}
\tab 
The Operator Product Expansion in  perturbative QCD      
provides a systematic procedure to compute the two-point Green 
functions at any order in $\alpha_s$ or operator 
dimension in the  deep euclidean regime $Q^2\equiv -q^2 \gg
\Lambda_{_{QCD}}^2$~\cite{OPEWilson,Shifman}. For
the spin--1 correlators one has
\be
\Pi(-Q^2)^{^{OPE}}\, \, = \,\, 
\,\,\,\, \bra{\cO}_{_{(0)}}\ket \,\, \,\, 
+ \,\,\,\, \sum_{m=2}^{\infty}\, \,
\Frac{\bra{\cO}_{_{(2m)}}\ket}{Q^{2m}} \,\, , 
\ee
where the coefficients   
$\bra \cO_{(2m)}\ket$ are provided by  the dimension--$(2m)$ operator 
in the OPE and they depend weakly on the momenta (only through
logarithms)~\cite{Shifman}.  In this work,   
the term $\bra \cO_{_{(0)}}\ket$ 
corresponds   to the identity operator in the OPE and yields the purely
perturbative QCD contribution (pQCD).   
The $V+A$ correlator  becomes   
${\bra \cO^{^{LL}}_{_{(0)}}\ket\stackrel{\alpha_s\to 0}{\longrightarrow} 
\Pi_{_{LL}}(-Q^2)^{^{free}}=-\frac{N_C}{12\pi^2}\ln{\frac{Q^2}{\mu^2}}}$   
in the free quark limit whereas 
for the $V-A$ Green function vanishes    for  any value of
$\alpha_s$ (${\bra \cO^{^{LR}}_{_{(0)}}\ket =\Pi_{_{LR}}(-Q^2)^{free}=0}$).

{\bf V-A correlator}
\\ \tab 
In the case of the $V-A$ correlator ($\Pi_{_{LR}}=\Pi_{_{VV}}-\Pi_{_{AA}}$) 
the OPE starts at the dimension six operator~\cite{Shifman}:
\be 
\Pi_{_{LR}}(-Q^2)^{^{OPE}}\,\,\,=
\,\,\,  \,\,\, \sum_{m=3}^\infty\Frac{\bra \cO_{_{(2m)}}^{^{LR}}\ket}{Q^{2m}} \, .  
\ee
At  \ high \  euclidean  \ momenta  \  the  \ correlator \  is  \ 
driven  \  by \   the  \ 
dimension \  six \  condensate,  \ 
with~${\bra \cO_{_{(6)}}^{^{LR}}\ket= \, - 4 \pi \alpha_s \bra \bar{q}q\ket^2}$ 
at large $N_C$.
We will not consider the anomalous dimensions of the condensates
$\bra \cO_{_{(2m)}}\ket$, which will be taken as constants. In this case, one gets 
for the moments a $1/Q^{2m}$ power structure, 
\be 
\mA_{_{LR}}^{^{(n)}}(Q^2)^{^{OPE}}\,\,\,=\,\,\, 
\,\,\,\,\,\,\, \sum_{m=3}^\infty \, \Frac{a^{^{LR}}_{_{(n,2m)}}}{Q^{2m}} \, , 
\qquad \mbox{ with } \quad 
a^{^{LR}}_{_{(n,2m)}}\, = \, 
\Frac{(m-1+n)!}{(m-1)!\,\, n!}\,\,  \bra \cO_{(2m)}^{^{LR}}\ket \, , 
\ee
and a similar thing happens for the components $\mB^{^{(k)}}(Q^2)$, 
\be
\mB_{_{LR}}^{^{(k)}}(Q^2)^{^{OPE}}\,\,\, \, = 
\,\,\, \, \,\,\,\,\sum_{m=3}^\infty \, \Frac{b_{_{(k,2m)}}^{^{LR}}}{Q^{2m}} \, ,
\qquad \mbox{ with } \quad 
b_{_{(k,2m)}}\,= \, \sum_{n=1}^k\, M^{^{\mB\mA}}_{k,n} \, \,\,
a_{_{(n,2m)}}^{^{LR}} \, , 
\ee
being the constants $M^{^{\mB\mA}}_{k,n}$ given by  
the basis transformation in Eq.~\eqn{eq.cambio} 
that relates $\mA^{^{(n)}}(Q^2)$ and $\mB^{^{(k)}}(Q^2)$. 
When considering  
truncated OPE series, both $\mA^{^{(n)}}_{_{LR}}(Q^2)$ 
and $\mB^{^{(n)}}_{_{LR}}(Q^2)$ diverge beyond some order $n$.

The spectral function,  related to the 
components $\mB^{^{(k)}}(Q^2)$  through  
Eq.~\eqn{eq.impispectral2}, formally shows the same power behaviour, 
\be
\Frac{1}{\pi}\mbox{Im}\Pi_{_{LR}}(z) \,\,\,
= \,\,\, \displaystyle{\sum_{m=3}^\infty} \, 
\Frac{ \frac{1}{\pi}\mbox{Im}\Pi^{^{LR}}_{_{(2m)}}}{z^{m}}\, ,
\qquad
\mbox{ with }\quad 
\frac{1}{\pi}\mbox{Im}\Pi^{^{LR}}_{_{(2m)}} \, = \, 
\displaystyle{\sum_{k=1}^\infty}\, b_{_{(k,2m)}}^{^{LR}}\,\cdot \, g_k(0)\, .
\ee
Unfortunately, within the OPE the terms $b_{_{(k,2m)}}$ in 
the coefficients 
$\frac{1}{\pi}\mbox{Im}\Pi^{^{LR}}_{_{(2m)}}$ go on growing 
as $k\to \infty$ and the summation diverges, 
preventing the theoretical determination of the spectral function.

{\bf  V+A correlator}
\\ 
\tab 
Considering \ the \  Renormalization  \ Group \  Equations  \ up \  to  \ $\cO(\alpha_s)$, 
 \ the  \ $V+A$  \ correlator  \  
(${\Pi_{_{LL}}=\Pi_{_{VV}}+\Pi_{_{AA}}}$)  
shows the structure~\cite{Shifman} 
\bel{eq.PILLRGE}
\Pi_{_{LL}}(-Q^2)^{^{OPE}} \,\, \, \, = 
-\, \Frac{N_C}{12\pi^2}\, \left[\, \ln{\Frac{Q^2}{\mu^2}} \, \, 
%\, + \,\,\, \Frac{4}{b}\, 
\, - \,\,\, 
\Frac{3 C_F}{2 \beta_1} \, 
\ln{\left(\Frac{\alpha_s(\mu^2)}{\alpha_s(Q^2)}\right)} \, \right] 
\,\,\, \, + \,\,\, \sum_{m=2}^\infty 
\Frac{\bra \cO_{(2m)}^{^{LL}}\ket}{Q^{2m}}\, ,
\ee
with  $C_F=\frac{N_C^2-1}{2\, N_C}$, 
and \   \ \ $\beta_1= - \frac{1}{6}(11\, N_C-2\, n_f)$  \  \ 
provided  \ \  by \  \  the  \  \  \ 
$\beta$--function \  \  \  at \  \  lowest \  \  order,  \  \  \ 
$\frac{d\ln{\alpha_s}}{d\ln{\mu}}=\beta(\alpha_s)
=\frac{1}{\pi}\beta_1\alpha_s+\cO(\alpha_s^2)$,  
being $\alpha_s(\mu^2)$ the  strong running coupling constant.

For pQCD, with all the condensates 
set to zero, one finds the moments 
\be
\mA_{_{LL}}^{^{(n)}}(Q^2)^{^{pQCD}}\,\,\,\, = \,\,\,\,
\Frac{1}{n} \, \Frac{N_C}{12\pi^2} \,
\left( 1\, + \, \Frac{3 C_F}{4} \, \Frac{\alpha_s(Q^2)}{\pi} \right) \,\,\,\, 
+ \,\,\, \cO(\alpha_s^2(Q^2)) \, , 
\ee
which provides the  components 
\bel{eq.compBLLpQCD}
\mB_{_{LL}}^{^{(k)}}(Q^2)^{^{pQCD}} \,\,\,\, = \,\,\,\, 
\sqrt{2} \, \Frac{N_C}{12\pi^2} \,
\left( 1\, + \, \Frac{3 C_F}{4} \, \Frac{\alpha_s(Q^2)}{\pi} \right) 
\,\, \delta^{k,1}\,\,\,\, 
+ \,\,\, \cO(\alpha_s^2(Q^2)) \, .
\ee
The perturbative expansion of pQCD  in powers of  $\alpha_s$  produces  
the spectral function 
\bel{eq.promPILLpQCD}
\Frac{1}{\pi}\mbox{Im}\Pi_{_{LL}}(z)^{^{pQCD}} \,\,\, \, 
= \,\,\,\, 
\sum_{k=1}^\infty \, \mB^{^{(k)}}(z)^{^{pQCD}} \, \cdot \, g_k(0) \,\,\,\, 
= \,\,\,\, \Frac{N_C}{12\pi^2} \,
\left( 1\, + \, \Frac{3 C_F}{4} \, \Frac{\alpha_s(z)}{\pi} \right) 
\,\, \, \, 
+ \,\,\, \cO(\alpha_s^2(z)) \, .
\ee
Nevertheless, once the higher dimension operators  are taken into account  
($\bra \cO_{(4)}^{^{LL}}\ket= \frac{1}{12\pi} \alpha_s \bra G_{\mu\nu}^a
G^{\mu\nu}_a\ket$,    
$\bra \cO_{(6)}^{^{LL}}\ket= \frac{8}{9}\pi \alpha_s\bra \bar{q} q\ket^2, ...$)
the series of $\mB^{^{(k)}}_{_{LL}}(z)$  becomes divergent  when
$k\to\infty$, as it happened before in the $V-A$ case.

\subsection{Divergence of the components $\mB^{^{(k)}}(Q^2)$ 
within the OPE framework}
\tab
Since the moments  of the truncated OPE amplitudes diverge at some point, 
it is important to study how and when this divergence occurs.   
Since  we are now interested on dealing with  finite spectral functions,  
the pion pole is removed from the correlators for the analysis in this section: 
\be
\Pi_{_{LR}}(-Q^2)^{^{no-\pi}}\,\,\, \equiv \,\,\, 
\left[\Pi_{_{LR}}(-Q^2)\, +\, \Frac{F^2}{Q^2}\right] 
\quad , \qquad \qquad 
\Pi_{_{LL}}(-Q^2)^{^{no-\pi}}
\,\,\, \equiv \,\,\, \left[\Pi_{_{LL}}(-Q^2)\, - \, \Frac{F^2}{Q^2}\right]\, , 
\ee
which yields the moments
\be
\mA^{^{(n)}}_{_{LR}}(Q^2)_{_{no-\pi}}=
\left[\mA_{_{LR}}^{^{(n)}}(Q^2)+\Frac{F^2}{Q^2}\right] 
\quad , \qquad \qquad 
\mA^{^{(n)}}_{_{LL}}(Q^2)_{_{no-\pi}}=
\left[\mA_{_{LL}}^{^{(n)}}(Q^2)-\Frac{F^2}{Q^2}\right]\, , 
\ee
with $F\simeq F_\pi=0.0924$~MeV. This allows working with 
squared integrable spectral functions $\sigma_z(x)^{^{no-\pi}}$,    
owning a convergent series of components $\mB^{^{(k)}}(z)_{_{no-\pi}}$.

In order to estimate the  physical components 
$\mB^{^{(k)}}(Q^2)_{_{no-\pi}}$, we show in 
Figs.~\eqn{fig.momOPELR}~and~\eqn{fig.momOPELL} 
the components obtained from the experimental ansatz, 
\be
\mbox{Im}\Pi(t)^{^{ansatz}}_{_{no-\pi}} \,\,\,\, = \,\,\,\, 
\mbox{Im}\Pi(t)^{^{exp}}_{_{no-\pi}} \, \cdot \, \theta(t_0-t)\,\,\, 
+ \,\,\, 
\mbox{Im}\Pi(q^2)^{^{pQCD}} \,\cdot \,  \theta(t-t_0) \,\, ,
\ee
being Im$\Pi(t)^{^{exp}}_{_{no-\pi}}$ the experimental data 
with the pion pole removed~\cite{Aleph} 
and $t_0=3$~GeV$^2$.  
In order to recover global duality one cannot take an arbitrary 
matching point~\cite{DRgaussian,matchingDR}, 
though the duality becomes better and
better as $t_0$ is increased and the oscillations in the spectral function 
vanish.   
Although the experimental data only reach up to $t=M_\tau^2$   
in the $\tau$--decay experiments~\cite{Aleph,CLEO,OPAL}, 
one knows from the $e^+ e^-$ experiments that 
pQCD provides an appropriate 
description of the vector spectral
function~\cite{QCDPich}. Even the $\tau$ data for the $V+A$ correlator 
already show this regularity for $t\gsim 2$~GeV$^2$. 
Our experimental ansatz is however less accurate in the $V-A$~channel 
where the fluctuations are wider than in 
$\frac{1}{\pi}$Im$\Pi_{_{LL}}(t)$ and they still remain 
sizable at the $\tau$--threshold.
Nonetheless,  for $Q^2\sim t_0$, the $\mB^{^{(k)}}(Q^2)$ sum rules suppress
the transition region $t\sim t_0$ 
as $k$ grows, providing this ansatz a first estimate of
the physical components.

We   can   see   in Fig.~\eqn{fig.momOPELR} that,   although the leading order  
  OPE   contribution     governs   the very first components,   
these    eventually    diverge.      
We    have    taken    the    value      
${\bra \cO_{_{(6)}}^{^{LR}}\ket \simeq 
-3.8 \cdot 10^{-3}}$~GeV$^6$ in order to 
illustrate the behaviour of the 
$\mB_{_{LR}}^{^{(k)}}(Q^2)_{_{no-\pi}}$~\cite{narisonpid}.   
Adding the contribution from  some   
higher dimension operators does not solve the problem of the convergence at 
low  energies ($Q^2=2$~GeV$^2$),  
as one can see in the first plot of Fig.~\eqn{fig.momOPELR}.  
We have used the values 
$\bra \cO_{(8)}^{^{LR}}\ket\simeq 6.0\cdot 10^{-3}$~GeV$^8$,  
$\bra\cO_{(10)}^{^{LR}}\ket\simeq -9.1\cdot 10^{-3}$~GeV$^{10}$, 
$\bra\cO_{(12)}^{^{LR}}\ket\simeq 11.9\cdot 10^{-3}$~GeV$^{12}$
from the review in Ref.~\cite{narisonpid}, although there is still some
controversy about the value of the higher dimension 
condensates~\cite{OPEVincenzo,narisonpid,SRFriot,SRBijnens}.
\\
\tab 
However, the situation improves drastically once we go to the deep euclidean
regime. At  $Q^2= 10$~GeV$^2$ (second plot in Fig.~\eqn{fig.momOPELR}),  
the addition of higher dimension
contributions produces appreciable improvements in the convergence of the 
$\mB_{_{LR}}^{^{(k)}}(Q^2)_{_{no-\pi}}$. In Fig.~\eqn{fig.momOPELR} and next
figures,   
all the plots are normalized such that 
$\frac{1}{\pi}$Im$\Pi_{_{LL}}(t)^{^{pQCD}}\stackrel{t\to
\infty}{\longrightarrow}1$.

A   \ similar  \  result  \ is   \ obtained   \ for  \  the   \ $V+A$  \  correlator, \ 
where   \  the   \ inclusion  \ of higher  dimension operators   \ 
improves  \ slightly the description     \ 
in the deep euclidean region  \  ($Q^2=10$~GeV$^2$, second plot  \ 
in \  Fig.~\eqn{fig.momOPELL}).   \ 
We  \  have  \  considered   \  the   \ chiral   \ limit   \ and  \  used   \  the  \ 
values   \  $\alpha_s(Q^2)=-2\pi/  
\left[\beta_1 \ln{(Q^2/\Lambda_{_{QCD}}^2)}\right]$,    \    \ 
with $\Lambda_{_{QCD}}\simeq 0.235$~GeV   \ and   \ $n_f=3$~\cite{QCDPich},   \  \ 
${\bra\cO_{_{(4)}}^{^{LL}}\ket 
= \frac{1}{12\pi}\alpha_s\bra G_{\mu\nu}^a G^{\mu\nu}_a\ket 
\simeq 1.2\cdot 10^{-3}}$~\cite{YndurainAGG}, 
$ {\bra\cO_{_{(6)}}^{^{LL}}\ket= \frac{8\pi}{9}\alpha_s \bra \bar{q} q\ket^2 
\simeq 0.85\cdot 10^{-3}}$~\cite{narisonpid}.
\begin{figure}[t]
\begin{center}
\includegraphics[width=7cm,clip]{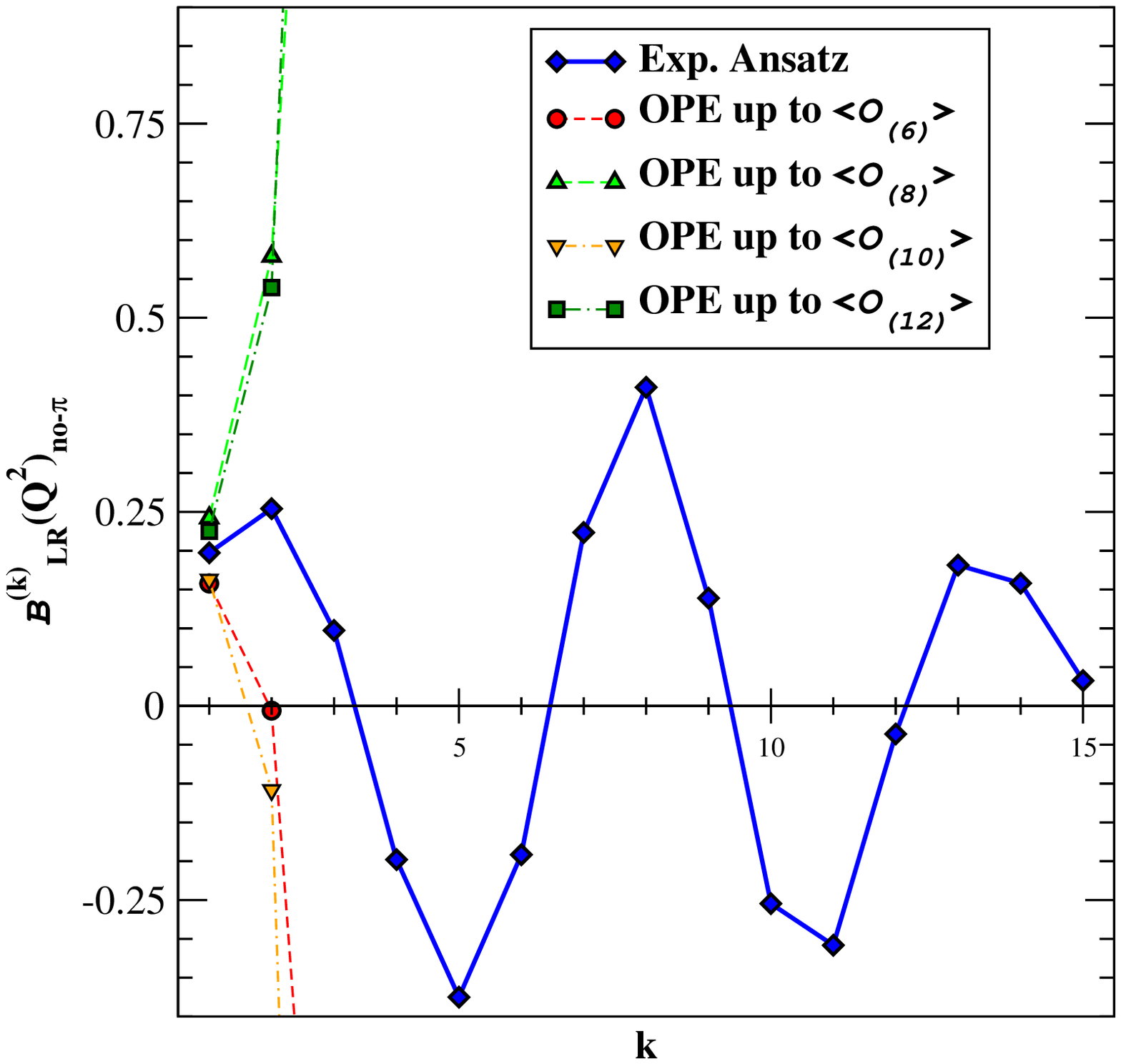}
\hspace*{0.5cm}
\includegraphics[width=7.5cm,clip]{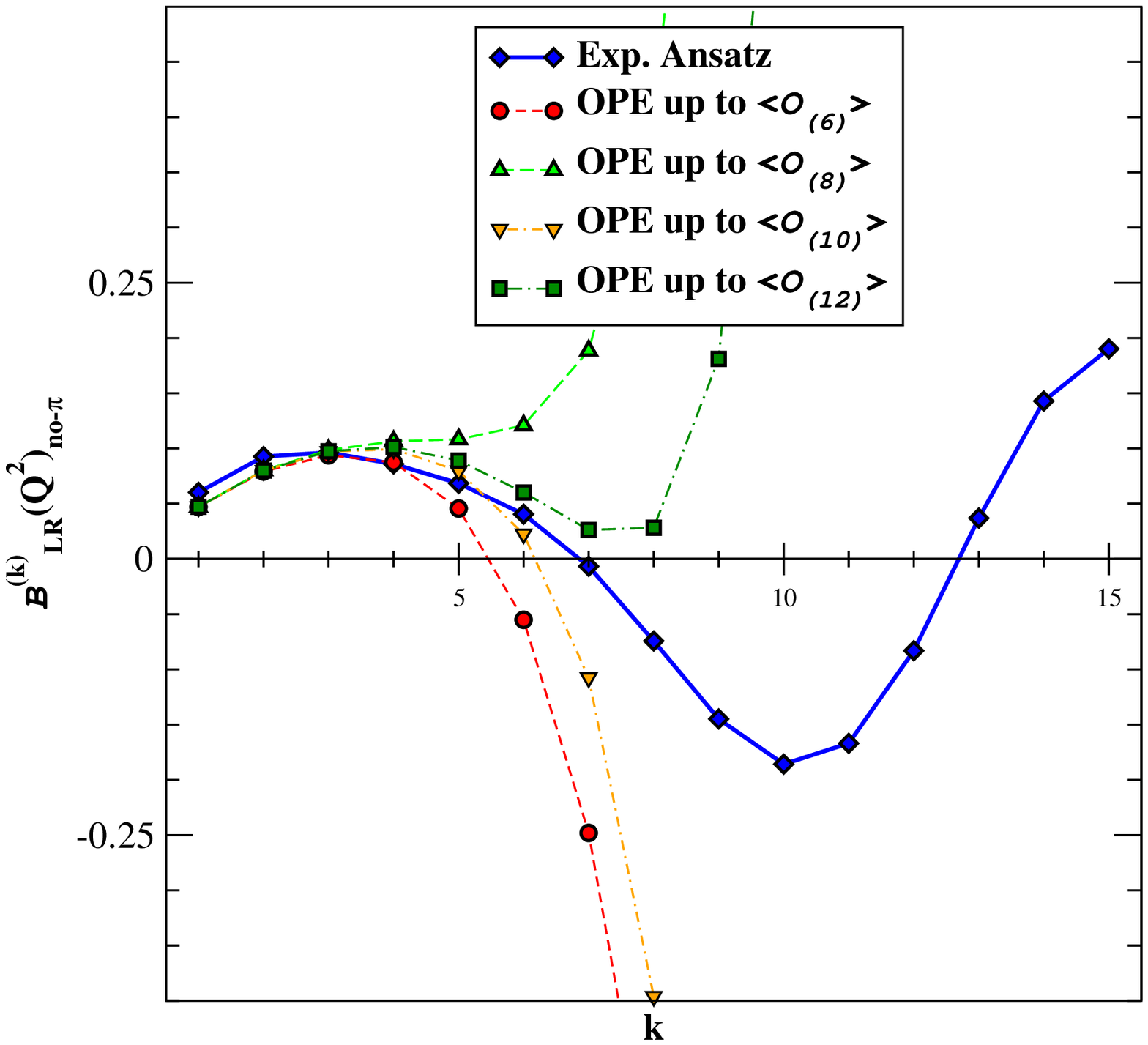}
\caption{Components 
$\mB_{_{LR}}^{^{(k)}}(Q^2)_{_{no-\pi}}$ for the $V-A$ correlators
provided by the OPE at $Q^2=2$~GeV$^2$ and $Q^2=10$~GeV$^2$ 
(left and right-hand-side, respectively). 
It is compared to the components derived from the experimental ansatz. 
From now on, 
the amplitudes in the plots are normalized such that
$\frac{1}{\pi}$Im$\Pi_{_{LL}}(t)^{^{pQCD}}\stackrel{t\to
\infty}{\longrightarrow}1$.}
\label{fig.momOPELR}
\end{center}
\end{figure}
\begin{figure}[t!]
\begin{center}
\includegraphics[width=7.5cm,clip]{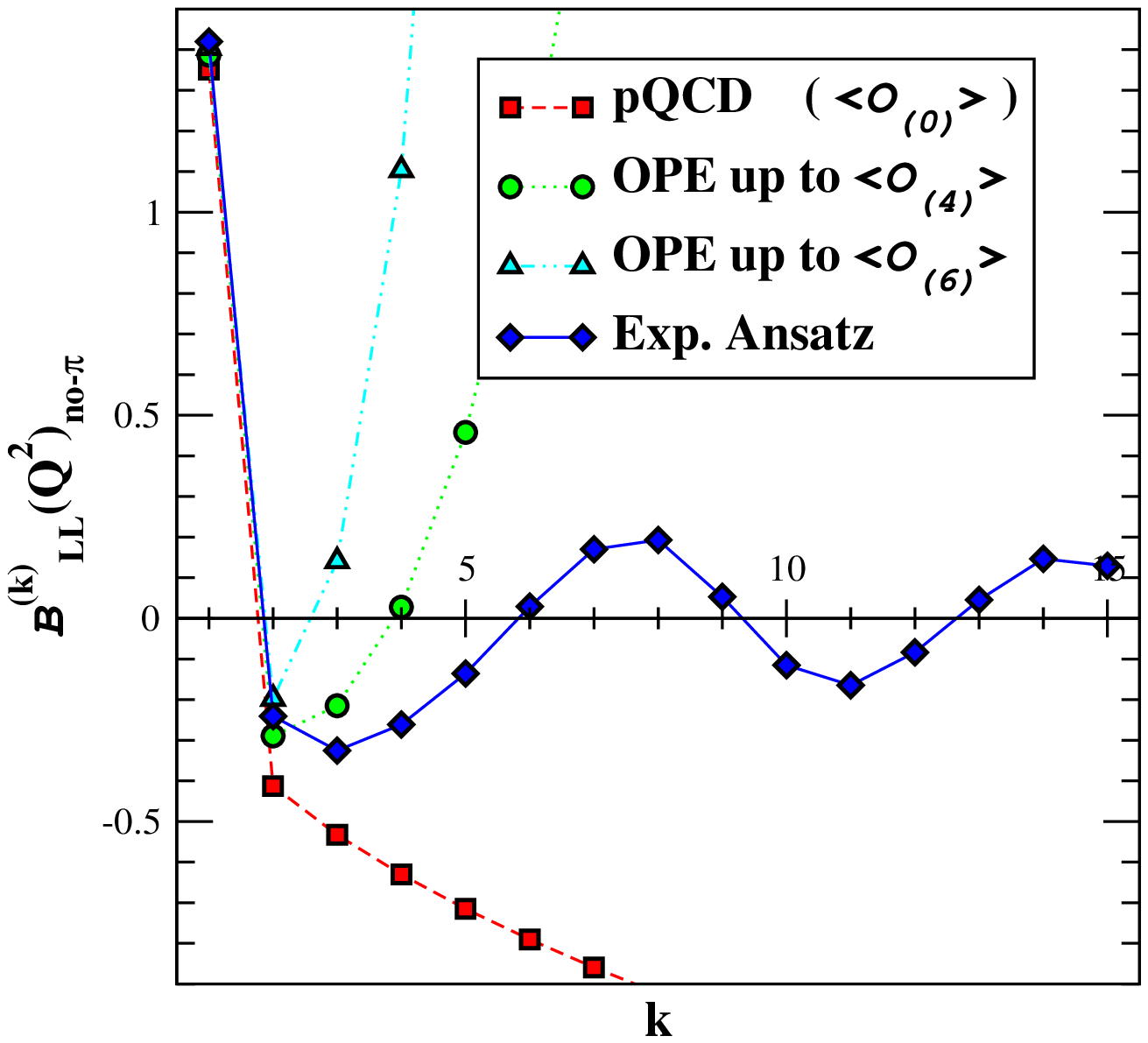}
\hspace*{0.5cm}
\includegraphics[width=7cm,clip]{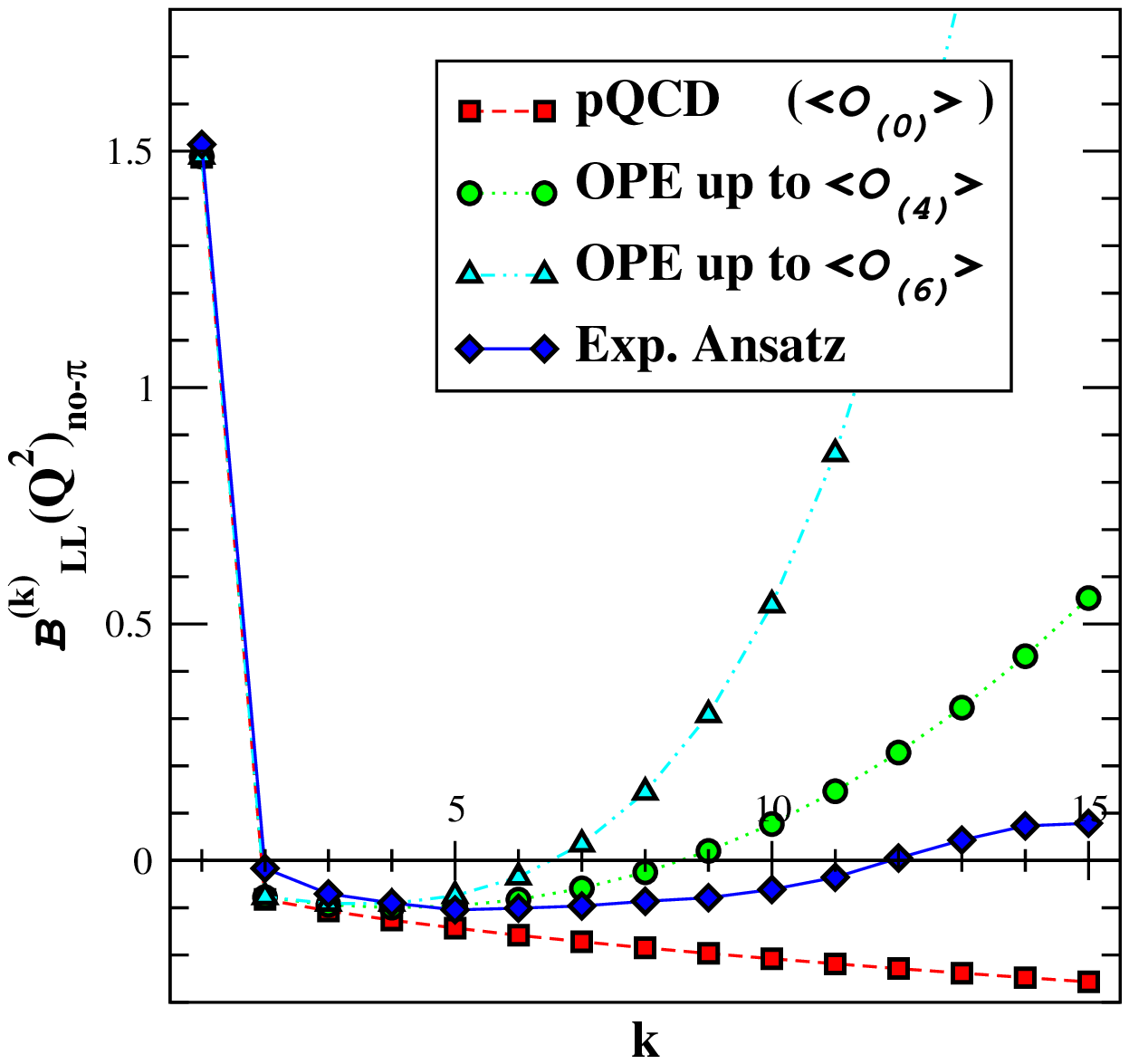}
\caption{Components $\mB_{_{LL}}^{^{(k)}}(Q^2)_{_{no-\pi}}$ 
for the $V+A$ correlators
provided by the OPE at $Q^2=2$~GeV$^2$ and $Q^2=10$~GeV$^2$.}
\label{fig.momOPELL}
\end{center}
\end{figure}

\subsection{Averaged correlators within the OPE}
\tab
We have seen that the OPE remains reliable only up to a given finite order $k$ 
in the
moments. Through the consideration of a convenient distribution function
$\xi_a(x)$, we can nevertheless remove 
the influence of the components $\mB^{^{(k)}}(Q^2)$  of order $k>a+1$. 
The averaged OPE amplitude is obtained from the calculation of the moments in
the euclidean domain and it is compared 
to the averaged amplitude coming from the
experimental ansatz in the minkowskian region.

The transform of the $1/Q^{2m}$ terms, e.g. in the $V-A$~correlator,  
produces again an OPE-like power series:
\be
\Pi_{_{LR}}(-Q^2)\,=\, \sum_{m=3}^\infty 
\Frac{\bra\cO^{^{LR}}_{_{(2m)}}\ket}{Q^{2m}} 
\qquad \longrightarrow \qquad
\Frac{1}{\pi}\mbox{Im}\overline{\Pi}_{_{LR}}(z)^{^{\xi_a}}\,=\, 
\sum_{m=\frac{a}{2}+1}^\infty 
\Frac{\frac{1}{\pi}\mbox{Im}\overline{\Pi}^{^{LR, \, \xi_a}}_{_{(2m)}}}{z^{m}} 
\,\, ,
\ee 
with the coefficients  
$\frac{1}{\pi}$Im$\overline{\Pi}^{^{LR,\, \xi_a}}_{_{(2m)}}\, = \, 
(-1)^{\frac{a}{2}}\, \,\bra\cO^{^{LR}}_{_{(2m)}}\ket
\,\, \,\, \left[ \frac{2^{a+1}\,\, 
\Gamma\left(\frac{a}{2}+\frac{3}{2}\right)}{
\sqrt{\pi}\, \Gamma(a+2)\, 
\Gamma\left(\frac{a}{2}+1\right)} \, \,
\frac{\Gamma\left(m+\frac{a}{2}+1\right)}{
\Gamma\left(m-\frac{a}{2}\right)}\right]$, and 
where the condensates $\bra\cO_{_{(2m)}}^{^{LR}}\ket$ 
have been taken as a constant.

For a given $\xi_a(x)$, the average of a rational term  
$\Pi(-Q^2)\sim  \Frac{\bra\cO_{_{(2m)}}\ket}{Q^{2m}}$ 
is then zero whenever $2m\le a$, so  
the averaged amplitude is never sensitive 
to the presence of the pion pole.  
This means that, in the short distance region where the 
${V-A}$~correlator is expected to be  
governed by the $\Frac{\bra \cO^{^{LR}}_{_{(6)}}\ket }{Q^6}$ 
term,  the averaged spectral function   
is zero for $a\geq 6$.   
Likewise,  
$\frac{1}{\pi}$Im$\overline{\Pi}_{_{LR}}(z)^{^{\xi_a}}$ accepts the same 
number of Weinberg sum-rules as the original amplitude  
$\frac{1}{\pi}$Im$\Pi_{_{LR}}(t)$. 
Its  non-zero experimental value for $\xi_8(x)$ 
(first plot in Fig.~\eqn{fig.averageOPE})  hints  that for  
$z\lsim 5$ GeV$^2$ the corrections  from  
$\bra \cO_{_{(2m)}}\ket$ with $m\geq 5$ produce rather relevant effects.

In $\Pi_{_{LL}}(-Q^2)^{^{pQCD}}$ up to 
$\cO(\alpha_s)$,  one finds from 
Eqs.~\eqn{eq.compBLLpQCD}~and~\eqn{eq.promPILLpQCD} 
that for any $\xi_a(x)$,  
\be
\Frac{1}{\pi}\mbox{Im}\overline{\Pi}_{_{LL}}(q^2)^{^{pQCD}} \,\,\,\,
= \,\,\,\, \left. 
\Frac{1}{\pi}\mbox{Im}\Pi_{_{LL}}(q^2)^{^{pQCD}} \right|_{_{\cO(\alpha_s)}}
\,\,\,\, + \,\,\,\, \cO(\alpha_s^2) \,\, .
\ee
It would be interesting to study how  this identity is modified at higher
orders in $\alpha_s$.  
The comparison of the theoretical $V+A$ averaged
correlator to our  experimental ansatz  is
provided on the second plot from Fig.~\eqn{fig.averageOPE},  
showing a very good
agreement within  $\cO(\alpha_s^2)$ uncertainties --solid band--. 
For $\xi_8(x)$, the average is independent of $\bra \cO_{_{(4)}}^{^{LL}}\ket$, 
$\bra \cO_{_{(6)}}^{^{LL}}\ket$ and $\bra \cO_{_{(8)}}^{^{LL}}\ket$, being governed by the pQCD
distribution. On the contrary to
$\Pi_{_{LR}}(-Q^2)$, the contributions from the condensates 
$\bra \cO_{_{(2m)}}^{^{LL}}\ket$ with $m\geq 5$ 
seem  to be much more suppressed.

\begin{figure}[t!]
\begin{center}
\includegraphics[width=7.5cm,clip]{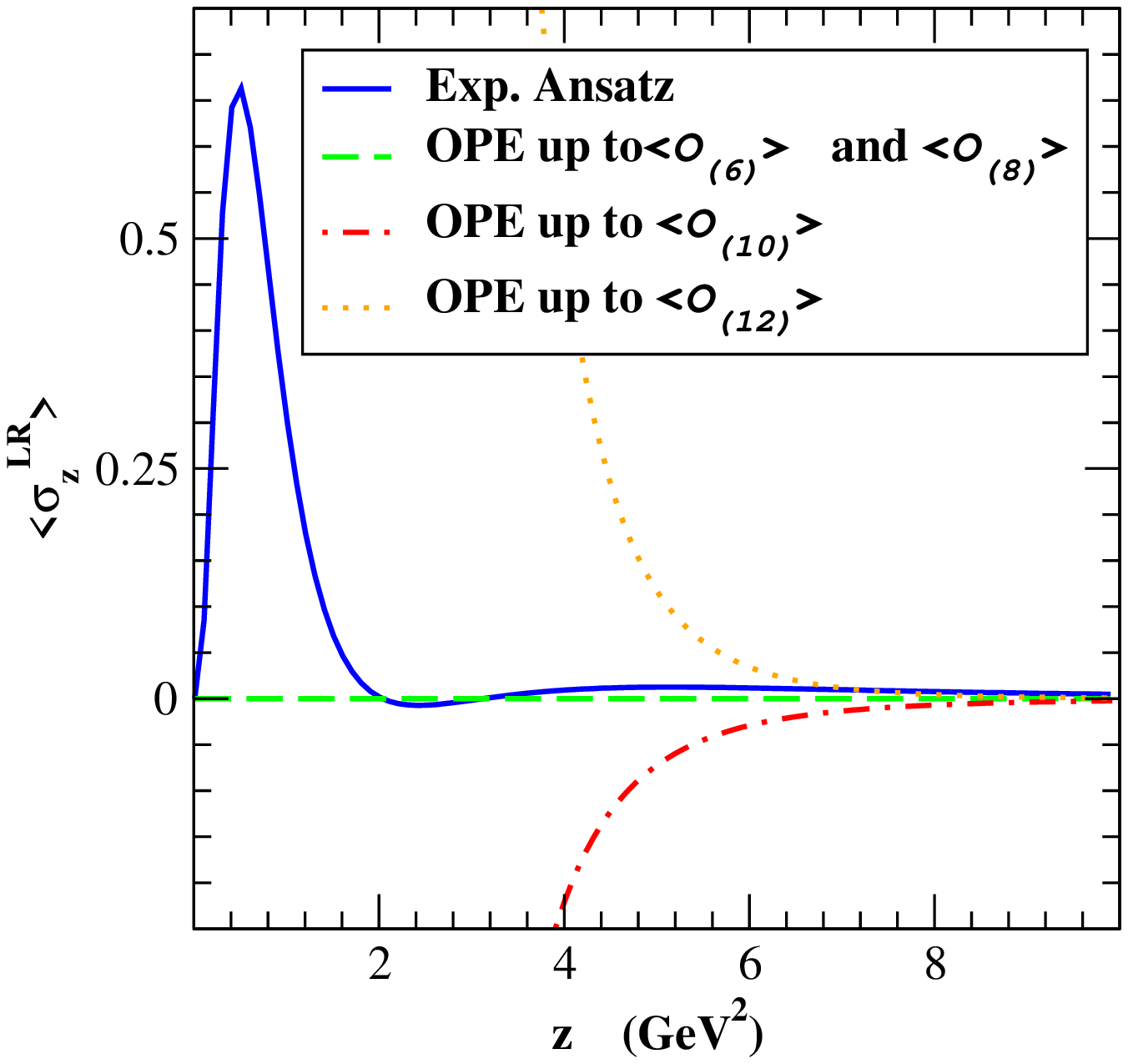}
\includegraphics[width=7cm,clip]{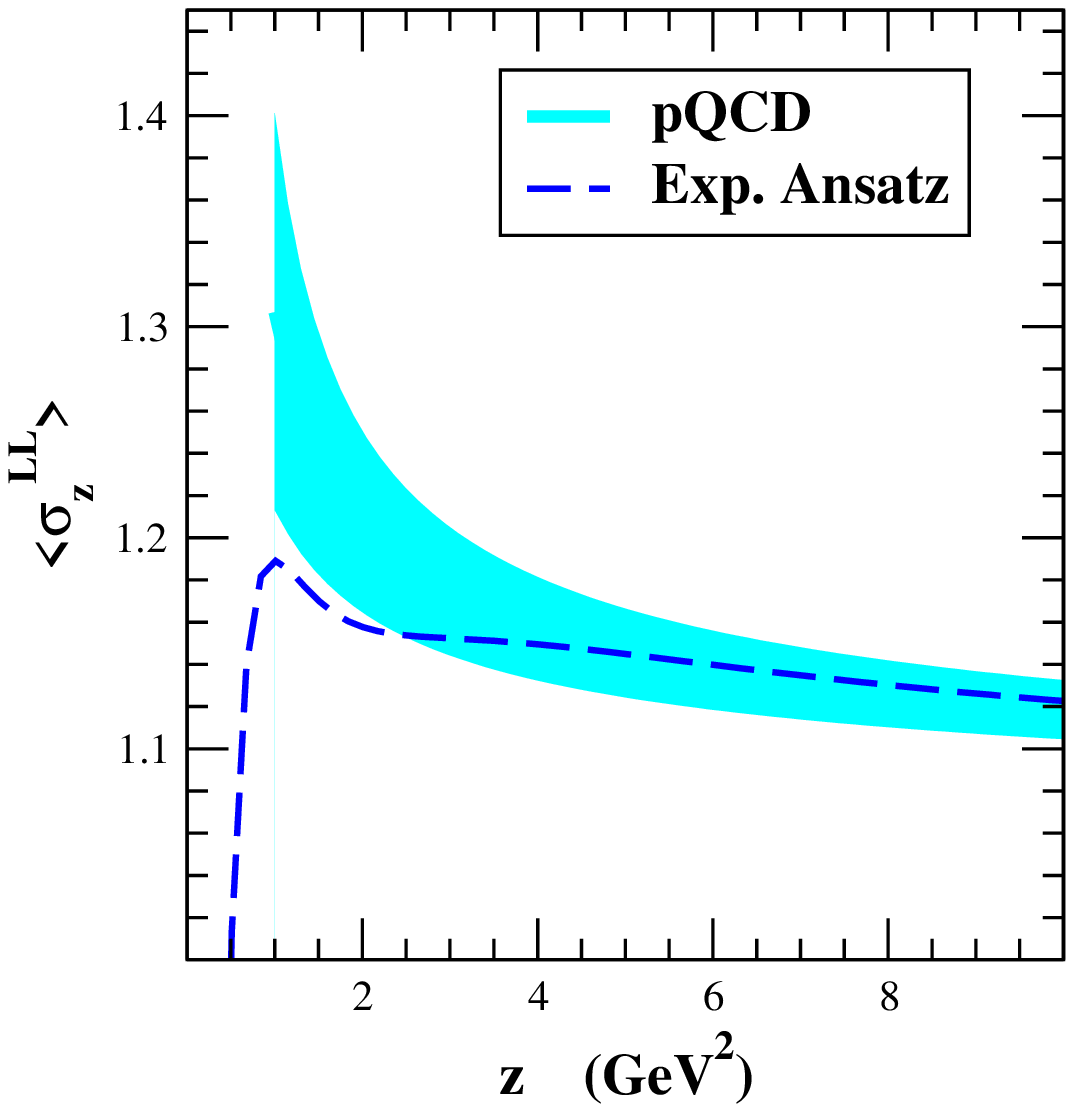}
\caption{$V-A$ and $V+A$ correlators $\sigma_z(x)$  averaged by the
distribution  $\xi_8(x)$ 
(${\bra
\sigma_z\ket_{_{\xi_8}}=\frac{1}{\pi}\mbox{Im}\overline{\Pi}(z)^{^{\xi_8}}}$). 
The solid band on the second plot 
represents the $\cO(\alpha_s^2)$ uncertainties.}
\label{fig.averageOPE}
\end{center}
\end{figure}

By considering appropriate distributions $\xi_a(x)$ one has, in addition,  
the possibility of recovering $\alpha_s$ and 
the vector condensates $\bra \cO_{_{(2m)}}^{^{VV}}\ket$ 
through a direct analysis of the $e^+e^-$ experimental 
measurements~\cite{Novo2000,PIVV}.  
Nonetheless, this analysis escapes the aim of this article and 
it is left for  a next work. 
The situation in the $V-A$ case seems not so favourable  since the 
energy range of  data is much more reduced 
($t\leq M_\tau^2\simeq 3$~GeV$^2$), mainly laying within the long distance
region where the OPE stops being valid.

\section{QCD correlators in resonance theory at large $N_C$}
\tab
In order to inspect the range of energies 
where the OPE breaks down it is necessary to add some
extra ingredients to our theory. One has to add a mechanism that fully
reproduces the OPE expressions at high euclidean 
momenta, being able, in addition, to provide the right low
energy  theory ($\chi PT$) and the minkowskian 
description (experimental data, resonance structure...). 
We will see how  a theory with explicit resonances suits  this picture.

The exact recovering of perturbative 
QCD requires the full pile of hadronic states. 
However, the infinite 
summation  of resonance from large $N_C$ QCD is by itself a limit.  
One of the available methods to regularise the series is through 
an ultraviolet cut-off.  
Hence, one constructs a set of   quantum field 
theories, $R\chi T^{(n)}$,  with a finite   number $n$ of resonances  
to describe the observables at some given momentum
$p^\mu$~\cite{Beanecutoff,Periscutoff,PerisRegge}.     
$QCD_\infty$ would be  recovered 
by taking the limit $n\to\infty$ keeping the external momenta fixed 
($R\chi T^{(n)}[p^\mu] \stackrel{n\to \infty}{\longrightarrow} 
R\chi T^{(\infty)}[p^\mu]=QCD_\infty[p^\mu]$).

Accommodating $R\chi T^{(\infty)}$ and  
QCD at large $N_C$ is not trivial. 
The first approach must be focused on recovering  pQCD  in the deep euclidean
domain. The appearance at $N_C\to\infty$ 
of non-perturbative condensates in the OPE produces
small corrections into the pQCD correlator at $Q^2\gg \Lambda_{_{QCD}}^2$. 
However, due to the non-perturbative effects at large $N_C$,  
the smooth pQCD spectral function  turns into a meromorphic function with an
infinite number of real poles.

It is important to recall the possible existence of duality violating
terms. The Green-functions $\bra T\{J_\mu(x)J^\dagger_\nu(0)\}\ket$ 
are not fully determined by the singularities at $x^2=0$~\cite{dualviolation}. 
For instance, one may have
terms of the form $\frac{1}{x^2+\rho^2}$, whose Fourier transform 
falls off in the deep euclidean as $\exp{\left(-Q \,\rho \right)}$ 
but becomes oscillating, $\sin{\left( q\,\rho  \right)}$,  in the
minkowskian region ($Q\equiv\sqrt{-q^2}$ and $q\equiv \sqrt{q^2}$).  
It is actually interesting that some resonance models  
generate explicitly this kind of 
duality violating terms~\cite{Son,Com}, pointing out the fact that a resonance
theory is fully able to produce both the OPE and non-OPE QCD singularities.  
We will 
derive the OPE condensates in this section by neglecting   
the duality violating terms in the deep euclidean domain although  
further works should focus on the estimate of these terms~\cite{Cataduality}.

In real world,  the resonance  poles become complex and 
get shifted to unphysical Riemann sheets due to Dyson-Schwinger resummations at
all orders in $1/N_C$; once the higher $1/N_C$ corrections
are taken into account and 
the resonance gain 
their physical non-zero widths the amplitude becomes again
smooth, as it is found experimentally. The width can be accurately  
computed in some 
situations~\cite{anchura,GP:97,Palomar},
although its derivation is, in general,  non-trivial.

\subsection{$V+A$ correlator in resonance theory}
\tab
At large $N_C$, the spin--1 correlators are meromorphic functions characterised
by the position and residues of their poles. The $V+A$~spectral function is 
given by 
\be
\Frac{1}{\pi} \mbox{Im}\Pi_{_{LL}}(t)^{^{N_C\to\infty}}\,\, = \,\, 
\,\, 
F^2 \,  \delta(t)\, \, + 
\,\, 
\displaystyle{\lim_{M_\infty^2\to\infty} }\,\,\, 
\displaystyle{\sum_{M_j^2\leq M_\infty^2}}\,  
 F_{j}^2\,\,\, \delta( t-M_{j}^2) \, ,   
\ee
which generates the moment integrals 
given by the positive meromorphic functions
\bel{eq.RChT1}\ba{rl}
\mA^{^{(n)}}_{_{LL}}(Q^2)^{^{N_C\to\infty}}\, \,=& \,  \,
\,\, 
\Frac{F^2}{Q^2} \, \, + 
\,\, 
\displaystyle{\lim_{M_\infty^2\to\infty} } \,\,\, 
\displaystyle{\sum_{M_j^2\leq M_\infty^2}}\,  
\Frac{F_{j}^2\, Q^{2n}}{(M_{j}^2+Q^2)^{n+1}} 
\, ,  
\ea
\ee 
with $M_\infty$ denoting  the mass of the highest resonance in 
the resonance theory.  
the constants $M_j$ and $F_j$ are the mass and coupling constant of the vector
and axial-vector states at LO in $1/N_C$ and $F$ is the pion decay constant. 
In the limit when $M_\infty^2\to\infty$, this upper 
mass acts like an ultraviolet  
cut-off of the large $N_C$ infinite resonance 
summation~\cite{EspriuRegge,Beanecutoff,Periscutoff,PerisRegge}.   
Taking the $Q^2\to \infty$ limit is not trivial at all   
since there is always an infinite set of resonance above any 
considered $Q^2$, whose effects  are in general not so clearly negligible.  
From now on, the limit $M_\infty^2\to \infty$ will be always implicitly assumed.

\subsubsection{Perturbative and non-perturbative contributions 
in pQCD}
\begin{figure}[t]
\begin{center}
\includegraphics[width=7cm,clip]{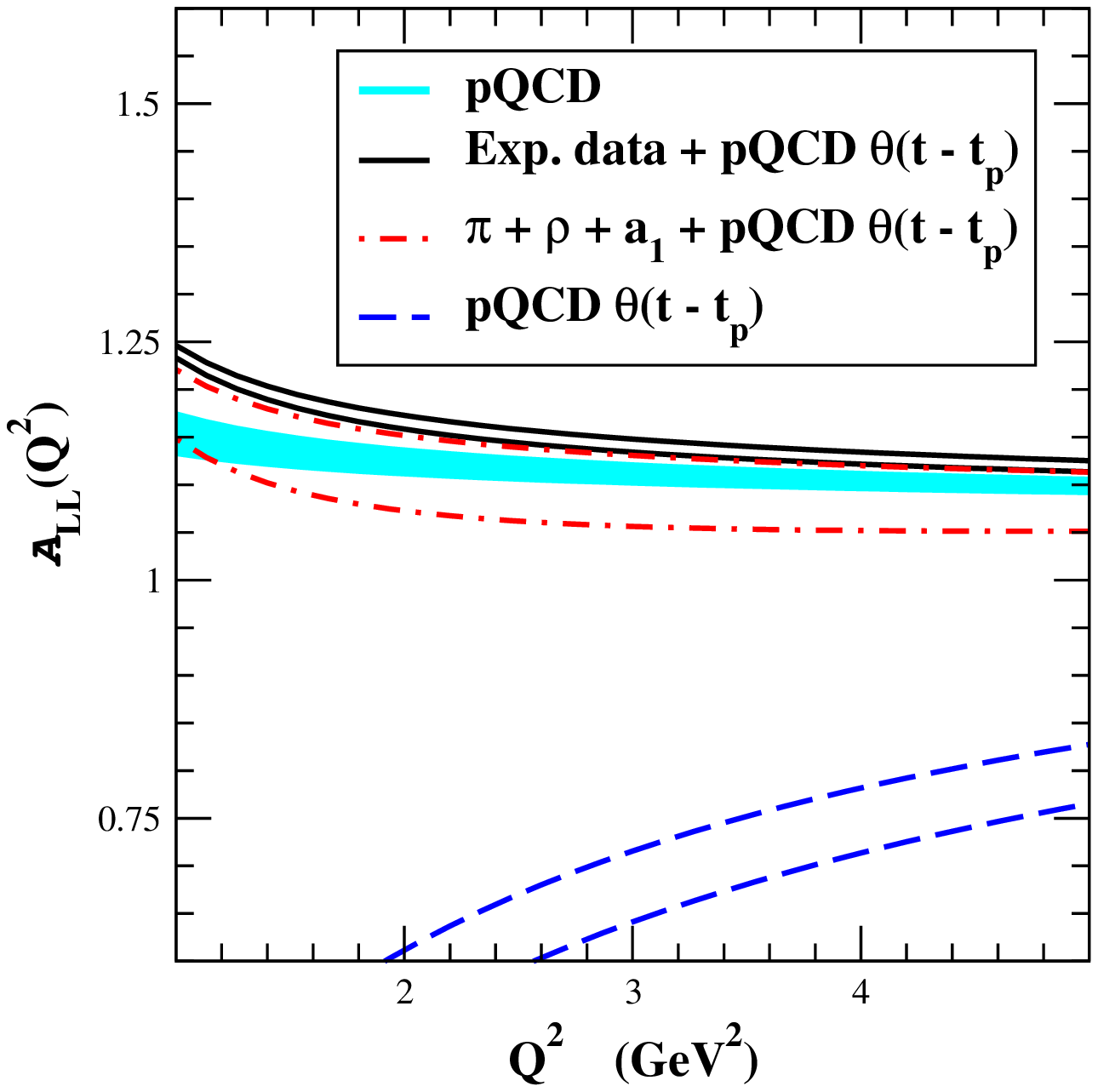}
\hspace*{1cm}
\includegraphics[width=7cm,clip]{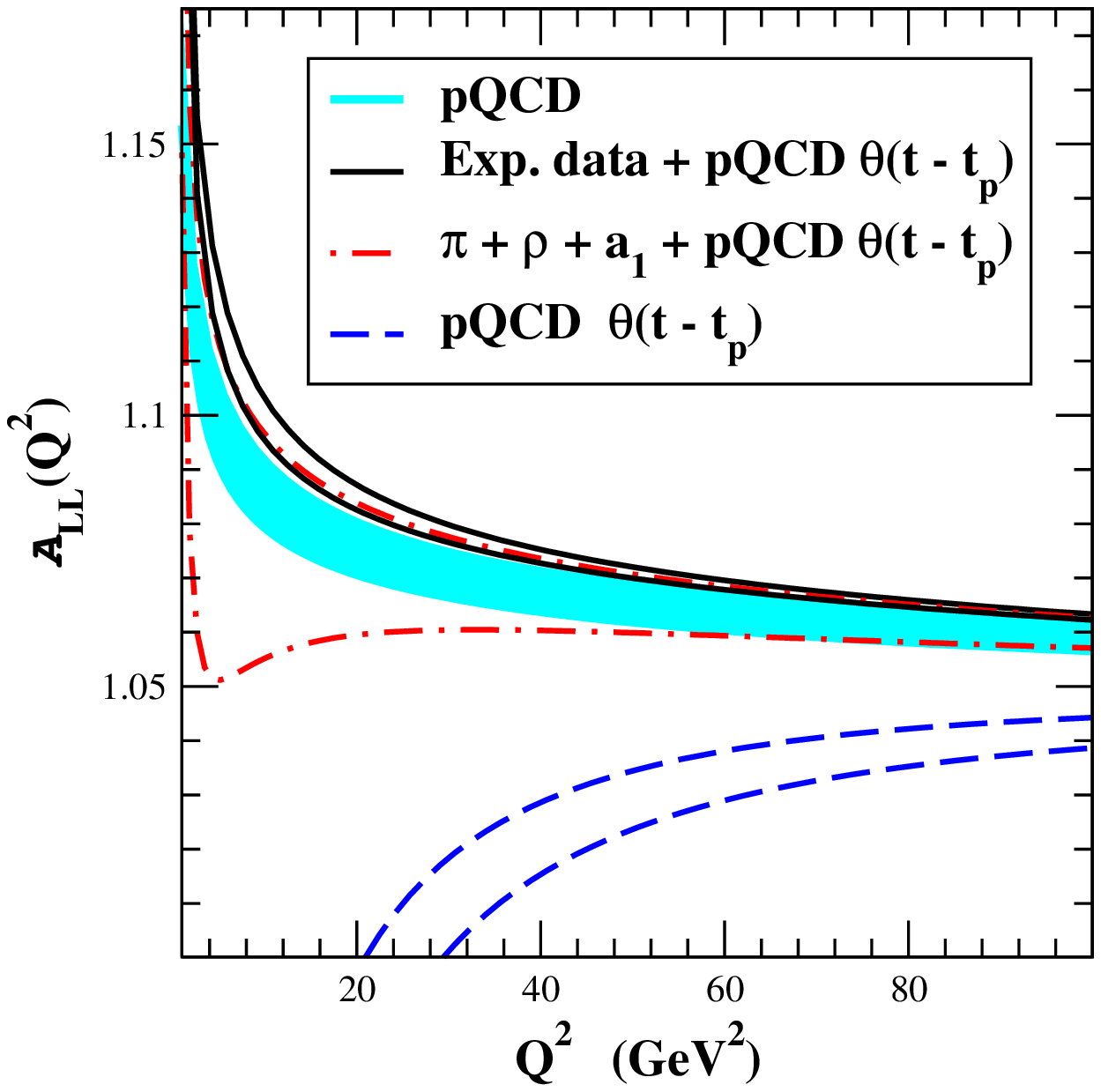}
\caption{$V+A$ Adler function from the pQCD space-like calculation up to 
$\cO(\alpha_s)$, where the band represents the 
$\cO\left(\alpha_s^2\right)$ uncertainties. 
It is compared with $\mA_{_{LL}}(Q^2)_{_{pert.}}^{^{pQCD}}$,   
derived through dispersion relations 
from considering the spectral function 
$\frac{1}{\pi}$Im$\Pi_{_{LL}}(t)^{^{pQCD}}\, \cdot\, \theta(t-t_p)$ 
(dashed curve, ${t_p=(1.25}$~GeV$)^2$ and 
$t_p=(1.45$~GeV$)^2$ for the upper an lower curves respectively). 
The dashed--dotted curve includes also the contributions from the pion,
$\rho(770)$ and $a_1(1260)$ whereas the solid lines consider instead 
the experimental data~\cite{Aleph} for the spectral function at $t<t_p$.
}
\label{fig.pQCDpert}
\end{center}
\end{figure}
\tab 
In this section we make an analysis of pQCD which will be relevant for the
resonance calculation.  We study how the  
pQCD amplitude is recovered in the deep
euclidean thanks to a spectral function that behaves like 
$\frac{1}{\pi}$Im$\Pi_{_{LL}}(t)^{^{pQCD}}$ at high energies.

We will consider an auxiliary spectral function with  the  form 
\be
\Frac{1}{\pi} \mbox{Im}\Pi_{_{LL}}(t)^{^{pQCD}}_{_{pert.}}
\,\,\,\, \equiv \,\,\,\, 
\Frac{1}{\pi} \mbox{Im}\Pi_{_{LL}}(t)^{^{pQCD}}
\,\, \cdot \,\, \theta(t-t_p) \,\, , 
\ee
where the low
energy region has been removed. It yields the moments 
\be 
\mA_{_{LL}}^{^{(n)}}(Q^2)^{^{pQCD}}_{_{pert.}}\,\,\,\, = \,\,\,\, 
\Int_{t_p}^\infty \, \,\Frac{dt \,\, Q^{2n}}{(t+Q^2)^{n+1}} \,\, 
\Frac{1}{\pi} \mbox{Im}\Pi_{_{LL}}(t)^{^{pQCD}}
\,\, .
\ee 
In the deep euclidean, these moment integrals recover the pQCD 
moments $\mA_{_{LL}}^{^{(n)}}(Q^2)^{^{pQCD}}$ with   
the proper $\alpha_s$ running up to subdominant $1/Q^{2m}$ power corrections.

The corresponding Adler function up to
$\cO(\alpha_s)$ is plotted in Fig.~\eqn{fig.pQCDpert}. 
The solid band covers an uncertainty of 
$\pm (\frac{\alpha_s}{\pi})^2$.  
The value of $t_p$  was varied  between the values $t_p=(1.26$~GeV$)^2$ and 
$t_p=(1.45$~GeV$)^2$. 
However, the asymptotic behaviour 
$\mA_{_{LL}}(Q^2)^{^{pQCD}}_{_{pert}}=
\left. \mA_{_{LL}}(Q^2)^{^{pQCD}}\right|_{_{\cO(\alpha_s)}}
+\cO(\alpha_s^2)$ 
is  only achieved 
at  energies $Q^2\gg 100$~GeV$^2$,   
by far much larger than the expected  $Q^2\sim 2$~GeV$^2$.  
These discrepancies are much easier to understand  through the
observance of  the Adler function at $\cO(\alpha_s^0)$: 
\bel{eq.pQCDpert}
\mA_{_{LL}}(Q^2)^{^{pQCD}}_{_{pert.}}=
\Frac{N_C}{12\pi^2}\left[\Frac{Q^2}{Q^2+t_p}\right] + \cO(\alpha_s) 
=  \Frac{N_C}{12\pi^2}\left[\, 1\, \, - \,\,  \Frac{t_p}{Q^2} 
\,\,\, + \,\,\, 
\cO\left(\Frac{t_p^2}{Q^4}\right) + \cO(\alpha_s)\right] \, ,
\ee
where the $\Frac{t_p}{Q^2}$ term generates deviations of the order of 1\% 
at $Q^2=100$~GeV$^2$, larger than those coming from the $\cO(\alpha_s^2)$
corrections.

This points out that although the high energy logarithmic dependence of 
Im$\Pi_{_{LL}}(t)^{^{pQCD}}_{_{pert.}}$ 
provides the proper asymptotic behaviour of the pQCD moment integrals, 
the non-perturbative region owns crucial information about how fast this limit
is reached. If one includes the contribution of the resonances laying on
$t<t_p$ (Goldstones from the chiral symmetry breaking, $\rho(770)$ and
$a_1(1260)$) the asymptotic behaviour is again reached within the expected range of
energies. Alternatively, one may consider the experimental data  for  
Im$\Pi_{_{LL}}(t)$ at  $t<t_p$~\cite{Aleph}, getting similar results due 
to the moderate size of the width of these states ($M_j \Gamma_j
\lsim \Frac{1}{N_C}\, \cdot \, M_j^2 $). 
For Fig.~\eqn{fig.pQCDpert} we have use the inputs 
$M_\rho=0.77$~GeV, $M_{a_1}=1.26$~GeV, $F=92.4$~MeV, 
$F_\rho=154$~MeV and $F_{a_1}=123$~MeV~\cite{therole,PDG}.

\subsubsection{Perturbative and non-perturbative contributions in $QCD_\infty$}
\tab 
We analyse in this section the conditions to recover pQCD in the deep
euclidean region of momenta through a resonance theory.   
pQCD will naturally arise  when summing up  the infinite series of
resonances, being the higher dimension contributions in the OPE 
a mere  remnant from the discrete summation of poles.

First \  of  \ all,  \ it \  becomes  \ necessary \  to \  split \  the \  
resonance  \  series \   into  \  two \  
sub-series,    \  
${\Pi(-Q^2)^{^{N_C\to\infty}}=\Delta\Pi(-Q^2)^{^{N_C\to\infty}}_{_{pert.}}
+\Delta\Pi(-Q^2)^{^{N_C\to\infty}}_{_{non-p.}}}$: a perturbative part 
(denoted as $\Delta\Pi_{_{pert.}}^{^{N_C\to\infty}}$) 
where the resonances already lay on the perturbative QCD
regime, with $M_{j}^2 > M_p^2\sim 2$~GeV$^2$, which will provide 
the asymptotic behaviour of the moments; 
and a non-perturbative part 
(namely $\Delta\Pi_{_{non-p.}}^{^{N_C\to\infty}}$) 
with the resonance masses laying within the
non-perturbative regime of QCD,
$M_{j}^2 \leq M_p^2$, which will drive  how fast the moments 
converge to the pQCD description. 
$M_p$ is the mass of the last multiplet included in 
$\Delta\Pi_{_{non-p.}}^{^{N_C\to\infty}}$. We will assume that the masses $M_j$ are
on increasing order, $M_1\leq M_2\leq...$

The non-perturbative sub-series 
$\Delta\mA_{_{LL}}^{^{(n)}}(Q^2)_{_{non-p.}}^{^{N_C\to\infty}}=
\displaystyle{\sum_{j=1}^p 
%\sum_{M_{j}^2\leq M_p^2} 
} 
\, \, 
\frac{F_{j}^2  \, Q^{2n}}{(M_{j}^2+Q^2)^{n+1}}$ 
is finite and it can be analytically
expanded in powers of $\Frac{M_j^2}{Q^2}$, with $M_j^2\leq M_p^2\sim 2$~GeV$^2$. 
Hence, the pQCD behaviour and the logarithmic $\alpha_s$ corrections are  
only recovered through the perturbative part of the series, which, 
in addition, generates  
extra contributions to the $\cO\left(\Frac{1}{Q^{2m}}\right)$ OPE terms.

We need to  transfer the information from 
$\frac{1}{\pi}$Im$\Pi_{_{LL}}(t)^{^{pQCD}}$ 
to the discrete summation of infinite terms in $QCD_\infty$. 
At this point one needs to make an assumption on the asymptotic structure of the
mass spectrum at high energies. We will consider some smooth $M_j^2=f(j)$
dependence for $j\geq p+1$. 
The pQCD spectral function 
is discretized in the perturbative region $[M_p^2,\infty)$ 
through the  step-like function $H(t)$:
\be
H(t)\, \,\,\, = 
\,\,\,\,
\sum_{j=p+1}^\infty \,\, 
[\theta(M_{j}^2-t)\, - \, \theta(M_{j-1}^2-t)]\, \,\,\,
\Frac{1}{\pi} \,  \mbox{Im}\Pi_{_{LL}}(M_j^2)^{^{pQCD}}
\, ,
\ee
providing the step-like interpolation 
$H(M_j^2)=\frac{1}{\pi}$Im$\Pi_{_{LL}}(M_j^2)^{^{pQCD}}$ for any $j\geq p+1$ 
and $H(t)=0$  for $t\leq M_p^2$.  
The step function is defined as $\theta(y)=\left\{\ba{ll} 0& \,\, \mbox{for } 
y<0   \\ 1 & \,\, \mbox{for } y\geq 0\ea \right.$ .

The \ difference \  between  \ the  \ moments  \ 
$\mA^{^{(n)}}_{_{LL}}(Q^2)^{^H}$ \  of \  this  \ function  \ and  \ the
  \ original  \  ones   \ 
$\mA^{^{(n)}}_{_{LL}}(Q^2)^{^{pQCD}}_{_{pert.}}$ is
given by the expression 
\be
\mA^{^{(n)}}_{_{LL}}(Q^2)^{^H}\,\,\,\,
= \,\,\,\,
\Int_{M_p^2}^\infty \,\,  \Frac{dt \,\, Q^{2n}}{(t+Q^2)^{n+1}} \,\, H(t)
\,\,\,\, =  \,\,\,\, 
\mA^{^{(n)}}_{_{LL}}(Q^2)^{^{pQCD}}_{_{pert.}} \, \cdot \, 
\left[1\, - \, \Delta^{^H}_n\right]
\ee
with  $\mA^{^{(n)}}_{_{LL}}(Q^2)^{^H}\geq 0$ 
and $\Delta^{^H}_n$ a number  in the interval  
\be
0\quad \leq \quad \Delta^{^H}_n 
\quad \leq \quad     
\Frac{|\beta_1|\, \alpha_s^2(M_p^2)}{2\pi^2} 
 \, \Frac{\delta \hat{M}^2}{Q^2}\,\, \cdot \,\, \left[ 
 (n+1)\, \Frac{M_p^2}{Q^2}\, + \, \ln{\Frac{Q^2}{M_p^2}} \, 
 + \, \cO\left(\Frac{M_p^4}{Q^4}\right)\right] \, \, 
 \cdot \,\, \left[1+\cO(\alpha_s(M_p^2))\right]\,   \, , 
\ee
for $n\geq 1$, being $\delta \hat{M}^2=$max$\{\delta M_j^2\}_{j=p+1}^\infty$. 
This $\Delta^{^H}_n$ vanishes for any $n$ 
faster than any logarithmic  pQCD correction  at $Q^2\to\infty$.

The last step consists on converting the step-like function $H(t)$   
into a narrow-width spectrum through the prescription 
\bel{eq.FjIMPI}
F_{j}^2 \,\,\, \, = \,\,\,\, 
H(M_j^2)\,\, \cdot \,\, \delta M_j^2 
\,\,\,\, = \,\,\,\, 
\Frac{1}{\pi}\mbox{Im}\Pi_{_{LL}}(M_j^2)^{^{pQCD}}  
\,\, \cdot \,\, \delta M_j^2 \, ,
\ee
for any $j\geq p+1$, which produces the moments 
\bel{eq.pertV}
\ba{rl}
\Delta\mA_{_{LL}}^{^{(n)}}(Q^2)_{_{pert.}}^{^{N_C\to\infty}}
\, \, \, \, =& \,\,\, \, \mA_{_{LL}}^{^{(n)}}(Q^2)^{^H} 
\,\,\, \cdot \,\,\,\left[1 - \Delta_n^{^{N_C}}\right]^{n+1}\,
\, ,
\ea
\ee
with $\Delta_n^{^{N_C}}$ a number in the range 
$\left[0, \frac{\delta\hat{M}^2}{Q^2+M_p^2+\delta \hat{M}^2}\right]$, vanishing
faster than any log.

Hence,  for high energies $M_p^2\ll Q^2 \ll M_\infty^2$,  
the contribution  coming from the perturbative part of the
series takes the form
\bel{eq.OPEpert}
\ba{rl}
\Delta\Pi_{_{LL}}(-Q^2)_{_{pert.}}^{^{N_C\to\infty}}
\, \, \, \, =& \,\,\, \, 
\Pi_{_{LL}}(-Q^2)^{^{pQCD}}
 \,\,\, \, +  \,\,\, \, 
\displaystyle{\sum_{m=1}^\infty} \, \Frac{\Delta\bra
\cO_{_{(2m)}}^{^{LL}}\ket^{^{pert.}}}{Q^{2m}} \, \, ,
\ea
\ee
where, in addition, to the pQCD result  one gets
$\Frac{1}{Q^{2m}}$ power  terms which vanish faster than $\alpha_s(Q^2)$.  
The extraction of  
$\Delta\bra \cO_{_{(2m)}}^{^{LL}}\ket^{^{pert.}}$ cannot be 
done by trivially expanding the
resonance propagators in powers of $\frac{M_R^2}{Q^2}$, since there are always 
infinite states with masses $M_R^2\geq Q^2$. 
They can be analytically computed just in some  cases.  
In this work, they are recovered through numerical simulations.

On the other hand, the non-perturbative part of the series produces just 
$1/Q^{2m}$ power terms at $M_p^2\ll Q^2$:
\bel{eq.OPEnp}
\ba{rl}
\Delta\Pi_{_{LL}}(-Q^2)_{_{non-p.}}^{^{N_C\to\infty}}
\, \, \, \, =& \,\,\, \, 
\displaystyle{\sum_{m=1}^\infty} \, \Frac{\Delta\bra
\cO_{_{(2m)}}^{^{LL}}\ket^{^{non-p.}}}{Q^{2m}} \, \, ,
\ea
\ee
with $\Delta\bra \cO_{_{(2m)}}^{^{LL}}\ket^{^{non-p.}}=(-1)^m 
\,\sum_{j=1}^p  F_j^2 \,(M_j^2)^{m-1}$.

Matching the sum of the  perturbative and non-perturbative sub-series  
with the OPE yields the constraints
\bel{eq.condeOPELL}
\ba{rcl}
\bra \cO_{_{(2)}}^{^{LL}}\ket \quad =& \quad \Delta\bra
\cO_{_{(2)}}^{^{LL}}\ket^{^{pert.}}
\,\,\, + \,\,\, F^2 \,\,\, + \,\,\, \displaystyle{\sum_{j=1}^p \, F_j^2}
\,\,\, &= \,\,\, 0 \, , 
\\
\bra \cO_{_{(4)}}^{^{LL}}\ket \quad =& \quad\Delta\bra
\cO_{_{(4)}}^{^{LL}}\ket^{^{pert.}}
\,\,\,\,\,\, - \,\,\,\,\,\, \displaystyle{\sum_{j=1}^p \,  F_j^2  \, M_j^2}
\,\,\, &= \,\,\,  \Frac{1}{12\pi} \, \alpha_s \bra G_{\mu\nu}^a
G^{\mu\nu}_a\ket \,\,\, > \,\,\, 0  \, .
\ea
\ee

A priori one may think of  two kind of structures for the spectrum: 
one may have a chaotic spectrum where the values of the masses 
$\{M_n^2\}_{n=1}^\infty$ do not follow any particular law; or 
we may have an ordered spectrum where the masses 
$M_n^2$  are ruled by some asymptotic
expression when $n\to \infty$. Quark model and Regge theory studies 
seems to point out most likely the second option. 
In this case, 
once an asymptotic structure of the 
spectrum $M_n^2=f(n)$   
is assumed , the prescription for the coupling $F_n^2$  
from Eq.~\eqn{eq.FjIMPI} 
is shown to be  enough to recover the pQCD result. If one  
assumes  besides that  the couplings $F_n^2$ also behave  smoothly 
for high values of $n$, then this condition becomes also necessary up to power
corrections in Eq.~\eqn{eq.FjIMPI}: 
\bel{eq.FNnec}
F_n^2 \,\,\, = \,\,\, \delta M_n^2 \,\, \cdot \,\, 
\left[
\, \Frac{1}{\pi}\mbox{Im}\Pi_{_{LL}}(M_n^2)^{^{pQCD}} \,\,\,
\, + \,\,\,\, \cO\left(\Frac{1}{M_n^2}\right) \,  
\right] \, .
\ee
Actually, a variation of the choice from 
$\Frac{1}{\pi}\mbox{Im}\Pi_{_{LL}}(M_n^2)^{^{pQCD}}$ 
to $\Frac{1}{\pi}\mbox{Im}\Pi_{_{LL}}(M_{n-1}^2)^{^{pQCD}}$ 
in Eq.~\eqn{eq.FNnec} 
shows that both forms are  equivalent since they  differ by a  term
$\cO\left(\alpha_s^2\Frac{\delta M_n^2}{M_n^2}\right)$.
If the logarithm running  of 
$\frac{1}{\pi}$Im$\Pi_{_{LL}}(t)^{^{pQCD}}$ in Eq.~\eqn{eq.FNnec} 
is replaced 
by a different log,  then the corresponding deep euclidean  amplitude    
follows a   completely different logarithmic behaviour  to that from pQCD.

It is also possible \  to study the vector and axial correlators as separated  \ 
entities.  \ 
One could consider uncorrelated spectrums, such that they follow different  \ 
asymptotic behaviours.  There would be   \ 
two different expressions $\delta M_{V_j}^2$ and $\delta M_{A_j}^2$ for the \ 
squared mass inter-spacing between vectors and axial-vectors,  \ 
respectively, for high values of the masses.   \ 
This would imply that, instead Eq.~\eqn{eq.FNnec}, the couplings would be  \ 
given by  \ 
${F_{V_n}^2\simeq \delta M_{V_n}^2\cdot 
\frac{1}{\pi}\mbox{Im}\Pi_{_{VV}}(M_{V_n}^2)^{^{pQCD}}}$  
and  \ 
${F_{A_n}^2\simeq \delta M_{A_n}^2
\cdot \frac{1}{\pi}\mbox{Im}\Pi_{_{AA}}(M_{A_n}^2)^{^{pQCD}} }$,  
in order to recover $\Pi_{_{VV}}(-Q^2)^{^{pQCD}}$ 
and $\Pi_{_{AA}}(-Q^2)^{^{pQCD}}$ at high $Q^2$.  
The current analysis corresponds to the
case where the vectors and axial-vectors follow a similar law, although it can
be extended to more general frameworks in a straight-forward way.

\subsection{$V-A$ correlator}
\tab
The left-right correlator and their moments are  given 
in $QCD_\infty$ by the limit:
\be
\Frac{1}{\pi}\mbox{Im}\Pi_{_{LR}}(t)^{^{N_C\to\infty}} 
\,\,\,\, = \,\,\,\, 
-\, F^2 \, \delta(t) \,\,\,\, 
+ \,\,\,\, \displaystyle{\lim_{M_\infty^2\to\infty}\, 
\sum_{M_j^2\leq M_\infty^2 } }
\,\, F_j^2 \,\, [-\pi_j]\,\,\delta(t-M_j^2)\, \, ,
\ee
which provides the moments 
\bel{eq.LRcorre1}
\ba{rl}
\mA^{^{(n)}}_{_{LR}}(Q^2)^{^{N_C\to\infty}}\,\,\,=& \,\,\,-\,  \Frac{F^2}{Q^2} 
\,\,\,+ 
\,\,\, \displaystyle{\lim_{M_\infty^2\to\infty}\, 
\sum_{M_j^2\leq M_\infty^2 }}\, \, 
\Frac{ F_j^2\,\,[-\pi_j] \,\,\,\, \, Q^{2n}}{(M_j^2\,\, +\,\, Q^2)^{n+1}} \, , 
\ea
\ee
with $\pi_j$ the parity  of the  $j$--th multiplet, $(-1)$ for vectors and $(+1)$
for axial-vectors.  
This expressions are provided by the sum-rule 
integration in the circuit $C=C_{in}+C_{out}$ shown in Fig.~\eqn{fig.circuit}, 
\begin{figure}[t!]
\begin{center}
\includegraphics[width=5cm,angle=-90]{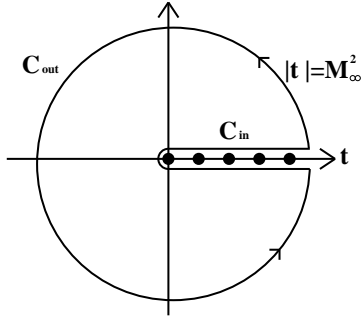}
\caption{Integration circuit for  the moment integrals.}
\label{fig.circuit}
\end{center}
\end{figure}
\bel{eq.LRintegral}
\mA^{^{(n)}}_{_{LR}}(Q^2)\,\, \,\, 
= \,\,\, \, \lim_{M_\infty^2\to\infty} 
\Frac{1}{2\pi i} 
\Oint_{C} \,\,\, \Frac{Q^{2n}\,\, dt}{(t+Q^2)^{n+1}}
\,\, \Pi_{_{LR}}(t) \, \,\,\, 
= \,\,\, \, 
\lim_{M_\infty^2\to \infty} \Int_0^{M_\infty^2}
\Frac{Q^{2n}\,\, dt}{(t+Q^2)^{n+1}}\,\, \, 
\Frac{1}{\pi}\mbox{Im}\Pi_{_{LR}}(t) \, ,
\ee
where the integration in $C_{out}$ is assumed to vanish as
$M_\infty^2\to\infty$.

The $V-A$ Green function will be also split into a perturbative and
non-perturbative sub-series, where the couplings and masses where formerly
fixed in the $V+A$~analysis.    
In this case, the summation is not positively defined and one needs to specify
the parity $\pi_j$ of every state. We will consider  a spectrum 
of multiplets with alternating parity where the lowest state is a vector, 
this is, $\pi_j=(-1)^{j}$.  
At high euclidean momenta $M_p^2\ll Q^2\ll M_\infty^2$, 
the perturbative sub-series  shows the structure  
\bel{eq.OPEpertLR}
\ba{rl}
\Delta\Pi_{_{LR}}(-Q^2)_{_{pert.}}^{^{N_C\to\infty}}
\, \, \, \, =& \,\,\, \, 
\displaystyle{\sum_{m=1}^\infty} \, \Frac{\Delta\bra
\cO_{_{(2m)}}^{^{LR}}\ket^{^{pert.}}}{Q^{2m}} \, \, ,
\ea
\ee
where no dimension-zero term arises since both the vector and axial correlators
reproduce the pQCD expression.  
The non-perturbative part of the
series can be analytically expanded without further problems 
and the summation of both contributions must match the OPE through the
constraints
\bel{eq.condeOPELR}
\ba{rcl}
\bra \cO_{_{(2)}}^{^{LR}}\ket \quad =& \quad \Delta\bra
\cO_{_{(2)}}^{^{LR}}\ket^{^{pert.}}
\,\,\, - \,\,\, F^2 \,\,\, 
+ \,\,\,\displaystyle{\sum_{j=1}^p} \, [-\pi_j] \, F_j^2
\,\,\, &= \,\,\, 0 \, , 
\\
\bra \cO_{_{(4)}}^{^{LR}}\ket\quad =& \qquad\Delta\bra
\cO_{_{(4)}}^{^{LR}}\ket^{^{pert.}}
\,\,\,\,\,\, 
- \,\,\,\,\,\, 
\displaystyle{\sum_{j=1}^p} \, [-\pi_j] \,  F_j^2  \, M_j^2
\,\,\, &= \,\,\, 0  \, .
\\
\bra \cO_{_{(6)}}^{^{LR}}\ket\quad =& \qquad\Delta\bra
\cO_{_{(6)}}^{^{LR}}\ket^{^{pert.}}
\,\,\,\,\,\, 
+ \,\,\,\,\,\, 
\displaystyle{\sum_{j=1}^p} \,[-\pi_j]  \,  F_j^2  \, M_j^4
\,\,\, &= \,\,\,  - \, 4 \, \pi\,  \alpha_s \, \bra \bar{q} q \ket^2 
\,\,\, < \,\,\, 0  \, .
\ea
\ee

\subsection{Study of some hadronical models for the spectrum}
\tab
In this section, we will study some of the current models for the
meson spectrum. 
The resonances with masses in the low energy region 
($M_n^2\lsim 2$~GeV$^2$) are 
susceptible to stronger deviation with respect to the expected 
asymptotic  behaviour $M_n^2=f(n)$ for high values of $n$.  
Below $2$~GeV$^2$, only two resonance multiplets are found,  
corresponding to the vector $\rho(770)$ and the axial--vector 
$a_1(1260)$~\cite{PDG}. The mass spectrum $M_n^2$ with
$n\geq 3$ is interpolated through 
the first two multiplets above
2~GeV$^2$, namely  $\rho(1450)$ with 
$M_3^2=M_{\rho'}^2\simeq 1.45^2$ and $a_1(1640)$ with
$M_4^2=M_{a_1'}^2\simeq 1.64$~GeV$^2$~\cite{PDG}.
We will include in the perturbative sub-series the multiplets  
with  $n> p=2$, leaving $M_1^2=M_\rho^2$ and $M_2^2=M_{a_1}^2$ for the
non-perturbative part of the resonance series.

In order to obtain the contribution to the condensates from the perturbative
sub-series we will analyse the $V+A$ and $V-A$ Adler functions 
from $\Delta \Pi^{^{N_C\to\infty}}_{_{pert.}}$ minus the
pQCD amplitude,  
$(\Delta\mA(Q^2)^{^{N_C\to\infty}}_{_{pert.}}
-\mA(Q^2)^{^{pQCD}})$.  
We will work only up to $\cO(\alpha_s)$ in the pQCD distribution so 
$\cO(\alpha_s^2)$ uncertainties are expected in the derivation of the 
condensates.  
Considering the Adler function  avoids 
complications like  
renormalization scale dependences or absolute convergence of
the resonance series. Nevertheless, the $V-A$ measurement is utterly improved
by a cross-check analysis of $\Pi_{_{LR}}(-Q^2)$.

We perform the interpolation 
$(\Delta\mA(Q^2_\ell)_{_{pert.}}^{^{N_C\to\infty}}
-\mA(Q^2_\ell)^{^{pQCD}})= \displaystyle{\sum_{m=1}^8} 
\Frac{m \,\, \Delta\bra\cO_{_{(2m)}}\ket^{^{pert.}}}{Q^{2m}_\ell}$ 
through eight points 
$Q^2_\ell=Q^2_1+\Delta Q^2\, (\ell-1)$, 
with $Q^2_1=4$~GeV$^2$, $\Delta Q^2=0.025$~GeV$^2$ and 
$\ell=1,2...8$  
From these eight constraints we extract the perturbative contribution to the
first eight condensates  $\Delta\bra\cO_{_{(2m)}}\ket^{^{pert.}}$.   
The interpolating point are varied to 
${Q^2_1=2,\, 3, \, 5, \, 6}$~GeV$^2$ in order to check the consistence of the
procedure and the size of the uncertainties.  
It is not possible to go to higher or lower energies since, respectively, 
on one hand,  the
$\cO(\alpha_s^2)$ corrections become comparable to the $\cO(1/Q^6)$ terms and, 
on the other, the OPE series breaks down within the very low energy range.
The central values of the  results represent the output for
$Q^2_1=4$~GeV$^2$ whereas the error bars express the spreading coming from  the
analysis at different energies, providing an estimate of the $\cO(\alpha_s)$
corrections.  
The errors 
in the condensates of dimension two, four and six  coming from neglecting
the $1/Q^{2m}$ terms with dimension beyond $2m=16$  
and from  the duality violating terms  are  expected to be rather
suppressed.  
We will make some considerations about the duality violations 
in the analysis of the five dimensional model although a more exhaustive
analysis can be found in Ref.~\cite{Cataduality}.

\subsubsection{$QCD_\infty$ in $1+1$ dimensions}
\tab 
In the pioneer work on large $N_C$~\cite{NC1}, QCD was study at
$N_C\to\infty$ in a configuration space of dimension $1+1$. 
In this case the spectrum could be solved and 
followed an asymptotic behaviour with constant
squared mass inter-spacing,  $\delta M_n^2=\delta \Lambda^2$.  
A similar type of  spectrum appears as well in different QCD 
models~\cite{EspriuRegge,LVS,PerisRegge}.     
We will consider $M_n^2=\Lambda^2 +n \,\delta \Lambda^2$ 
for $n\geq 3$, with $\delta \Lambda^2=(M_{a_1'}^2-M_{\rho'}^2)$ and 
$\Lambda^2=M_{\rho'}^2-3 \, \delta \Lambda^2$.

If one remains at $\cO(\alpha_s^0)$ the imaginary part of the pQCD  
correlator is just a constant and therefore all the couplings are equal to  
$F_n^2=\frac{N_C}{12\pi^2}\, \cdot  \, \delta \Lambda^2 \simeq (122$~MeV$)^2$. 
Thence, the perturbative sub-series 
contributions to the condensates derived from the numerical analysis are    
\bear
%\ba{rllcrll}
%%%\ba{rll}
\Delta\bra\cO_{_{(2)}}^{^{LL}}\ket^{^{pert.}} \quad =& \quad
 \left(-45.816^{+0.040}_{-0.011}\right)\cdot 10^{-3} \mbox{ GeV}^2
\quad & \sim \quad\,\,- \, 5\, F^2 \, ,
%& &
\nn \\ 
\Delta\bra\cO_{_{(4)}}^{^{LL}}\ket^{^{pert.}} \quad =& \quad
\left(41.00^{+0.22}_{-0.40}\right)\cdot 10^{-3}\mbox{ GeV}^4
\quad& \sim \quad\,\, \, 8\, F^2 \, M_\rho^2\, ,
%& &
\nn \\
\nn \\
\Delta\bra\cO_{_{(2)}}^{^{LR}}\ket^{^{pert.}} \,\,
\quad =& \quad\left(7.4349^{+0.0011}_{-0.0060}\right)\cdot 10^{-3}\mbox{ GeV}^2
\quad&\sim \quad\,\, F^2 \, ,
\nn \\
\Delta\bra\cO_{_{(4)}}^{^{LR}}\ket^{^{pert.}}
\quad =&\quad\left(-13.43^{+0.05}_{-0.12}\right)\cdot 10^{-3}\mbox{ GeV}^4
\quad&\sim \quad\,\, -\, 3 \, F^2 \, M_{\rho}^2\, ,
%&& &&
\nn \\
\Delta\bra\cO_{_{(6)}}^{^{LR}}\ket^{^{pert.}}
\quad =&\quad\left(23.4^{+0.9}_{-1.1}\right)\cdot 10^{-3}\mbox{ GeV}^6
\quad&\sim\quad \,\, 8 \, F^2 \, M_{\rho}^4\, .
%%%\ea
\eear

In this model, one may actually compute the exact value of the condensates
coming from the perturbative sub-series~\cite{Cataduality,toymodel}:
\bear
\Delta\bra\cO_{_{(2m)}}^{^{LL}}\ket^{^{pert.}}\,\,&=& \,\, 
\Frac{(-1)^m}{m}  
\,\, \Frac{N_C}{12\pi^2} \left(\delta \Lambda^2\right)^m 
\, \, B_m\left(\frac{M_{\rho'}^2}{\delta \Lambda^2}\right) \, ,
\nn  \\
\Delta\bra\cO_{_{(2m)}}^{^{LR}}\ket^{^{pert.}}\,\,&=& \,\, 
\Frac{(-1)^m}{m}  
\,\, \Frac{N_C}{24\pi^2} \left(2\, \delta \Lambda^2\right)^m 
\, \, \left[\, 
B_m\left(\frac{M_{\rho'}^2}{2\, \delta \Lambda^2}\right) 
\, - \, 
B_m\left(\frac{M_{a_1'}^2}{2\, \delta \Lambda^2}\right) 
\, \right]
\, ,
\eear
being \  \  $B_m(x)$ \  \  the  \  \ Bernoulli   \ \ polynomials \  \  ($B_1(x)=x-\frac{1}{2}$,
 \  \ $B_2(x)=x^2-x+\frac{1}{6}$,  \  \  ${B_3(x)=x^3-\frac{3}{2}x^2+\frac{1}{2}x,\, ...}$).
For our choice of parameters one finds 
\bear
\Delta\bra\cO_{_{(2)}}^{^{LL}}\ket^{^{pert.}}\, \,\,\,
=&\, \, -\, \Frac{N_C}{12\pi^2}\,M_{\rho'}^2\, 
\left[1-\frac{\delta\Lambda^2}{2\, M_{\rho'}^2}\right] \,\,   
&=\, \,\,\, -45.821\cdot 10^{-3}~\mbox{GeV}^2\, , 
\nn 
\\ 
\Delta\bra\cO_{_{(4)}}^{^{LL}}\ket^{^{pert.}}\, \,\,\, 
=&\, \, \,\Frac{1}{2}\,  \Frac{N_C}{12\pi^2}\,M_{\rho'}^4\, 
\left[1-\frac{\delta\Lambda^2}{M_{\rho'}^2}
+\frac{(\delta\Lambda^2)^2}{2\, M_{\rho'}^4}\right] \,\,   
&= \,\,\, \, 41.08\cdot 10^{-3}~\mbox{GeV}^4\, , 
\nn 
\\
\\
\Delta\bra\cO_{_{(2)}}^{^{LR}}\ket^{^{pert.}}\, \,\,\, =
& \,\, \Frac{N_C}{24\pi^2}\,\delta\Lambda^2\,\,   
&= \,\,\, \, 7.4357\cdot 10^{-3}~\mbox{GeV}^2\, , 
\nn 
\\  
\Delta\bra\cO_{_{(4)}}^{^{LR}}\ket^{^{pert.}}\, \,\,\,=
&\,\, - \, \Frac{N_C}{24\pi^2}\,\delta\Lambda^2\, M_{\rho'}^2\,   
\left[1-\frac{\delta\Lambda^2}{2\, M_{\rho'}^2}\right] \,\,   
&=\,\,\,\, -13.45\cdot 10^{-3}~\mbox{GeV}^4\, , 
\nn 
\\ 
\Delta\bra\cO_{_{(6)}}^{^{LR}}\ket^{^{pert.}}\, \,\,\, =
&\,\,  \Frac{N_C}{24\pi^2}\,\delta\Lambda^2\, M_{\rho'}^4\,   
\left[1-\frac{\delta\Lambda^2}{M_{\rho'}^2}\right] \,\,   
&=\,\,\,\, 23.7\cdot 10^{-3}~\mbox{GeV}^6\, , 
\nn 
\eear
in total agreement with our former numerical calculation. This  
provides a check of the accuracy that will make our 
next numerical calculations reliable.  
The $V-A$ condensates depend linearly on $\delta M_j^2=\delta \Lambda^2$ 
and both  
$\Delta\bra\cO_{_{(2m)}}^{^{LL}}\ket^{^{pert.}}$  and 
$\Delta\bra\cO_{_{(2m)}}^{^{LR}}\ket^{^{pert.}}$ have a strong dependence on   
the mass of the lowest state included in the perturbative
sub-series, $M_{\rho'}$. 
In the limit  $\frac{\delta M_j^2}{M_{\rho'}^2}\to 0$, one
recovers exactly the pQCD  
expression $\Pi(-Q^2)^{^{pQCD}}_{_{pert.}}$ at
$\cO(\alpha_s^0)$ from Eq.~\eqn{eq.pQCDpert}, 
with $\Delta\bra\cO_{_{(2m)}}^{^{LL}}\ket^{^{pert.}}
=\Frac{(-1)^m}{m} \, t_p^m$ and 
$\Delta\bra\cO_{_{(2m)}}^{^{LR}}\ket^{^{pert.}}=0$, being $t_p=M_{\rho'}^2$.  
The condensates result naturally of the expected size 
without imposing further
constraints, pointing out the close relation between the different QCD scales, 
hadronic masses, mass inter-spacings, couplings  and condensates.

We will  take  $F=92.4$~MeV, $M_\rho=0.77$~GeV, the asymptotic spectrum 
$M_n^2=\Lambda^2 + n \,\delta \Lambda^2$ for $n\geq 3$ and  
the pQCD correlator $\frac{1}{\pi}$Im$\Pi_{_{LL}}(q^2)$ 
as inputs,   and the parameters     
$F_\rho,\, F_{a_1}$ and $M_{a_1}$ will be left 
as free parameters. They can be recovered 
by demanding  that the  
condensates $\bra\cO_{_{(2)}}^{^{LL}}\ket$, 
$\bra\cO_{_{(2)}}^{^{LR}}\ket$ and  $\bra\cO_{_{(4)}}^{^{LR}}\ket$ are zero. 
For a general resonance 
model, Eqs.~\eqn{eq.condeOPELL}~and~\eqn{eq.condeOPELR}  yield the relations
\be
\ba{rl}
F_\rho^2 \,\, =&\,\,   -  \,\, \Frac{1}{2}\,\, \left(
\Delta\bra\cO_{_{(2)}}^{^{LL}}\ket^{^{pert.}} 
+ \Delta\bra\cO_{_{(2)}}^{^{LR}}\ket^{^{pert.}}\right) \, ,
\\
\\
F_{a_1}^2\,\, =&\,\,   \,\,-\,\, \Frac{1}{2}\,\, \left(
\Delta\bra\cO_{_{(2)}}^{^{LL}}\ket^{^{pert.}} 
- \Delta\bra\cO_{_{(2)}}^{^{LR}}\ket^{^{pert.}} \right) \,\,\,\, - \,\,\,\, F^2 \, ,
\\
\\
M_{a_1}^2 \,\, =&
\Frac{F_\rho^2 \, M_\rho^2 \,
- \, \Delta\bra\cO_{_{(4)}}^{^{LR}}\ket^{^{pert.}}}{F_{a_1}^2} =   
 M_\rho^2  
\left\{
\Frac{\, - \,\Frac{1}{2}\,
\left(\Delta\bra\cO_{_{(2)}}^{^{LL}}\ket^{^{pert.}} 
+ \Delta\bra\cO_{_{(2)}}^{^{LR}}\ket^{^{pert.}} \right) \,
- \, \Frac{\Delta\bra\cO_{_{(4)}}^{^{LR}}\ket^{^{pert.}} 
}{M_\rho^2}
}{ \,\,-\,\, \left(\Delta\bra\cO_{_{(2)}}^{^{LL}}\ket^{^{pert.}} 
- \Delta\bra\cO_{_{(2)}}^{^{LR}}\ket^{^{pert.}}\right)\,\,\,\, - \,\,\,\, F^2 \,\,
}
\right\}  .
\ea
\ee    
In our case at $\cO(\alpha_s^0)$ one finds 
\be
F_\rho=138.5 \mbox{ MeV}\, , \qquad 
F_{a_1}=134.5 \mbox{ MeV} \, , \qquad 
M_{a_1} \,\, = \,\, 1172  \mbox{ MeV} \, . 
\ee
The couplings remain of the order of the asymptotic value $F_\rho\sim
F_{a_1}\sim F_{j\geq 3}$, falling both resonance couplings 
and axial-vector mass within the expected
range of values obtained in former phenomenological 
analysis~\cite{therole,anchura,GPP:04,PDG}. 
These light states, laying by  the non-perturbative QCD regime, 
suffer slight deviations from the asymptotic behaviour 
in such a way that the OPE is exactly recovered.

The short distance matching  fixes all the parameters in our analysis, 
providing a prediction for the condensates of higher dimension 
through Eqs.~\eqn{eq.condeOPELL}~and~\eqn{eq.condeOPELR}:
\be
\bra\cO_{_{(4)}}^{^{LL}}\ket \,\,\, =\,\,\,  
4.87 \cdot 10^{-3}  \mbox{ GeV}^4
\, ,
\qquad 
\bra\cO_{_{(6)}}^{^{LR}}\ket
\,\,\, = \,  \,\,-\, 3.63 \cdot 10^{-3}  \mbox{ GeV}^6\, . 
\ee
They \ already \  fall  \ pretty  \ near \  the \  usual  \ determinations  \ 
$\bra\cO_{_{(4)}}^{^{LL}}\ket
\simeq 1.2\cdot 10^{-3}$~GeV$^4$~\cite{YndurainAGG}  \ 
and  \ 
${\bra\cO_{_{(6)}}^{^{LR}}\ket \simeq 
- 3.8\cdot 10^{-3}}$~GeV$^6$~\cite{narisonpid}.

\begin{table}
\begin{center}
\begin{tabular}{|c|c|c|c|c|}
\hline
\rule[-0.7em]{0em}{1.9em}
 & 1+1 $QCD_\infty$ &
5D--spectrum&
Conv. WSR1 &
Conv. WSR2
\\
 \hline\rule{0em}{1.2em}
$\Delta\bra\cO_{_{(2)}}^{^{LL}}\ket^{^{pert.}}$ ($10^{-3}$~GeV$^2$)& 
$-46.9^{+1.7}_{-1.9}$ & 
$-68^{+5}_{-4}$ & 
$-37.4^{+2.1}_{-3.0}$  & 
$-28\pm 6$  
\\
 \hline\rule{0em}{1.2em}
$\Delta\bra\cO_{_{(4)}}^{^{LL}}\ket^{^{pert.}}$ ($10^{-3}$~GeV$^4$) & 
$8^{+23}_{-18}$ & 
$140^{+70}_{-60}$  & 
$-30\pm 40$   & 
$-80^{+70}_{-110}$  
\\
 \hline\rule{0em}{1.2em}
$\Delta\bra\cO_{_{(2)}}^{^{LR}}\ket^{^{pert.}}$ ($10^{-3}$~GeV$^2$) & 
$8.3771^{+0.0021}_{-0.0070}$ & 
$6.832^{+0.003}_{-0.006}$   & 
$10.902^{+0.005}_{-0.008}$   & 
$14.058^{+0.008}_{-0.011}$  
\\
 \hline\rule{0em}{1.2em}
$\Delta\bra\cO_{_{(4)}}^{^{LR}}\ket^{^{pert.}}$ ($10^{-3}$~GeV$^4$) & 
$-15.15^{+0.14}_{-0.08}$ & 
$-12.34^{+0.12}_{-0.11}$   & 
$-19.90^{+0.18}_{-0.21}$    & 
$-29.97^{+0.30}_{-0.25}$  
\\
 \hline\rule{0em}{1.2em}
$\Delta\bra\cO_{_{(6)}}^{^{LR}}\ket^{^{pert.}}$  ($10^{-3}$~GeV$^6$)
& 
$26.5\pm 1.3$ & 
$21.7^{+2.0}_{-1.1}$   & 
$34.9^{+4.0}_{-1.6}$   & 
$46.2^{+6.0}_{-2.2}$  
\\
\hline\end{tabular}
\caption{Contributions from the perturbative sub-series to the 
condensates within different 
hadronical models. The columns with Converging WSR1 and WSR2 corresponds to 
squared mass inter-spacings of the form ${\delta M_n^2\sim 1/\sqrt{n}}$ and 
${\delta M_n^2\sim 1/n}$, respectively. All the amplitudes are considered up to
$\cO(\alpha_s)$.} 
\vspace*{-1em}
\label{tab.condepert}
\end{center}
\end{table}
\begin{table}
\begin{center}
\begin{tabular}{|c|c|c|c|c|}
\hline
\rule[-0.7em]{0em}{1.9em}
& 1+1 $QCD_\infty$ &
5D--spectrum&
Conv. WSR1 &
Conv. WSR2
\\
 \hline\rule{0em}{1.2em}
$F_\rho$ (MeV) & 
$139\pm 3$  & 
 $174^{+5}_{-7}$  & 
 $115^{+7}_{-5}$   & 
 $83^{+16}_{-20}$   
\\
 \hline\rule{0em}{1.2em}
$F_{a_1}$ (MeV) &  
$138\pm 3 $ & 
$169^{+6}_{-8} $ &
$125^{+6}_{-4} $  & 
$111^{+12}_{-14} $  
\\
 \hline\rule{0em}{1.2em}
$M_{a_1}$ (MeV) & 
$1180^{+17}_{-19}$ & 
 $1029^{+19}_{-13}$  & 
 $1330\pm 40$   & 
 $1560^{+180}_{-120}$   
\\
 \hline\rule{0em}{1.2em}
$\bra\cO_{_{(4)}}^{^{LL}}\ket$   ($10^{-3}$~GeV$^4$)& 
$-30^{+22}_{-30}$ & 
$100^{+70}_{-60}$  & 
$-70\pm 40$  & 
$-120^{+70}_{-110}$  
\\
 \hline\rule{0em}{1.2em}
$\bra\cO_{_{(6)}}^{^{LR}}\ket$   ($10^{-3}$~GeV$^6$)
& 
$-3.7^{+0.8}_{-0.3}$ & 
$0.3^{+2.2}_{-1.5}$   & 
$-9.7^{+1.5}_{-0.7}$ & 
$-24^{+7}_{-8}$ 
\\
\hline\end{tabular}
\caption{Predictions for the two first resonance multiplets --$\rho(770)$ and
$a_1(1260)$-- within each model  for the inputs $F=92.4$~MeV and
$M_\rho=0.77$~GeV. We consider the amplitudes up to $\cO(\alpha_s)$. } 
\vspace*{-1em}
\label{tab.predictions}
\end{center}
\end{table}

The calculation can be taken  up to $\cO(\alpha_s)$ so 
the $\alpha_s(Q^2)$ running in $\mA^{^{(n)}}_{_{LL}}(Q^2)$ is recovered. 
The next-to-leading order computation is relevant in order to
improve the determinations of the condensates 
$\bra\cO_{_{(4)}}^{^{LL}}\ket$ and 
$\bra\cO_{_{(6)}}^{^{LR}}\ket$, checking the impact coming from  
the perturbative QCD corrections
in $\alpha_s$. 
The contributions to the condensates at this order   
that come from the perturbative sub-series may be found
in the first column of  Table~\eqn{tab.condepert},  
where we have considered the same values for $\alpha_s(Q^2)$ and the pQCD 
correlator as in Section~3.
In Table~\eqn{tab.predictions}, it is possible to find  
the corresponding $\rho(770)$ 
and $a_1(1260)$ couplings and masses, and  
the value of the condensates  
$\bra\cO_{_{(4)}}^{^{LL}}\ket$ and  $\bra\cO_{_{(6)}}^{^{LR}}\ket$ 
derived from the short distance matching 
$\bra\cO_{_{(2)}}^{^{LL}}\ket=\bra\cO_{_{(2)}}^{^{LR}}\ket=
\bra\cO_{_{(4)}}^{^{LR}}\ket=0$.       
Small variations arise with respect  
to the $\cO(\alpha_s^0)$ result, as expected if the
asymptotic behaviour of  resonance parameters depends smoothly on $\alpha_s$. 
However, the $\cO(\alpha_s^2)$ uncertainties increase 
the error in the $V+A$ condensates by two orders of
magnitude, where $\bra\cO_{_{(4)}}^{^{LL}}\ket$ becomes now negative, 
pointing out the necessity of working at that order if more accurate
determinations are required.

\subsubsection{Five--dimensional spectrum}
\tab 
Another available scenario that has appeared recently 
is the one provided by models in five dimensions~\cite{Son,Com}. 
Here the resonances appear as Kaluza-Klein modes from the quantization of the
momentum in the fifth dimension, producing a four dimensional   
effective spectrum with the dependence 
$M_n^2\sim n^2$, this is, $\delta M_n^2\sim n \sim \sqrt{M_n^2}$.   
We will use the interpolation $M_n=\Lambda+ n \, \delta \Lambda$, 
with $\delta \Lambda=(M_{a_1'}-M_{\rho'})$ and $\Lambda=
M_{\rho'}-3\delta\Lambda$.
The corresponding contributions to condensates coming  from 
the perturbative sub-series are    shown in Table~\eqn{tab.condepert}. 
Taking these values and the inputs $F$ and $M_\rho$ one derives the 
resonance parameters 
and the predictions for the condensates shown in Table~\eqn{tab.predictions}.
One finds an acceptable value for the $a_1(1260)$ mass although the values of
the resonance couplings go high above the usual determinations. 
The predictions for the condensates $\bra\cO^{^{LL}}_{_{(4)}}\ket$ and 
$\bra\cO^{^{LR}}_{_{(6)}}\ket$ appear  slightly shifted with respect to
1+1~$QCD_\infty$ though  the positive sign for 
$\bra\cO^{^{LL}}_{_{(4)}}\ket$ is properly restored.

It is important to remark that this is not exactly a five dimensional
calculation  but an
effective study of its spectrum.    
The series of infinite  resonances   
in the 5D--theory are not regulated in the way of our cut-off, yielding some
extra features as the existence of infinite Weinberg sum rules~\cite{Com}.     
In the strict five dimensional
calculation one must handle  the full
Kaluza-Klein pile (including also the first two resonances) 
with its precise dependence. For instance, one has     
$M_n^2=\frac{\pi^2}{4\, l_1^2}\left(n+\frac{1}{2}\right)^2$ 
and $F_n^2=\frac{N_C}{24\pi^2}\cdot \frac{\pi^2}{2\,l_1^2}
\left(n+\frac{1}{2}\right)$ for the RS1~metric in
Ref.~\cite{Com},     
being $l_1= 3\pi/(4\, M_\rho)$ the position of the infrared brane.  
This exact structure of the spectrum 
generates a $V-A$ amplitudes without condensates,
\bel{eq.duaviola}
\Pi_{_{LR}}(-Q^2)\,\,\,\,
\stackrel{Q^2\gg M_\rho^2}{\simeq }\,\,\,\, 
- \, \Frac{N_C}{12\pi} \,\, \exp{
\left[\, -\, \Frac{3\pi\, Q}{2\, M_\rho}\, \right]} \, .
\ee

For the perturbative QCD range of euclidean momenta  
the exponential factor produce a huge suppression   
($\cO(10^{-4})$ already for $Q^2=2$~GeV$^2$ and $\cO(10^{-9})$ for 
$Q^2=10$~GeV$^2$). 
Variations on the lightest 
Kaluza-Klein modes due to $\cO(\alpha_s)$ corrections (present analysis) 
or  perturbations on the
metric~\cite{Com} result 
into the appearance of $1/Q^{2m}$ power terms.
The duality violating term in Eq.~\eqn{eq.duaviola}  affects our numerical  OPE 
interpolation  beyond  the dimension--sixteen condensate, so the 
determinations are still safe.     
Nevertheless, the exponential may become enhanced with respect to the $1/Q^{2m}$
terms when considering moment
integrals $\mA^{^{(k)}}(Q^2)$ or components $\mB^{^{(k)}}(Q^2)$ 
for large values of $k$.  Eventual contributions might produce observable
modifications to the usual OPE calculations~\cite{Cataduality}.

\subsubsection{Converging Weinberg sum rules}
\tab
For  \  a   general  \ spectrum  \ and  \ within  \ our   considered \  cut-off regularization,  \ 
large   $N_C$   QCD  \ do   not   obey  \  in  general     the two Weinberg sum rules (WSR),  \ 
$\Int_0^\infty dt \, \frac{1}{\pi}$Im$\Pi_{_{LR}}(t)=0$  \  and  \ 
${\Int_0^\infty dt \,\, t\, \frac{1}{\pi}\mbox{Im}\Pi_{_{LR}}(t)=0}$, since the
infinite summations of resonances are not convergent.   
The fulfilling of these  
two WSR  immediately implies a dominant behaviour 
$\Pi_{_{LR}}(-Q^2)\sim \Frac{1}{Q^6}$ 
in the deep euclidean.     
%although  the inverse is not true at all;    
%for instance,  t'~Hooft's large $N_C$ model,  1+1~$QCD_\infty$~\cite{NC2},  
%has a proper 
%euclidean OPE expansion for the $V-A$ correlator but its WSR are not
%convergent. The direct identification of some sum rules with the 
%condensates may lead to wrong determinations if one does not take into account
%the infinite resonance structure of QCD. 
%Likewise, 
%the infinite hadron interactions  generate duality violating terms and      
%the effects due to non-zero  resonance 
%widths are not  trivial~\cite{Cataduality}. 
%
%Although there is no argument that ensures that the WSR are obeyed 
%when $N_C\to \infty$, 

One may consider the covergence of the two WSR 
as an available hypothesis for the
model building. In this case, the summations 
\bel{eq.sumrules}
\Int_0^\infty \, dt \,\,\, t^m \,\,\, \Frac{1}{\pi}\mbox{Im}\Pi_{_{LR}}(t)
\,\,\,\, = \,\,\,\, 
\,\,\,\,- \,\,  F^2 \, \delta^{m,0} \,\, + \,\, 
\displaystyle{\sum_{j=1}^\infty }  \, [-\pi_j] \,\, F_j^2 \,\, 
(M_j^2)^m \, ,
\ee 
must be zero (and therefore converging) for $m=0$ and  \ 
$m=1$~\cite{Weinbergsumrules}. This demands  \ 
that $F_n^2\stackrel{n\to\infty}{\longrightarrow}0$ and  \ 
$F_n^2\, M_n^2\stackrel{n\to\infty}{\longrightarrow}0$.   \ 
If we desire  just two Weinberg sum rules,  \ 
then $F_n^2\, (M_n^2)^m$ must not vanish when $n\to \infty$ for $m\geq 2$.  \  
If the asymptotic value of the couplings is fixed through  \ 
${F_n^2\simeq \delta M_n^2\,\cdot\, 
\frac{1}{\pi}\mbox{Im}\Pi_{_{LL}}(M_n^2)^{^{pQCD}}\sim \delta M_n^2}$, 
one finds in Eq.~\eqn{eq.sumrules}  
that   $F_n^2 \, (M_n^2)^{m}\sim  \delta M_n^2 \cdot (M_n^2)^m$ must vanish when 
$M_n^2\to\infty$  just for $m=0$ and $m=1$;  the inter-spacing    
$\delta M_n^2$ must tend to zero as $n$ goes to infinity faster than 
$\delta M_n^2\sim \frac{1}{\sqrt{n}} \sim \frac{1}{M_n^2}$  and 
slower than $\delta M_n^2\sim \frac{1}{n} \sim e^{-M_n^2}$.

We will analyse both extreme cases through two different modelings of the
spectrum.  \  On one hand,  \  we \  will \  consider \  the model WSR1,  \ 
owning \  the  \  spectrum  \ 
${M_n^2=\Lambda^2 +\sqrt{n} \,\delta \Lambda^2}$  with  \ 
${\delta \Lambda^2=(M_{a_1'}^2-M_{\rho'}^2)/(\sqrt{4}-\sqrt{3})}$ and  \ 
$\Lambda^2=M_{\rho'}^2-\sqrt{3} \, \,\delta\Lambda^2$.   \ 
This  \ provides  \ the    limit \  case   \  
${\delta M_n^2\sim \frac{1}{\sqrt{n}} \sim \frac{1}{M_n^2}}$.   \ 
For the other limit \  case we consider the model WSR2, \  with  \ 
spectrum  \ 
${M_n^2=\Lambda^2 + \delta \Lambda^2 \ln{(n)}}$, 
with $\delta \Lambda^2=(M_{a_1'}^2-M_{\rho'}^2)/(\ln{(4)}-\ln{(3)})$ and 
$\Lambda^2=M_{\rho'}^2-\ln{(3)} \, \, \delta\Lambda^2$.
This model owns  the high mass  behaviour 
$\delta M_n^2\sim \frac{1}{n} \sim e^{-M_n^2}$.

The results for both kinds of spectrums up to $\cO(\alpha_s)$ 
can be found in the third and fourth
columns of Tables~\eqn{tab.condepert}~and~\eqn{tab.predictions}.
The axial-vector mass goes beyond the usual  range 
{$1$~GeV$\lsim M_{a_1}\lsim
1.26$~GeV,} whereas  the resonance couplings 
decrease up to unphysical values, what seems to rule out this kind of 
converging--WSR models.

\subsection{Truncated resonance theory and 
 Minimal Hadronical Approximation}
\tab 
The contribution \ to  \ the  \ OPE  \ condensates  \ coming \  
from  \ the \  high \ 
mass  \ multiplets  \ vary \  depending  \ on  \ the  \ model \  for  \ the hadronic spectrum.  \ 
However,  \  all  \ of them \  reproduce 
pQCD  \  and  \ generate   \ 
contributions  \ to  \ the \  condensates  \ 
of \  the \  right \  size  \ --the  \ standard \  hadronical \  parameters  \ $F^2$ and $M_\rho^2$--  
 \ without  \ demanding  \ any  \ further  \ fine  \ tuning,   \ 
${ \left|\Delta \bra\cO_{_{(2m)}}^{^{LL}}\ket^{^{pert}}\right|\, , \,\,\, 
\left|\Delta\bra\cO_{_{(2m)}}^{^{LR}}\ket^{^{pert}}\right|\,
\quad \sim \quad \cO\left(F^2 (M_\rho^2)^{m-1} \right)}$.

The matching with the OPE can be improved  through  
a more exhaustive scanning of the resonance parameters (e.g. 
$\Lambda^2$ and $\delta
\Lambda^2$ in the 1+1~$QCD_\infty$ spectrum)  
and analysing possible  corrections to the asymptotic dependence of  $M_n^2$.   
Likewise, the second vector multiplet,  with $M_{\rho'}^2\sim
2$~GeV$^2$, lies by the border of the non-perturbative QCD region and it may
suffer still sizable variations with respect to the asymptotic behaviour. 
A deeper study should  consider the three resonances  
$\rho(770)$, $a_1(1260)$ and $\rho(1450)$ apart,  
within the non-perturbative sub-series.

Nonetheless, in many situations the inclusion of the full infinite  pile of 
hadronic states is quite involved.  
The information about higher mass states is rather poor, so one cannot yield a
precise determination of the perturbative sub-series 
$\Delta\Pi(-Q^2)_{_{pert.}}^{^{N_C\to\infty}}=\Pi(-Q^2)^{^{pQCD}}+\displaystyle{\sum_{m=1}^\infty} 
\Frac{\Delta\bra\cO_{_{(2m)}}\ket^{^{pert.}}}{Q^{2m}}$.
This forces to truncate the resonance  tower and to consider just a finite
number of states, those in $\Delta\Pi(-Q^2)_{_{non-p.}}^{^{N_C\to\infty}}$, setting 
$\Delta\Pi(-Q^2)_{_{pert.}}^{^{N_C\to\infty}}=0$.  
  The  separation
between perturbative and non-perturbative series is not clear 
and, formally, one may consider as many states as desired within 
$\Delta\Pi(-Q^2)_{_{non-p.}}^{^{N_C\to\infty}}$.  Moreover, in order 
to reproduce the right  long/short--distance behaviour of QCD ($\chi PT$/OPE) 
one needs to include at least 
a minimal number of multiplets in the resonance theory. This is denoted as 
Minimal Hadronical Approximation (MHA)~\cite{KPdR}. 
Strictly speaking,  MHA refers only to Green-functions which 
are order parameters 
although  some ansate replace $\Delta\Pi(-Q^2)_{_{pert.}}^{^{N_C\to\infty}}$    
by a pQCD continuum~\cite{matchingDR,711}.

In the MHA analysis of the $V-A$ correlator, 
one needs to keep at least the Goldstones from the chiral symmetry
breaking,  together with the first vector and axial-vector multiplets:
\be
\Pi_{_{LR}}(-Q^2)^{^{N_C\to\infty}}\,\, = \,\, -\, \Frac{F^2}{Q^2}\,\, 
+ \,\, \Frac{F_\rho^2}{M_\rho^2+Q^2} 
\,\,- \,\, \Frac{F_{a_1}^2}{M_{a_1}^2+Q^2} \,\,\,\, 
+ \,\,\,\, \cO\left(\Frac{\delta M^2}{Q^2}\right)\,\, . 
\ee
The terms $\cO\left(\frac{\delta M^2}{Q^2}\right)$  from
$\Delta\Pi_{_{LR}}(-Q^2)_{_{pert.}}^{^{N_C\to\infty}}$ are dropped out in MHA.   
Through a short--distance matching  to the dominant power behaviour of the OPE 
($\Pi_{_{LR}}(-Q^2)\sim 1/Q^6$),   
the values of the lightest resonance parameters are constrained, 
getting the familiar relations~\cite{Weinbergsumrules,spin1fields,PI:02},
\bel{eq.WSR}
\widetilde{F}_\rho^2\, - \, \widetilde{F}_{a_1}^2\, =\, F^2 \, , 
\qquad \qquad \widetilde{F}_\rho^2\, \widetilde{M}_\rho^2\, 
- \, \widetilde{F}_{a_1}^2 \, \widetilde{M}_{a_1}^2 \, = 0\, .
\ee
The couplings and masses are now substituted by some effective parameters 
($F_{\rho/a_1}\to \widetilde{F}_{\rho/a_1}$ and 
$M_{\rho/a_1}\to \widetilde{M}_{\rho/a_1}$) where the information of the
perturbative sub-series is also encoded. They suffer 
corrections proportional to $\Delta\bra \cO_{_{(2m)}}^{^{LR}}\ket^{^{pert.}}$ 
with respect to their  
original values in the full theory and some information is lost since
the number of hadronic parameters decreases considerably. 
For the experimental values $\widetilde{M}_\rho=0.77$~GeV and  
$\widetilde{F}_\rho\simeq 154$~MeV, coming from the 
decay $\rho^0\to e^+ e^-$~\cite{therole}, one finds the predictions  
$\widetilde{F}_{a_1}=\sqrt{\widetilde{F}_\rho^2-F^2}\simeq 123$~MeV,  
$\widetilde{M}_{a_1}=\widetilde{F}_\rho
\widetilde{M}_\rho/\widetilde{F}_{a_1}\simeq 964$~MeV and 
${\bra \widetilde{\cO}_{_{(6)}}^{^{LR}}\ket=
-\Frac{\widetilde{F}_\rho^2}{\widetilde{F}_{a_1}^2}\, F^2\,
\widetilde{M}_\rho^4\simeq -4.7\cdot 10^{-3}}$~GeV$^6$. 
\\
\tab
A further analysis of the vector and axial form 
factors~\cite{spin1fields} leads  to  
${\widetilde{F}_\rho=\sqrt{2}\, F \simeq 131}$~MeV,   
${\widetilde{F}_{a_1}=F\simeq 92.4}$~MeV,    
$\widetilde{M}_{a_1}=\sqrt{2}\, \widetilde{M}_\rho\simeq 1089\mbox{ MeV}$ and 
$\bra\widetilde{\cO}_{_{(6)}}^{^{LR}}\ket=-2 F^2 \widetilde{M}_\rho^4 \simeq 
- 6\cdot 10^{-3}$~GeV$^6$.   One must be aware of the intrinsic uncertainty
laying in the truncated resonance theory  (which must not be misleadingly taken
as model dependence).   
Performing a short--distance matching for 
a wider set of  matrix elements modifies the value of the parameters and one
risks to reach inconsistencies between the different constraints, 
eventually requiring the introduction the
next resonance multiplet in the MHA description.

The \  difference between  \ the QCD parameters   \ 
in MHA  \ and those derived \  through other techniques     \ 
is \  just  \ due  \ to  \ the absence of the perturbative part of the resonance series.  \ 
The \  uncertainty \  in \  the  \ MHA  \ short-distance \  matching  \ to  \ OPE  \ is  \ 
given \  by \  the \  size  \ of  \  ${\Delta \bra\cO_{_{(2m)}}^{^{LR}}\ket^{^{pert.}}\sim 
\cO\left(\delta M^2 (M_{\rho'}^2)^{m-1}\right)}$.   
The smooth variation of the resonance parameters from MHA to
MHA+$\rho'$~\cite{SRFriot}  
hints its close relation with the full  large $N_C$ resonance theory.

The employment of this approximation has led to a very successful large $N_C$ 
phenomenology~\cite{PI:02}: 
study of vector, axial-vector and scalar form-factors~\cite{spin1fields, JOP}, 
determination of $\chi PT$ couplings~\cite{therole,spin1fields,PI:02,
3puntosandchiral,extrachiral1,extrachiral2}, 
description of 
two--point Green functions~\cite{SRFriot,
Weinbergsumrules,711,spin1fields,extrachiral1}  
and three-point QCD Green functions~\cite{3puntos,3puntosandchiral,
multipuntosextra}.  
In addition, once the theory is well founded at leading order in $1/N_C$,
unitarity imposes serious constraints on the next-to-leading order effects.
Thus, the final state interaction admits a well defined description 
within the  MHA
framework~\cite{HLS,anchura,GP:97,PP:01,JOP,Palomar,
ND,GPP:04,PaP:01}. Likewise, some calculations have affronted the
problem of the renormalization and the radiative corrections to the couplings
at next-to-leading order in
$1/N_C$~\cite{quantumloops,CP:01,Natxoprepara,BGT:98,ChR:98,
Meissnerloops,HLS}.

\begin{figure}[t!]
\begin{center}
\includegraphics[width=7cm,clip]{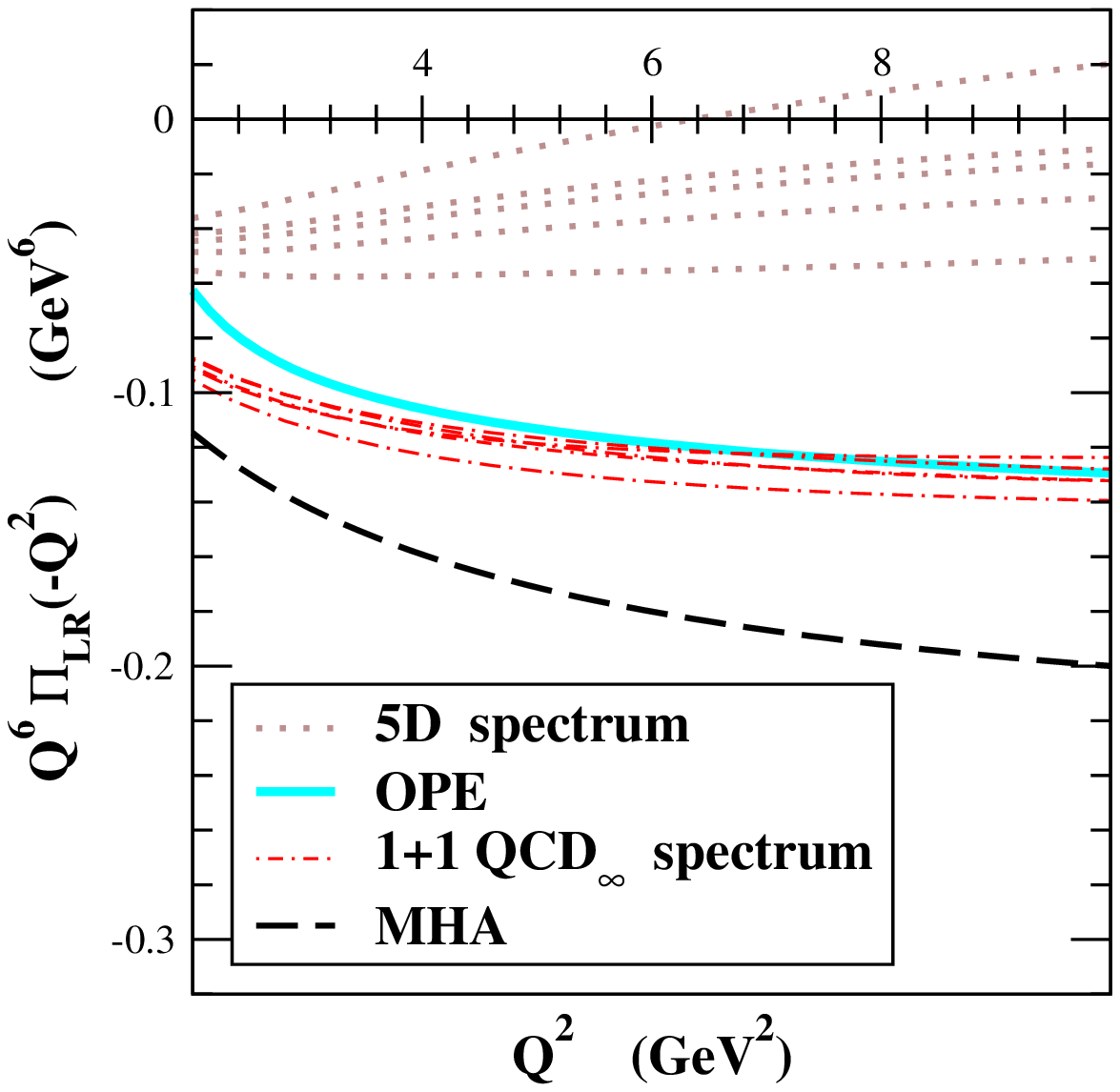}
\hspace*{1cm}
\includegraphics[width=7cm,clip]{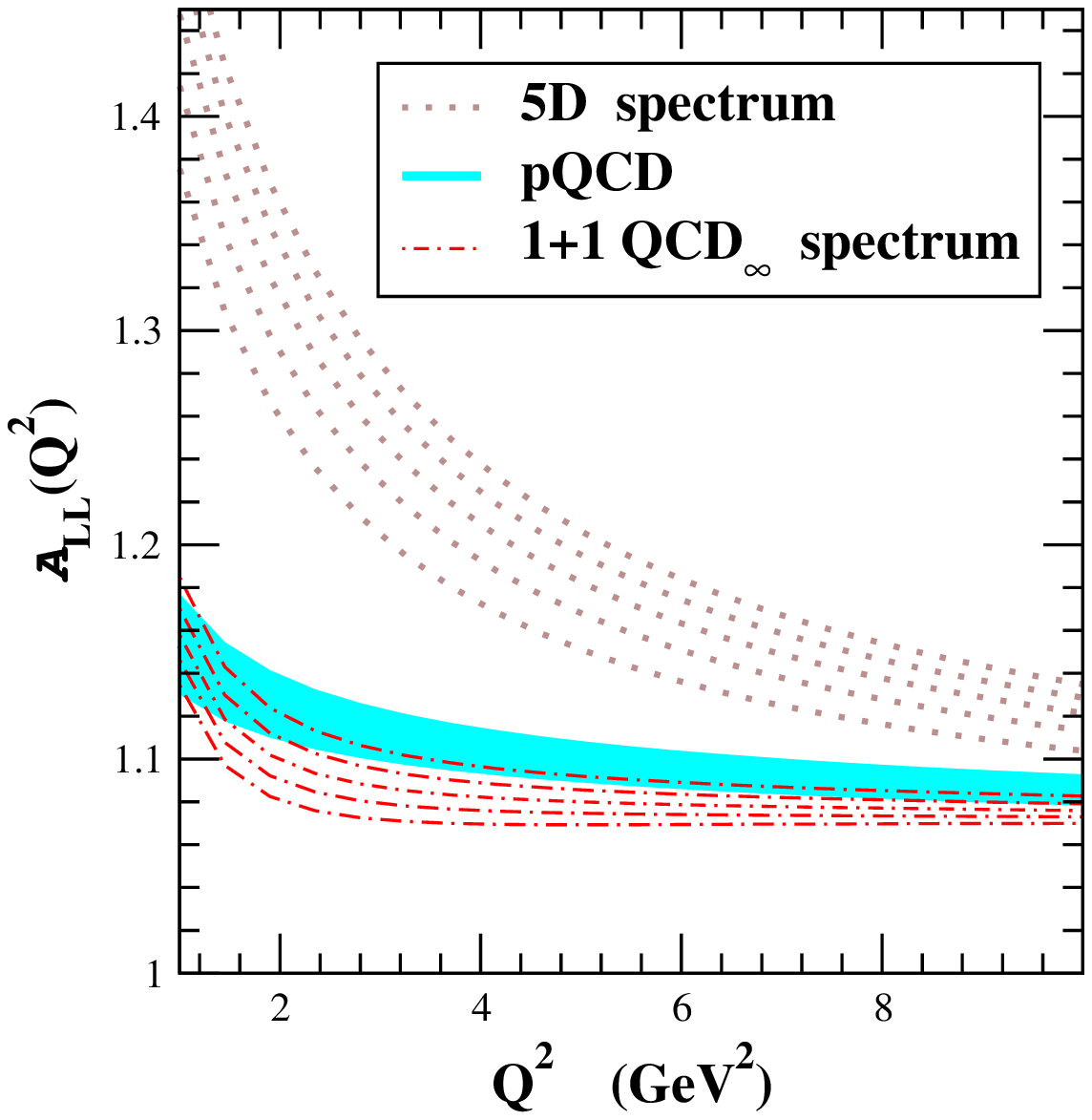}
\caption{$V-A$ and $V+A$ Adler functions in R$\chi T^{(\infty)}$ up to 
$\cO(\alpha_s)$  and within MHA. 
They are compared to the results from the OPE and pQCD at that order.}
\label{fig.adlerRT}
\end{center}
\end{figure}
\section{Space-like region and local duality:   resonance theory   vs. OPE}
\tab
In this section we show how the resonance description matches the 
OPE in the euclidean domain.  
The 1+1~$QCD_\infty$ model and the 5D-spectrum  are considered   
although the first one  yields the closest results to phenomenology.   
One can see on right-hand-side of 
Fig.~\eqn{fig.adlerRT}  the comparison between  the 
${V+A}$~Adler functions
coming from the OPE  and the large $N_C$ resonance theory. 
We have plotted  
$Q^6 \, \Pi_{_{LR}}(-Q^2)$ on the left-hand-side 
of Fig.~\eqn{fig.adlerRT} in order to show the short distance $1/Q^6$ 
behaviour in a more transparent way. 
For the OPE and pQCD, we have taken the values from  Section~3,  
with $V-A$ operators up to dimension twelve. 
The MHA expression for $\Pi_{_{LR}}(-Q^2)$ is plotted for the values     
$\widetilde{F}_\rho/\sqrt{2}=\widetilde{F}_{a_1}=F$ and 
$\widetilde{M}_{a_1}=\sqrt{2}\, \widetilde{M}_\rho$.

Although the $\cO(\alpha_s^2)$ uncertainties are still sizable and 
subleading $1/N_C$ effects need to be analysed,  the euclidean  
amplitudes are already sensitive to the different structures of the spectrum.
The 1+1~$QCD_\infty$  
spectrum recovers  quite accurately the OPE and
pQCD. The amplitude coming from the 
five dimensional model does not agree for the choice of parameters 
in this work, pointing out the necessity of a further 
tuning of  $M_{\rho'}$ and $M_{a_1'}$ in order to match the OPE at
short distances.  Finally, one can  see that, although MHA produces a
$V-A$~correlator of the right order of magnitude and $1/Q^6$ dependence,  
it does not completely  agree the OPE determination. 
A better agreement is found if the short distance constraints  of the
form-factors are relaxed and only the two WSR are kept~\cite{SRFriot}.   
The inclusion in the MHA of the next resonance multiplet, corresponding to 
the $\rho(1450)$,  
improves the $\rho+a_1$ calculation 
and increases the accuracy of the determination~\cite{SRFriot}.

Another way to study the properties of the amplitude in the euclidean region is
through its moments at a fixed energy. 
In Section 3, we showed how the components $\mB^{^{(k)}}(Q^2)$ 
suppress the dependence of the dispersive integral 
on the spectral function $\frac{1}{\pi}$Im$\Pi(t)$
around $t\sim Q^2$. The $\mB^{^{(k)}}(Q^2)$ 
of the physical amplitudes oscillate as $k$ grows  and eventually damp off.  
The OPE is able to follow the
physical $\mB^{^{(k)}}(Q^2)$ for the first values of $k$ but, for a fixed $Q^2$, 
it breaks down above some $k_{_{OPE-lim.}}$. Increasing the energy
or  considering higher dimension operators allows increasing 
the range of validity of the OPE, $k_{_{OPE-lim}}$.

The large $N_C$ resonance framework reaches 
a much further range, reproducing at the same time the good OPE results  
for $k\lsim k_{_{OPE-lim.}}$. In
Figs.~\eqn{fig.momRChTLR}~and~\eqn{fig.momRChTLL}, 
we can see the corresponding
components from the experimental ansatz in the $V+A$ and $V-A$ channels compared
with the resonance description from the  1+1~$QCD_\infty$ 
spectrum and the 5D spectrum, being  again 
the first one clearly favoured. The situation is  manifest
in the $V+A$ correlator, where the oscillation in the 5D-model is much wider 
and one even finds  opposite signs for the components of the two resonances 
models.  
In the $V-A$ case, 
it is also compared with the MHA expression, 
with ${\widetilde{F}_\rho/\sqrt{2}=\widetilde{F}_{a_1}=F}$ and 
$\widetilde{M}_{a_1}=\sqrt{2}\, \widetilde{M}_\rho$.
In general, the $\mB^{^{(k)}}(Q^2)$  in the large $N_C$ resonance theory  
follow a natural oscillating behaviour. It    
does not vanish for high values of $k$ since 
the states own zero width, whereas  the physical components 
eventually damp off due to the  non-zero widths of the hadronic states. 
Nonetheless, through a careful estimate of the first components 
and their uncertainties, one is already able to distinguish
between the different models for the spectrum and their couplings.

\begin{figure}[h!]
\begin{center}
\includegraphics[width=7.5cm,clip]{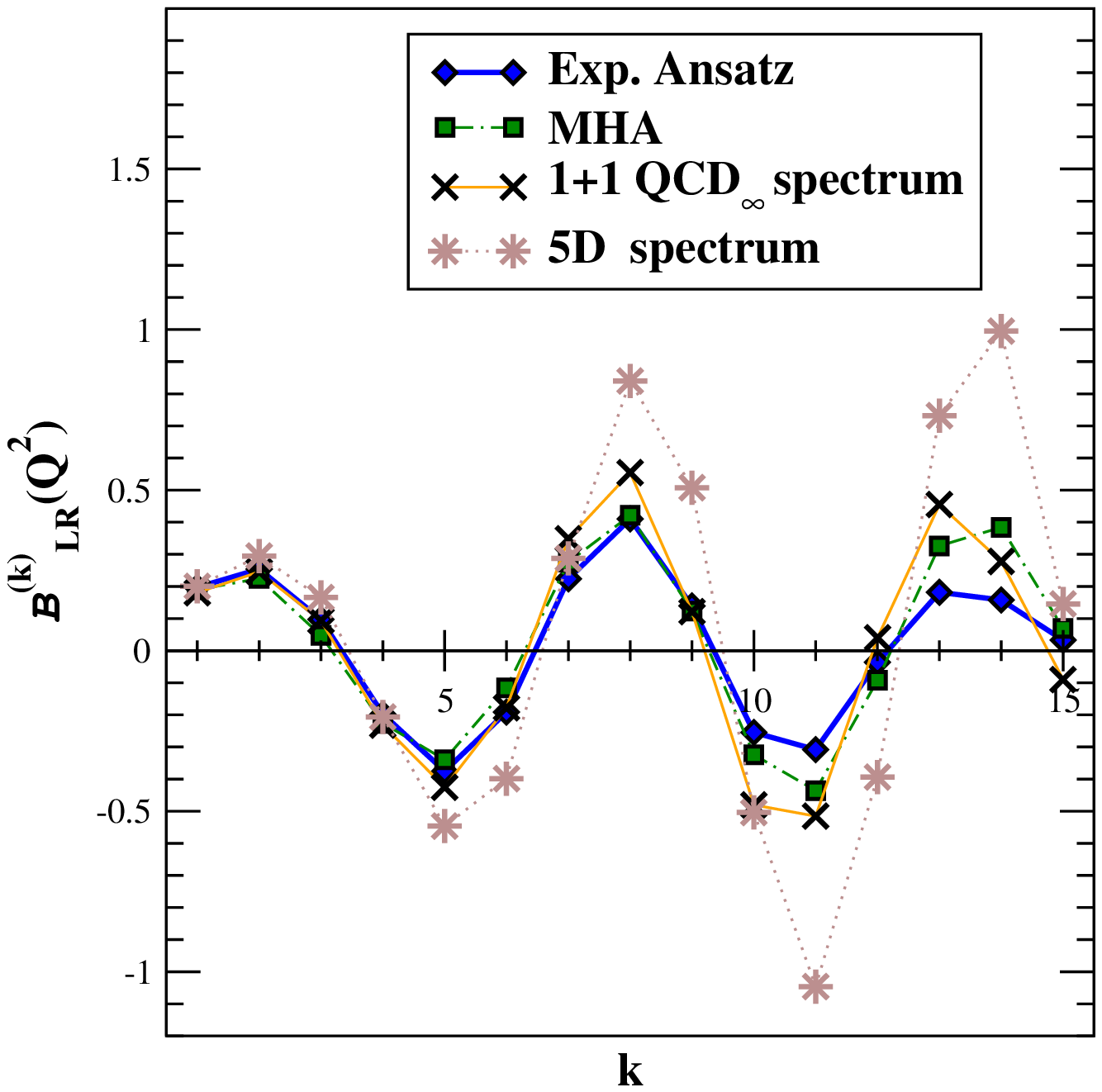}
\hspace*{0.5cm}
\includegraphics[width=7.5cm,clip]{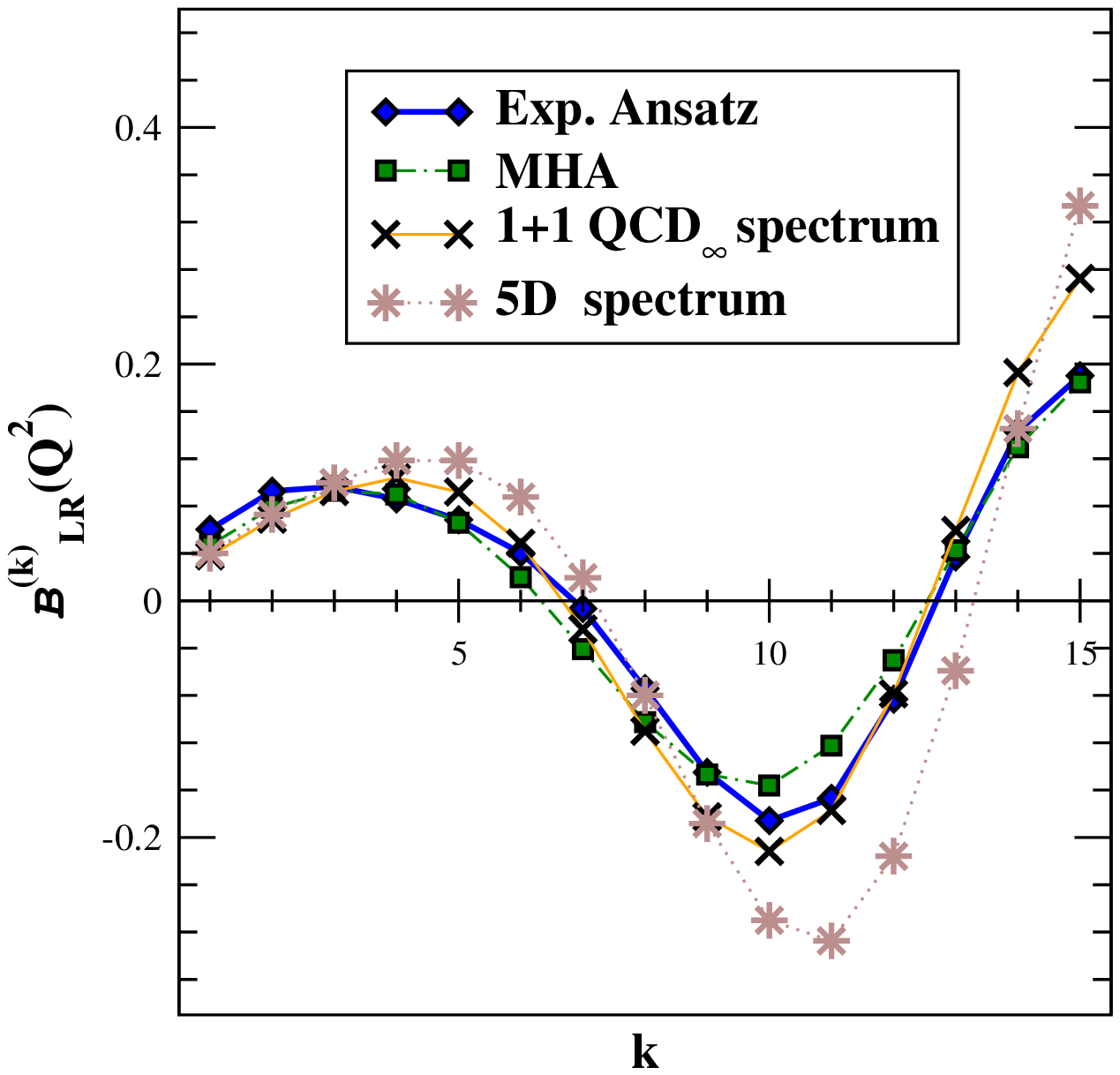}
\caption{Components $\mB^{^{(k)}}_{_{LR}}(Q^2)$ of 
 the $V-A$ correlator for $Q^2=2$~GeV$^2$ and $Q^2=10$~GeV$^2$. The 
experimental ansatz is compared to the  
large $N_C$ resonance theory determinations and MHA.}
\label{fig.momRChTLR}
\end{center}
\end{figure}
\begin{figure}[h!]
\begin{center}
\includegraphics[width=8cm,clip]{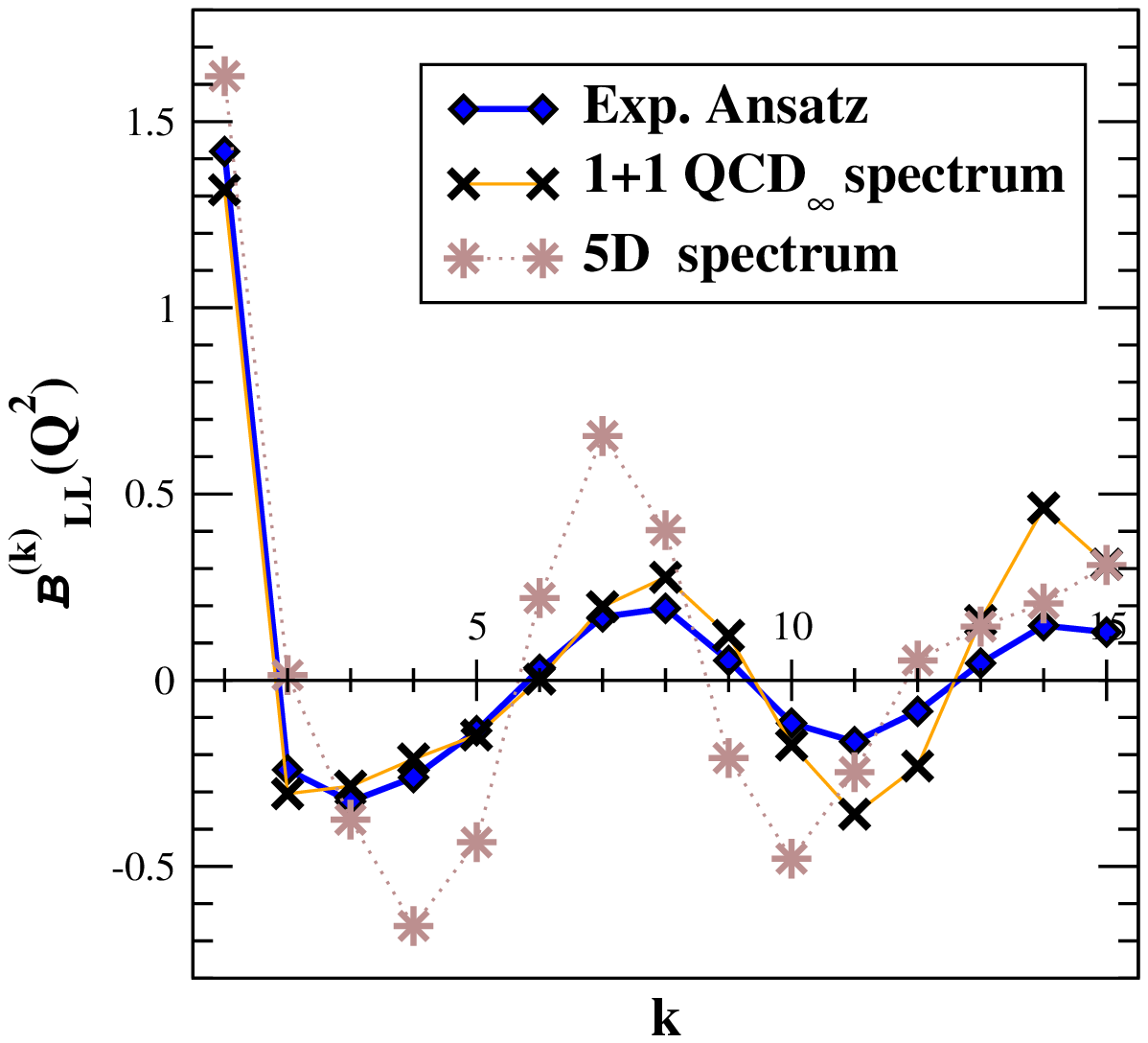}
\hspace*{0.5cm}
\includegraphics[width=7.5cm,clip]{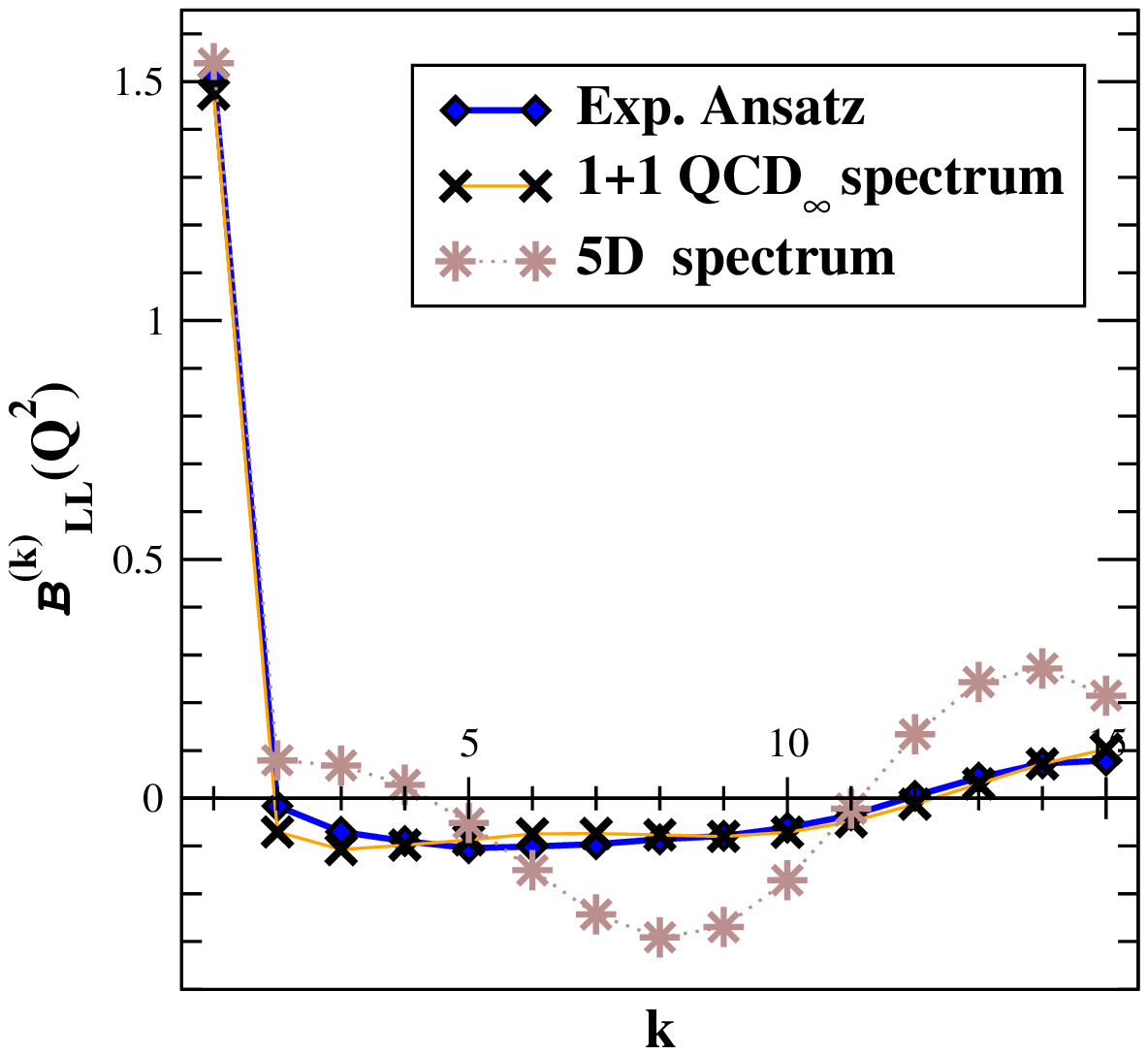}
\caption{Components $\mB^{^{(k)}}_{_{LL}}(Q^2)$ of 
 the $V+A$ correlator for $Q^2=2$~GeV$^2$ and $Q^2=10$~GeV$^2$. 
The  experimental ansatz is  compared to 
the resonance expression at large $N_C$.}
\label{fig.momRChTLL}
\end{center}
\end{figure}

\section{Time-like region and averaged spectral functions}
\tab
The QCD interaction at $N_C\to\infty$  is
so strong that distorts the smooth pQCD spectral functions into a series of
narrow-width resonances. However,  the higher order corrections in $1/N_C$
provide the hadronic states with non-zero widths, so the physical amplitudes
become again smooth. 
However, it is interesting to extract information from the experiment 
already at leading order 
in $1/N_C$, so  we will consider the averaged spectral functions  
introduced before in Section~2.3. Instead of working
with the spectral function $\frac{1}{\pi}$Im$\Pi(t)$ 
at a given positive 
energy $t$, it is rather convenient to employ the 
function $\sigma_z(x)$,  
with the mapping $x=\left(\frac{t-z}{t+z}\right)$.  
$\sigma_z(x)$ is then averaged by the  distribution 
$\xi_a(x)$, with 
dispersion  $(\Delta x)^2_{\xi_a}=\frac{1}{a+3}$. The average 
$\frac{1}{\pi}$Im$\overline{\Pi}(z)^{^{\xi_a}}=\bra
\sigma_z\ket_{_{\xi_a}}$ is essentially provided by the spectral function 
$\sigma_z(x)$ in the interval 
$|x|\lsim (\Delta x)_{\xi_a}$, i.e., by 
$\frac{1}{\pi}$Im$\Pi(t)$ in the interval 
$|t-z|\lsim 2 \, z \, (\Delta x)_{\xi_a}$.   
The interesting peculiarity here is that the averaged correlator 
only depends on
the first $a+1$ moments. As we saw in the former section, 
the first components $\mB^{^{k}}(z)$ are mainly
governed by the large $N_C$ 
resonance contributions, being the higher orders in $k$ ruled by the
subleading effects in $1/N_C$.

One may compare the
experimental and theoretical $\bra \sigma_z\ket_{_{\xi_a}}$   
pondered by distributions  
$\xi_a(x)$ with $a$ small enough.  
This procedure, suppresses the influence of the data away of the center of the
distribution, $t=z$. However, if $z$ lays near  a resonance peak and  
$a$ is taken too large,   then the distribution 
$\xi_a(x)$ may   result  too narrow and one has to consider   
the physical  non-zero width of the hadronic state.

This procedure could be useful for the analysis of the spectral functions
from $\tau$--decays, where the data reaches just $t\lsim 3$~GeV$^2$.   
At low enough  energies  (for $z\lsim 1.5$~GeV$^2$ 
and taking $\xi_{a}(x)$ with $a\sim 8$),  
the region  where
there are no experimental data remains on the tail of the distribution 
$\xi_{a}(x)$, yielding suppressed corrections.  
The range of energies around the $\rho$ resonance is specially interesting, 
where very accurate data~\cite{Aleph,CLEO,OPAL,Amendolia,Novo2000} 
and an exhaustive theoretical work 
on the resonance parameters 
already  exists~\cite{therole,spin1fields,
quantumloops,Sakurai,anchura,GP:97,PP:01,ND}.

On Fig.~\eqn{fig.promediaLOW}, 
we can see 
how  R$\chi T^{^{(\infty)}}$  with the 1+1~$QCD_\infty$ spectrum 
matches pQCD 
for $z\gsim 5$~GeV$^2$, where the slight  discrepancy below 
demands a deeper analysis of the contributions from the non-zero 
dimension operators   in the OPE.    Actually, the 
fine agreement  with the experimental ansatz around the $\rho(770)$ peak 
($z\lsim 1$~GeV$^2$) points out again that the values  
in Table.~\eqn{tab.predictions}  for the $\rho$ and $a_1$ parameters  lie 
on the proper range.   
For the set of parameters considered in this work,  
The 5D~spectrum shows again a slower convergence and large discrepancies 
with the experimental ansatz at $z\lsim 1$~GeV$^2$.      
MHA is also shown in the $V-A$ case   with the couplings 
$\widetilde{F}_\rho/\sqrt{2}= \widetilde{F}_{a_1}=F$ and masses
$\widetilde{M}_{a_1}=\sqrt{2}\,\widetilde{M}_\rho$. At $z\lsim 1$~GeV$^2$, 
it provides an acceptable approximation of the experimental ansatz averaged
amplitude.

For energies beyond $1.5$~GeV$^2$ 
the tail of the distribution $\xi_a(x)$ reaches also the resonances with 
squared masses around $2-3$~GeV$^2$ and both MHA and the experimental ansatz 
with just $\tau$ data loose reliability.  This calls out for the inclusion  of
higher energy $e^+ e^-$ data. A more exhaustive error analysis is also needed 
in order  to yield a precise and
accurate determinations of the $\rho(770)$ and $a_1(1260)$ parameters 
at leading order in $1/N_C$.

\begin{figure}[h!]
\begin{center}
\includegraphics[width=7.5cm,clip]{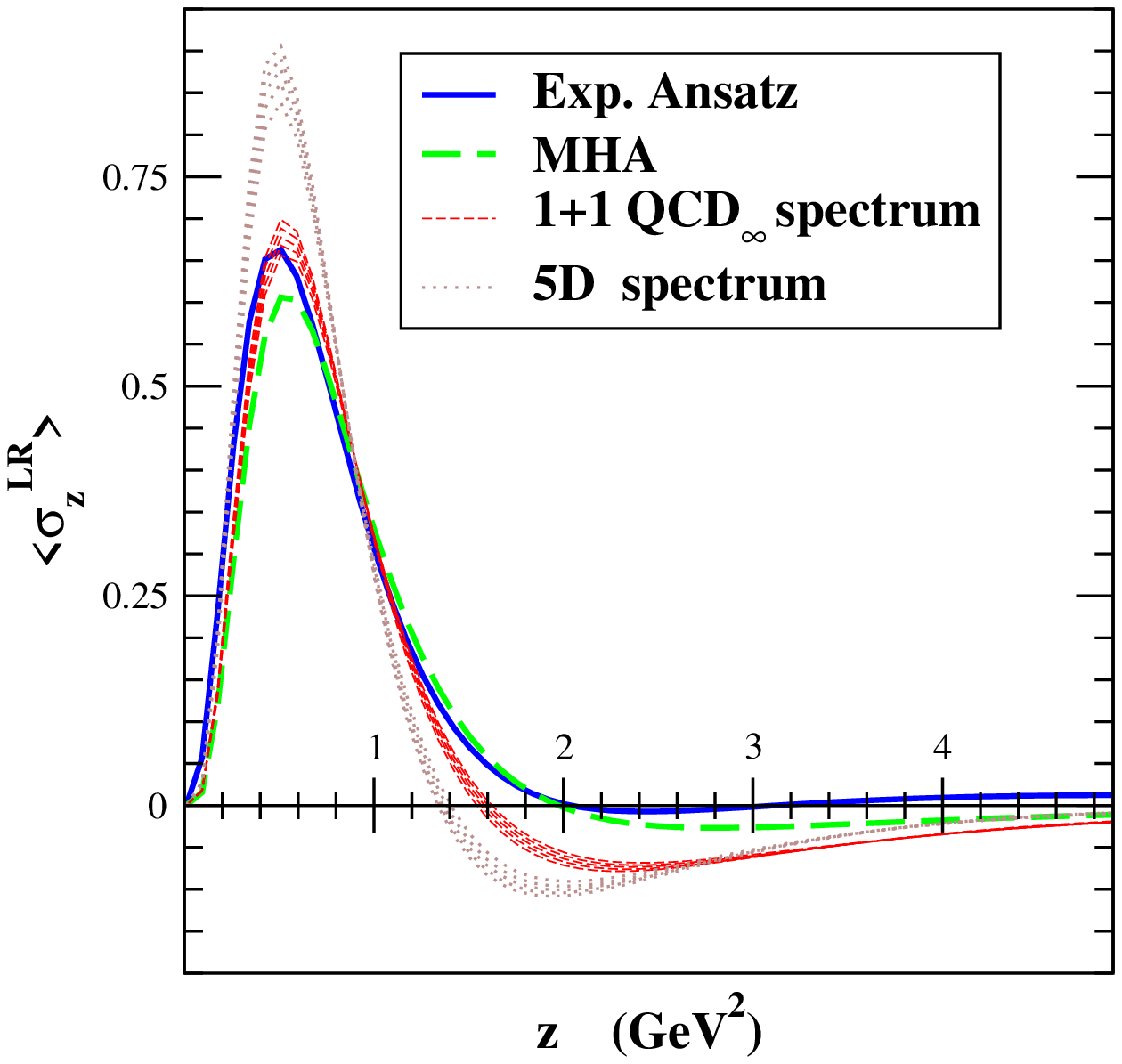}
\includegraphics[width=7cm,clip]{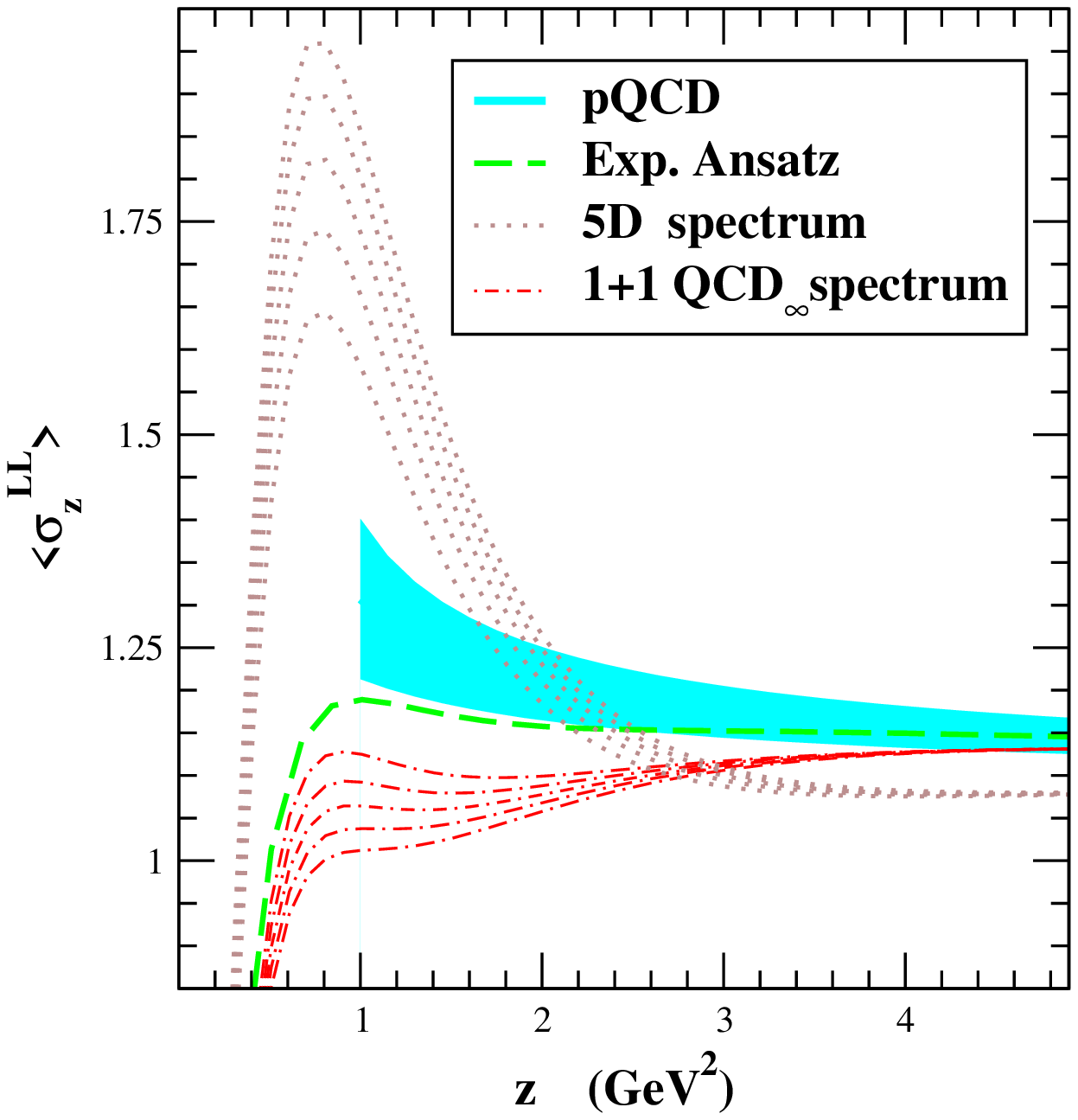}
\caption{Averaged spectral functions  
$\frac{1}{\pi}$Im$\overline{\Pi}(z)^{^{\xi_a}}=\bra \sigma_{z}\ket_{_{\xi_a}} $
for $a=8$.
}
\label{fig.promediaLOW}
\end{center}
\end{figure}

\section{Conclusions}
\tab 
This paper explores the large $N_C$ description of QCD   through  
a theory with an infinite number of hadronic states. 
We show how it is possible to recover the OPE up to order $\alpha_s$ in a
systematic way, obtaining the corresponding 
$\alpha_s(Q^2)$ running in $\Pi_{_{LL}}(-Q^2)$. 
Producing the precise anomalous dimensions in the
condensates  is still a hard task and  
the study at $\cO(\alpha_s^2)$ is relegated to next works.

The present analysis is based on three foundations. 
First, it is  assumed that, in the 
deep euclidean domain,  the QCD amplitudes and
the different moments  can be  fairly described through the OPE.  
Second, the amplitudes in the large $N_C$ limit 
accept a description in terms of an infinite exchange of
narrow resonances that embodies the OPE; the  correlators are meromorphic
functions determined by the positions and residues of an infinite set of 
real poles. Third, in order to handle 
the infinite tower of hadronic states 
one has  to assume a given asymptotic structure $M_n^2=f(n)$ 
for the spectrum at high energies and a smooth behaviour on $n$ for 
the resonance couplings. 
Although we still lack a definitive theoretical explanation   
of the  meson masses, we can nevertheless test (and eventually discard) 
some of the models currently considered. 1+1~$QCD_\infty$ ($M_n^2\sim n$) 
and  the 5D theories ($M_n^2\sim n^2$)  seem to own the most favoured 
spectrums by the phenomenology, being the first one in closer agreement.

The structure of the mass spectrum imposes constraints on 
the resonance couplings if we want to recover 
the pQCD expression for the $V+A$ correlator:
$$
F_n^2\,\,\,\, = \,\,\,\, \delta M_n^2\,\, \cdot\,\, 
\left\{
\Frac{1}{\pi}\mbox{Im}\Pi_{_{LL}}(M_n^2)^{^{pQCD}}\,\,\, 
+ \,\,\,\cO\left(\Frac{1}{M_n^2}\right)\right\} \, .
$$ 
The lightest mass states  cannot be fixed through this procedure
since pQCD breaks down and the asymptotic dependence 
of $M_n^2$ may suffer large
$\alpha_s$ corrections. 
The perturbative sub-series containing  the high mass resonances  is  
fixed by pQCD and our knowledge on $M_n^2$. 
As  the pQCD $V+A$ correlator is recovered, the $V-A$  amplitude 
shows an OPE structure of the form  $\Pi_{_{LR}}(-Q^2)\sim \Frac{1}{Q^{2m}}$.  
A final matching is
performed by fixing the values of the lightest  resonance parameters 
in order to ensure that 
$\Pi_{_{LR}}(-Q^2)$ goes like $1/Q^6$ and that the first
condensate in $\Pi_{_{LL}}(-Q^2)$  has dimension four. 
Although the deviations of $F_\rho$ and $F_{a_1}$ 
with respect to the asymptotic values of  $F_n$ are not large, 
they are essential to match $R\chi T^{(\infty)}$    
and the OPE at order $\alpha_s$. The 1+1~$QCD_\infty$ mass spectrum 
$M_{n\geq 3}^2=\Lambda^2+n\, \delta \Lambda^2$  yields the couplings 
$$ 
F_\rho\,=\, 139\pm 3 \mbox{ MeV }\, , 
\qquad  F_{a_1}\,=\, 138\pm 3 \mbox{ MeV }\, , 
\qquad  M_{a_1}\, = \, 1180^{+17}_{-19} \mbox{ MeV } \, .
$$
The prediction for $\bra\cO_{_{(6)}}^{^{LR}}\ket$  agrees  
standard OPE determinations but 
there are still large uncertainties in the $V+A$ sector where a deeper
$\cO(\alpha_s^2)$ study needs to be done.  
Considering pQCD just at $\cO(\alpha_s^0)$ 
produces small modifications on the resonance parameters, what ensures the
stability of the whole procedure under $\alpha_s$ corrections.

In order to isolate the peculiarities of  
the quark-gluon and resonance pictures we develop  
and adapt a set of sum rules that are specially sensitive 
to the resonance features.   
In $R\chi T^{^{(\infty)}}$,  the combinations of moments  
$\mB^{^{(k)}}(Q^2)=\sum_{l=1}^k M_{k,l}^{^{\mB\mA}}\, \mA^{^{(l)}}(Q^2)$ 
provided by the Legendre polynomials  
expose  a characteristic oscillation  which is also
experimentally observed. The different models for the spectrum
produce clearly different patterns of oscillation whose study 
can be used to discern the proper hadronical spectrum of QCD.  
The 1+1~$QCD_\infty$ model  reproduces the experimental 
estimate of the $\mB^{^{(k)}}(Q^2)$ up to $k\sim 10$ even for very low energies.  
For the $V-A$ amplitude, the simple MHA expression   
provides a description as good as the one  from 
the full $R\chi T^{^{(\infty)}}$, 
pointing out the fact that it encodes an important portion of the 
QCD information.

Another relevant technique 
for the determination of the resonance parameters is
the employment of averaging distributions peaked around some energy 
$t\sim z$.    
Former works on Gaussian sum rules illustrate the averaging  
procedure~\cite{DRgaussian}.  It  allows 
to isolate the data around a given energy region in order to study 
the parameters of a resonance laying in that interval or to perform checks of
duality violations.  In the present paper we have considered
the family of  
distribution $\xi_a(x)$ with a power--like dependence, which tend to
the Dirac delta when $a\to\infty$. It is specially useful for the 
$\rho(770)$ region where plenty of accurate experiments 
exist~\cite{Aleph,CLEO,OPAL,Novo2000,Amendolia,PIVV} 
and the lack of data  at high energies has little influence.  
A very good agreement between the experiment and the 1+1~$QCD_\infty$ model is
found for the averaged amplitudes 
$\frac{1}{\pi}$Im$\overline{\Pi}(z)^{^{\xi_a}}$ at 
$z\lsim 1$~GeV$^2$. The MHA provides as well a reasonable approximation.

Through the splitting of the resonance series into perturbative and
non-perturbative sub-series,  we have understood how 
the truncated resonance theories --and more exactly MHA--   
connect the full large $N_C$ theory.  
When  the contribution to the condensates coming  
from the perturbative sub-series is neglected,      
one introduces an error in the light resonance parameters.    
The terminology Minimal Hadronical ``\,Ansatz''\, 
is then misleading since what
we perform is simply an approximation with a  
 clearly defined  uncertainty. 
For the different  models considered in this work,  
the contributions  
$\Delta \bra \cO_{_{(2m)}}^{^{LR}}\ket^{^{pert.}}$ to the
condensates  are 
of the order of usual hadronical parameters,  
$F^2\,(M_\rho^2)^{m-1}$. The MHA parameters 
turn out to be of the same order of magnitude as in the full 
large $N_C$ resonance theory,  although  
even factor two discrepancies may arise, e.g., the value of the coupling 
$F_{a_1}$ or the condensate  $\bra \cO_{_{(6)}}^{^{LR}}\ket$. 
This kind of considerations should be taken seriously when
estimating the uncertainties in the MHA determinations.   
The inclusion  
of the next multiplet, $\rho(1450)$, can  improve
the approximation in a stable
way so the values of the condensates do no change drastically from 
MHA to MHA+$\rho'$~\cite{SRFriot}, finding  
a smooth convergence from MHA to $R\chi T^{^{(\infty)}}$.

To end with, I would like 
to comment some of the parallel work-lines arisen at the
study of $R\chi T^{^{(\infty)}}$.  As it was commented before, the
$\mB^{^{(k)}}(Q^2)$ sum rules may be also used to fix the values of the
condensates. 
They  enhance 
both the low and the high energy regions 
where we have, respectively,  experimental data and accurate theoretical
descriptions.     
Likewise,  by averaging the correlators  through  the distributions
$\xi_a(x)$, one is  able to remove the influence of the condensates up to the
desired dimension, leaving the pQCD contribution unchanged up to 
$\cO(\alpha_s)$.  
An analysis of the average  
$\frac{1}{\pi}$Im$\overline{\Pi}_{_{VV}}(z)^{^{\xi_a}}$ 
from the $e^+ e^-$ data would lead to an alternative  
determination of $\alpha_s$ and high dimension condensates.

The study under the OPE perspective has shown that,   
as far as the correlator shows a $1/Q^{2m}$  power--like structure in the deep
euclidean, all the
different functions present  a similar OPE--like structure: 
$$
\Pi(-Q^2)\, = \, \sum_m \, \Frac{\bra\cO_{_{(2m)}}\ket}{Q^{2m}} 
\qquad \Longrightarrow \qquad 
\left\{
\ba{rl}
\mA^{^{(n)}}(Q^2)\, =& \, \displaystyle{\sum_m} \, \Frac{a_{_{(n,2m)}}}{Q^{2m}} \, , 
\\
\mB^{^{(k)}}(Q^2)\, =& \, \displaystyle{\sum_m} \, \Frac{b_{_{(k,2m)}}}{Q^{2m}} \, ,
\\
\Frac{1}{\pi}\mbox{Im}\Pi(z) \, = & \, 
\displaystyle{\sum_m} \, \Frac{\frac{1}{\pi}\mbox{Im}\Pi_{_{(2m)}}}{z^{m}} \, ,
\\
\\
\Frac{1}{\pi}\mbox{Im}\overline{\Pi}(z)^{^{\xi_a}} \, = & \, 
\displaystyle{\sum_m} \, 
\Frac{\frac{1}{\pi}\mbox{Im}\overline{\Pi}^{^{\xi_a}}_{_{(2m)}}}{z^m} \, .
\ea
\right. 
$$
Up to $\cO(\alpha_s)$,  the averaged spectral function   
$\frac{1}{\pi}$Im$\overline{\Pi}(z)^{^{pQCD,\, \xi_a}}$  
is  just equal to the pQCD spectral function for any distribution $\xi_a(x)$. 
Possible 
duality violating terms in the QCD amplitudes could be eventually analysed
in a similar way.

All these techniques and considerations can be applied to other Green functions
(scalar correlators, ...) 
and observables (form-factors,...). The analysis of
exclusive channels would allow the extraction of the    
parameters of the  resonance lagrangian  at leading-order in $1/N_C$ relevant
for the different QCD matrix elements.   
These considerations are of particular interest 
in the case of more  complicated 
Green-functions, e.g. three-point Green-functions,  where     
the logarithmic behaviours are also originated by the infinite
exchange of resonances through the different external legs.

{\large \bf Acknowledgments}

I would like to thank J. Stern, J. Hirns and V. Sanz for their careful reading
of the manuscript. I also want to acknowledge the interesting 
discussions and useful criticisms from 
P.D. Ru\'\i z -Femen\'\i a, O. Cat\`a and S. Friot, and J. Portol\'es' aid with
the experimental data.  This work was  supported by    
EU~RTN~Contract~CT2002-0311.

\end{document}